\documentclass[aps,prc,showkeys,nofootinbib]{revtex4}

\usepackage{color} % Пакет для использования цвета для раскраски текста - см. описание
\usepackage{graphicx}% Include figure files
\usepackage{dcolumn}% Align table columns on decimal point
\usepackage{amsthm}
\usepackage{bm}% bold math

\usepackage{isotope}% isotope editing
\usepackage{braket}% Macros for Dirac bra–ket notation and sets {|}, Donald Arseneau: https://ctan.org/pkg/braket
\usepackage{physics}% physics packet, Sergio C. de la Barrera, 2012 (from ctan.org)

\newcommand{\sigmabf}{\mbox{\boldmath $\sigma$}}

\newcommand{\rhobf}{\mbox{\boldmath $\rho$}}

\begin{document}
%-----------------------------------------------------------------------------------------------------------------------

%-----------------------------------------------------------------------------------------------------------------------
\title{Enhancement of incoherent bremsstrahlung in proton-nucleus scattering in the $\Delta$-resonance energy region
% \vspace{2mm}
% \newline
% (2) Enhancement (Усиление) of incoherent $\gamma$-production in proton-nucleus scattering in the $\Delta$-resonance region
% \vspace{2mm}
% \newline
% (3) Anomalous magnetic moments of nucleons in nuclear matter and phenomenon of suppression of incoherent bremsstrahlung
}
% \author{Peng-Ming~Zhang$^{1}$}\email{zhangpm5@mail.sysu.edu.cn}%
% \author{Li-Ping~Zou$^{1}$}\email{zoulp@impcas.ac.cn} %
\author{Sergei~P.~Maydanyuk}\email{maidan@kinr.kiev.ua}%
%
% \affiliation{$^{(1)}$School of Physics and Astronomy, Sun Yat-sen University, Zhuhai, China}
% \affiliation{$^{(2)}$Institute of Modern Physics, Chinese Academy of Sciences, Lanzhou, 730000, China}
\affiliation{Institute for Nuclear Research, National Academy of Sciences of Ukraine, Kyiv, 03680, Ukraine}

\date{\small\today}
%-----------------------------------------------------------------------------------------------------------------------

%-----------------------------------------------------------------------------------------------------------------------
\begin{abstract}
We investigate emission of the bremsstrahlung photons in the scattering of protons off nuclei at the $\Delta$-resonance energy region.
% A focus is directed on question of how much the bremsstrahlung spectrum is changed after transition of one nucleon in nucleus to $\Delta$-resonance.
% We improve previous bremsstrahlung model,
Including properties of $\Delta$-resonance in the nucleus-target to the bremsstrahlung model,
we % estimate the spectra and
find the following.
%
%
% In these calculations we include coherent and incoherent bremsstrahlung contributions,
% we test this formalism and calculations for the \isotope[197]{Au} nucleus at $E_{\rm p}=190$~MeV on the experimental data.
%
(1) Ratio between incoherent emission and coherent emission is about $10^{6}$--$10^{7}$
for $p + \isotope[197]{Au}$ (without $\Delta$-resonance)
% at $E_{\rm p}=190$~MeV,
at energy of proton beam $E_{\rm p}$ of 190~MeV,
where the calculated full bremsstrahlung spectrum is in good agreement with experimental data.
This confirms importance of incoherent processes in study of $\Delta$-resonances in this reaction, which have never been studied yet.
% We calculate the bremsstrahlung spectra (including coherent and incoherent contributions) in the scattering of protons on the \isotope[12]{C}, \isotope[40]{Ca}, \isotope[208]{Pb} nuclei
% %v at energy of proton beam $E_{\rm p}$ of 800~MeV.
% at $E_{\rm p}=800$~MeV.
%
% This confirms our supposition that inclusion of incoherent processes in study of $\Delta$-resonance in proton-nucleus scattering is important (as can change picture of the studied process).
% This aspect has never been studied before, so it is one of aims of this paper.
%
% We show that increasing of energy of protons beam in nuclear scattering increases intensity of bremsstrahlung emission [see Fig.~\ref{fig.1}~(b) for comparison,
% test with experimental data].
% Bremsstrahlung emission is larger for heavier nuclei, difference between the spectra for light and heavy nuclei is essential
% (see Fig.~\ref{fig.3} for \isotope[12]{C}, \isotope[40]{Ca}, \isotope[197]{Au}, \isotope[208]{Pb},
% difference is about $10^{5}$ times between the spectra for \isotope[208]{Pb} and \isotope[12]{C} at $E_{\rm p} = 800$~MeV).
%
We estimate coherent and incoherent contributions, electric and magnetic contributions, full bremsstrahlung spectra for
% different nuclei and energies of proton beam,
the scattering of protons on the \isotope[12]{C}, \isotope[40]{Ca}, \isotope[208]{Pb} nuclei
%v at energy of proton beam $E_{\rm p}$ of 800~MeV.
at $E_{\rm p}=800$~MeV,
we find conditions for the most intensive bremsstrahlung emission.
% In the coherent bremsstrahlung, the magnetic emission
% is almost the same as electric emission [$\sigma_{\rm mag}^{\rm (coh)} / \sigma_{\rm el}^{\rm (coh)} = 3.3213$ for \isotope[197]{Au} at $E_{\rm p} = 190$~MeV for 10--180~MeV of photons;
% see Eq.~(\ref\label{eq.analysis.1.1})].
% In the incoherent bremsstrahlung, role of background emission based on $M_{k}$ is a little larger than magnetic contribution based on $M_{\Delta M}$
% ($\sigma_{\rm background}^{\rm (incoh)} / \sigma_{\rm mag}^{\rm (incoh)} = 4.04$ for \isotope[197]{Au} at $E_{\rm p} = 190$~MeV for 10--180~MeV of photons).
% Ratio between incoherent emission and coherent emission is increased at increasing of energy of photon emitted
% (see Fig.~\ref{fig.2}, for \isotope[197]{Au} at $E_{\rm p}=190$~MeV).
%
(2) Transition $p\,N \to \Delta^{+} N$
%bfrom proton of nucleus-target to $\Delta^{+}$-resonance ($p\,N \to \Delta^{+} N$) in the proton-nucleus scattering
in the nucleus-target
% we use scheme of the coherent processes described in Ref.~\cite{Gil_Oset.1998.PLB.v416}.
reinforces emission of bremsstrahlung photons in that reaction at $E_{\rm p}=800$~MeV.
% with one $\Delta$-resonance in nucleus-target
% (in comparison with the normal nucleus)
% (see Fig.~\ref{fig.4} for \isotope[12]{C}, \isotope[12][\Delta]{C} and \isotope[40]{Ca}, \isotope[40][\Delta]{Ca} at $E_{\rm p}=800$~MeV).
% This result is indication of reinforcement of bremsstrahlung in proton-nucleus scattering.
Difference between
the spectra for normal nuclei and nuclei with included $\Delta$-resonance is larger for more light nuclei,
but the spectra are larger for heavier nuclei.
% (for \isotope[12]{C}, \isotope[40]{Ca} in comparison with \isotope[12][\Delta]{C}, \isotope[40][\Delta]{Ca}).
% We provide corresponding calculations for \isotope[12]{C}, \isotope[40]{Ca} in comparison with \isotope[12][\Delta]{C}, \isotope[40][\Delta]{Ca}.
%
% In order to find nuclei with maximal reinforcement of bremsstrahlung due to transition $p\,N \to \Delta^{+} N$, we performed analysis and obtained condition~(\ref{eq.resultingformulas.2}) determining ratio between protons and neutrons for the normal nucleus.
% On the example for \isotope[10][\Delta^{+}]{C}, we show
% that difference between the spectra for \isotope[10]{C} and \isotope[10][\Delta^{+}]{C} is larger essentially
% than difference between the spectra for \isotope[12]{C} and \isotope[12][\Delta^{+}]{C} (see Fig.~\ref{fig.5}).
% However, stable nuclei (like \isotope[12]{C}, \isotope[40]{Ca}, \isotope[208]{Pb}) do not satisfy to that condition~(\ref{eq.resultingformulas.2}).
% As parameter $\bar{\mu}_{\rm pn}^{\rm (an)}$ has the highest influence on reinforcement of bremsstrahlung of such a type, it confirms importance of incoherent processes in this research
% (i.e., there is no enhancement of bremsstrahlung due to transition $p\,N \to \Delta^{+} N$, if incoherent processes are not included to the model).
%
(3) Taking into account shortly lived state of $\Delta$-resonance,
% it is formed in nucleus-target only during propagation of the scattered proton (from beam) through the space region of this nucleus.
% On such a basis, we calculate new bremsstrahlung spectra and find
% we will take into account only this space region of proton-nucleus system in calculation of matrix elements of emission and we will estimate the bremsstrahlung spectra.
we find that the spectrum with $\Delta$-resonance in the nucleus-target is essentially larger in the high energy photon region than the spectrum without this $\Delta$-resonance
(corresponding calculations for \isotope[12][\Delta]{C}, \isotope[40][\Delta]{Ca}, \isotope[208][\Delta]{Pb} in comparison with \isotope[12]{C}, \isotope[40]{Ca}, \isotope[208]{Pb} are provided).
Such an aspect is recommended for registration of $\Delta$-resonances in nuclei in possible future experiments.

% This effect has the same origin as phenomenon of destructive interference between emission from tunneling region and external region investigated for bremsstrahlung in the $\alpha$ decay.

% This property can be used for proposal for future experiments with measurements of photons,
% as tools to distinguish process of formation of $\Delta$-resonance in the nucleus-target.
%-----------------------------------------------------------------------------------------------------------------------

%-----------------------------------------------------------------------------------------------------------------------
\end{abstract}
%-----------------------------------------------------------------------------------------------------------------------

%-----------------------------------------------------------------------------------------------------------------------
% \textbf{PACS numbers:}
 % Multipole matrix elements (in electromagnetic transitions)

% 27.80.+w % A is greater than or equal to 190 and is less than or equal to 219 (properties of specific nuclei listed by mass ranges)
% 23.60.+e, % Alpha decay
% 23.70.+j, % Heavy-particle decay
% 25.70.-z, % Low and intermediate energy heavy-ion reactions
% 24.75.+i, % General properties of fission
% 25.85.Ca, % Spontaneous fission
% 25.70.Gh, % Compound nucleus


\maketitle
% *******************************************************************************************************************

% *******************************************************************************************************************
% \newpage
\section{Introduction
\label{sec.introduction}}

Non-nucleon degrees of freedom in nuclei were often studied in nuclear physics~\cite{Ahrens.1985.NPA,Gaarde.1991.ARNPS,Strokovskii.1993.PEPAN}.
% Here, baryon resonances, pions, quarks, gluons in nuclei are studied intensively.
Last years some attention has been focused on hadron excitation of $\Delta (1232)$-resonances in nuclear matter (for brevity, term $\Delta$-resonance in nuclei will be used in this paper)~\cite{Mukhin.1995.PhysUsp,Kondratyuk.1994.NPA}.
Proton, electron, photon and pion can be used as a probe to study such $\Delta$-resonances in nuclei.
But, the most easy way to study $\Delta$-resonances in nuclei is based on analysis of scattering of beams of pions on nuclei and photoinduced reactions~\cite{Nedorezov.2010.book}.
%
% However, interaction between virtual pion and one nucleon of nucleus-target in the proton-nucleus scattering can be useful
Last years the proton nucleus scattering has been used as a main reaction for this study,
where $\Delta$-resonances are formed in result of virtual pion interaction with one nucleon of nucleus-target~\cite{Igamkulov.2010.PEPAN}.
% such resonances are estimated from $p N \to \Delta N$ transition amplitude.
%
As it was studied in Ref.~\cite{Gil_Oset.1998.PLB.v416}, coherent photon production in the proton-nucleus scattering is another similar reaction
which can be used also for such a study.
%
% \vspace{1.0mm}
% \noindent
Interest to emission of photons is explained by that such photons can be measured and new information about physics of formation of $\Delta$-resonance in nuclear matter can be extracted.

\vspace{1.2mm}
However, there is another possibility to study $\Delta$-resonances in nuclei, that is to use analysis of bremsstrahlung photons which are also emitted during such complicated process of proton-nucleus scattering with formation of $\Delta$-resonances.
%
% \vspace{1.0mm}
% \noindent
% The bremsstrahlung emission of photons accompanying nuclear reactions is a traditional sector in nuclear physics, which has been causing much interest for a long time
% (see reviews~\cite{Pluiko.1987.PEPAN,Kamanin.1989.PEPAN}). % and books [3--5]).
% The bremsstrahlung emission of photons accompanying nuclear reactions provide rich independent information about the studied nuclear process.
Many effects, like dynamics of the nuclear process, interactions between nucleons, types of nuclear forces, structure of nuclei, quantum effects and anisotropy (deformations), etc. can be included in the model describing the bremsstrahlung emission
(for example, see
Refs.~\cite{Maydanyuk.2003.PTP,Maydanyuk.2006.EPJA,Maydanyuk.2008.EPJA,Maydanyuk.2008.MPLA} for general properties of bremsstrahlung in $\alpha$ decay,
% Maydanyuk.2010.PRC,Maydanyuk.2011.JPCS}
Ref.~\cite{Maydanyuk.2009.NPA} for extraction of information about deformation of nuclei in the $\alpha$ decay from analysis of experimental bremsstrahlung data,
%
% Refs.~\cite{Maydanyuk.2009.JPS} - DELETED,
%
Ref.~\cite{Maydanyuk.2011.JPG} for bremsstrahlung in the nuclear radioactivity with emission of protons,
Ref.~\cite{Maydanyuk.2010.PRC} for bremsstrahlung in the spontaneous fission of \isotope[252]{Cf},
Ref.~\cite{Maydanyuk.2011.JPCS} for bremsstrahlung in the ternary fission of \isotope[252]{Cf},
Ref.~\cite{Maydanyuk_Zhang_Zou.2018.PRC} for bremsstrahlung in the pion-nucleus scattering
from our research,
there are many investigations from other researchers).
Note perspectives on studying electromagnetic observables of light nuclei based on chiral effective field theory \cite{Pastore.2008.PRC}
(see also research~\cite{Eden.1996.PRC} for $pp$ bremsstrahlung).
Experimental measurements of such photons and their analysis provide necessary information about these aspects, model suitability can be therefore verified.
So, bremsstrahlung photons is an useful tool to investigate all above questions.
%
% \vspace{1.5mm}
% \noindent
This paper is devoted to study $\Delta$-resonances in nuclei by means of bremsstrahlung analysis.
% , with construction of needed model, corresponding calculations of the spectra with analysis.

% \vspace{1.2mm}
% Some investigations were performed in study of emission of photons produced due to transition $pN \to \Delta N$ in
% nucleus in the proton nucleus scattering [].
In research ~\cite{Gil_Oset.1998.PLB.v416} authors provided estimations of cross-sections, where coherent processes were included to the model and calculations.
%
% \vspace{1.0mm}
% \noindent
However, analysis of experimental study of bremsstrahlung in proton-nucleus scattering by TAPS collaboration~\cite{Goethem.2002.PRL} indicates that incoherent processes play an important role on the bremsstrahlung emission
($p + \isotope[197]{Au}$ at energy of proton beam of $E_{\rm p}=190$~MeV was studied).
There are different reasons to conclude that incoherent emission is more intensive than coherent one.
One of them is existence of so called ``plateau'' in the experimental data in Ref.~\cite{Goethem.2002.PRL} (in the middle part of spectrum).
That plateau can be explained if to add incoherent contribution to the formalism.
Without this incoherent contribution, the model has only coherent terms and gives spectrum with shape of logarithmic type (i.e., without plateau).
In Ref.~\cite{Maydanyuk_Zhang.2015.PRC} % which is the first paper about incoherent emission in our researches,
ratio between incoherent and coherent contributions was extracted on the basis of analysis of experimental data~\cite{Goethem.2002.PRL}.
It was concluded that incoherent bremsstrahlung is intensive (for heavy nuclei used in experiments).

In next research the unified formalism was constructed, % in Ref.~,
which includes both incoherent and coherent contributions.
That model allows to estimate ratios for different nuclei and energies of proton beam.
From calculations of such ratios we concluded that incoherent contribution is larger essentially for heavy nuclei, than coherent one.
Now we found that incoherent and coherent contributions have comparable values for light nuclei (like \isotope[4]{He}, etc.; those processes can be in stars).

After such an research and analysis, there is sense to remind researches~\cite{Clayton.1992.PRC,Clayton.1991.PhD}.
Measurements of bremsstrahlung photons in the proton-nucleus scattering are also presented in those works.
Here one can find the spectra with plateou.
However, accuracy of data~\cite{Clayton.1992.PRC,Clayton.1991.PhD} is smaller than data~\cite{Goethem.2002.PRL}.
So, before analysis described above, it was unclear about any incoherent emission.
So, works~\cite{Clayton.1992.PRC,Clayton.1991.PhD} could provide the first indications about important role of the incoherent bremsstrahlung in the proton-nucleus processes.

As it was shown in Refs.~\cite{Maydanyuk_Zhang.2015.PRC},
inclusion of relations between spin of the scattered proton and moments of nucleons of nucleus-target to the model allows to essentially improve agreement between theory and experimental data.
Without such an inclusion it is impossible to explain presence of plateau in experimental data~\cite{Goethem.2002.PRL}.
% So, inclusion of incoherent bremsstrahlung emission to the previous formalism~\cite{Maydanyuk_Zhang_Zou.2019.PRC.microscopy} should provide proper analysis of experimental data.
%
% As it was estimated, after inclusion of incoherent bremsstrahlung emission to the previous formalism~\cite{Maydanyuk_Zhang_Zou.2019.PRC.microscopy},
% level of agreement between calculations and experimental data is higher.
% analysis of internal structure of the scattered proton should be proper.
%
% However, according to previous study \cite{Maydanyuk_Zhang.2015.PRC}, incoherent processes have more essential role.
% In the scattering of protons off the \isotope[197]{Au} nuclei at energy of proton beam of $E_{\rm p}=190$~MeV
% scattering $p + \isotope[197]{Au}$ at energy of proton beam $E_{\rm p}$ of 190~MeV
For the studied reaction incoherent bremsstrahlung contribution is on $10^{6}-10^{7}$ times larger than coherent contribution in the full bremsstrahlung.

% \vspace{1.0mm}
% \noindent
As previously a detailed incoherent formalism of emission of bremsstrahlung photons during proton-nucleus scattering was constructed
%\cite{Maydanyuk.2012.PRC,
\cite{Maydanyuk_Zhang.2015.PRC,Maydanyuk_Zhang_Zou.2016.PRC,Liu_Maydanyuk_Zhang_Liu.2019.PRC.hypernuclei},
% see Ref.~\cite{Maydanyuk_Zhang_Zou.2019.PRC.microscopy} and reference therein),
in this paper we apply that model to a new situation where $\Delta$-resonance in nucleus-target can be produced.
In particular, we will analyze role of the incoherent processes, and how physics can be changed after its inclusion to analysis.

Photons emitted from incoherent processes have been studied in many topics of nuclear and particle physics.
%, using different theoretical models.
In general, physical picture of the studied reaction becomes more complete, after inclusion of such a type of photons.
%
% \noindent
For example, authors of Ref.~\cite{Zhu.2015.PRC} found not negligible role of incoherent photons in photoproduction of heavy quarks (charm, bottom) and heavy quarkonia [$J/\psi$, $\Upsilon(1S)$] in ultra-peripheral collisions of heavy nuclei Pb--Pb and proton-proton collisions at high energies
(see also Refs.~\cite{Zhu.2016.NPB,Ma_Zhu.2018.PRD}, review~\cite{Baur.2002.PhysRep}).
Incoherent photons have also an important place in study of interactions between dark matter and nuclear matter
\cite{Bell_Dent.2020.PRD}.
% (see also Refs.~\cite{Dent.2015.PRD}, reference therein).
% (see also Refs.~\cite{Dent.2015.PRD,Bell_Cai_Dent.2015.PRD,Anand.2014.PRC,Ibe.2018.JHEP}, reference therein).
%
% In many tasks, calculations of incoherent processes are essentially more complicated than coherent ones.
Generally it is difficult to realize calculations of the spectra including the incoherent contribution.
Researchers usually apply different approximations to calculate the coherent processes or incoherent ones~\cite{Remington.1987.PRC}.
So, in our research we focus on construction of unified formalism (with corresponding calculations and analysis) with joint description of the coherent and incoherent bremsstrahlung emission.
With detailed analysis, we find that the role of incoherent emission is important and the calculated spectra are changed essentially after taking this type of emission into account.

Important parameters in incoherent emission are magnetic moments of nucleons of nucleus and their spins.
Spin of $\Delta$-resonance is 3/2, while spins of protons and neutrons of nucleus are 1/2.
So, one can suppose that bremsstrahlung emission after inclusion of transition $NN \to \Delta N$ can be changed visibly.
%
% However, it is better to check this and to estimate how much is difference will be in emission of photons.
% Incoherent bremsstrahlung is directly related with magnetic moments of nucleons of nucleus-target.
% If to include transition to the formalism, than spin of $\Delta$ resonance is different from spin of nucleons.
% So, one can expect that this can change resulting spectrum of bremsstrahlung photons.
This is another motivation of our research in this paper.
We would like to check this and to estimate how much is difference will be in emission of photons.
Note that investigation in this paper of $\Delta$-resonances in nuclei on the basis of bremsstrahlung analysis (tested on existed experimental data for close processes) is performed at first time.

% \vspace{1.5mm}
The paper is organized in the following way.
In Sec.~\ref{sec.model} a new model of the bremsstrahlung photons emitted during proton nucleus scattering is presented.
In Sec.~\ref{sec.analysis} we give the results of study for the scattering of protons off the \isotope[12]{C}, \isotope[40]{Ca}, \isotope[208]{Pb} nuclei
at proton beam energy of 800~MeV with possible formation of $\Delta$-resonance.
We summarize conclusions in Sec.~\ref{sec.conclusions}.
Not published previously details of the model and some useful new calculations are presented in Appendixes~\ref{sec.app.short.2}--\ref{sec.app.3}.
% *******************************************************************************************************************

% *******************************************************************************************************************
% \newpage
\section{Model
\label{sec.model}}

\subsection{% Обобщенное уравнение Паули для нуклонов протон--ядерной системы с $\Delta$-изобарой и оператор излучения фотонов
Generalized Pauli equation for nucleons in the proton--nucleus system with $\Delta$-resonance and operator of emission of photons
\label{sec.2.1}}

% \subsection{Operator of emission of the $\alpha$ nucleus system
% \label{sec.2.2}}

Let us consider scattering of proton on nucleus in the laboratory frame, where nucleus is consisted on $A-1$ nucleons and one $\Delta$-resonance living short time.
We write hamiltonian
% of scattering of proton on nucleus with inclusion of emission of photons as
of such a system with inclusion of emission of photons as many-nucleon generalization of Pauli equation as
(obtained from Eq.~(1.3.6) in Ref.~\cite{Ahiezer.1981}, p.~33;
this formalism is along Refs.~\cite{Maydanyuk.2012.PRC,Maydanyuk_Zhang.2015.PRC,Maydanyuk_Zhang_Zou.2016.PRC,Maydanyuk_Zhang_Zou.2018.PRC,Maydanyuk.2011.JPG}, see reference therein)
\begin{equation}
\begin{array}{lcl}
\vspace{1mm}
  \hat{H} & = &
  \biggl\{
    \displaystyle\frac{1}{2m_{p}} \Bigl( \vu{p}_{p} - \displaystyle\frac{z_{p}e}{c} \vb{A}_{p} \Bigr)^{2} +
    z_{p}e\, A_{p,0} - \displaystyle\frac{z_{p}e \hbar}{2m_{p}c}\; \sigmabf \cdot \vb{rot\, A}_{p}
  \biggr\}\; + %\\
  % \biggl\{ \displaystyle\frac{1}{2m_{K}} \Bigl( \vu{p}_{K} - \displaystyle\frac{z_{K}e}{c} \vb{A}_{K} \Bigr)^{2} + z_{K}e\, A_{K,0} \biggr\}\; + \\

\vspace{1mm}
%  & + &
  \displaystyle\sum_{j=1}^{A-1}
  \biggl\{
    \displaystyle\frac{1}{2m_{j}} \Bigl( \vu{p}_{j} - \displaystyle\frac{z_{j}e}{c} \vb{A}_{j} \Bigr)^{2} +
    z_{j}e\, A_{j,0} - \displaystyle\frac{z_{j}e \hbar}{2m_{j}c}\; \sigmabf \cdot \vb{rot\, A}_{j}
  \biggr\}\; + \\

  & + &
  \biggl\{
    \displaystyle\frac{1}{2m_{\Delta}} \Bigl( \vu{p}_{\Delta} - \displaystyle\frac{z_{\Delta}e}{c} \vb{A}_{\Delta} \Bigr)^{2} +
    z_{\Delta}e\, A_{\Delta,0} -
    f_{\Delta} \cdot \displaystyle\frac{z_{\Delta}e \hbar}{2m_{\Delta}c}\; \sigmabf \cdot \vb{rot\, A}_{\Delta}
  \biggr\} +

  V(\vb{r}_{1} \ldots \vb{r}_{A-1}, \vb{r}_{\Delta}, \vb{r}_{\rm p}).
\end{array}
\label{eq.pauli.2}
\end{equation}
Here, $m_{i}$ and $z_{i}$ are mass and electric charge of nucleon with number $i$,
$\vu{p}_{i} = -i\hbar\, \vb{d}/\vb{dr}_{i} $ is momentum operator for nucleon with number $i$,
$V(\vb{r}_{1} \ldots \vb{r}_{A-1}, \vb{r}_{\Delta}, \vb{r}_{\rm p})$ is a general form of the potential of interactions between nucleons of nucleus, $\Delta$-resonance and the scattered proton,
$\sigmabf$ are Pauli matrixes,
$A_{i} = (\vb{A}_{i}, A_{i,0})$ is a potential of electromagnetic field formed by moving nucleon with number $i$ or $\Delta$-resonance,
$A$ in summation is mass number of the nucleus-target.
Also we have introduced a new parameter $f_{\Delta}$, related with that Dirac equation is for fermion with spin 1/2, while $\Delta$-resonance has spin 3/2.
For $\Delta$-resonance we use analog of Pauli equation with different term of spin
(in this paper we assume $f_{\Delta}=3$, as spin of $\Delta$-resonance is 3 times larger than for nucleon).

% Здесь $m_{i}$ и $z_{i}$ --- масса и электромагнитный заряд нуклона с номером $i$,
% $\vu{p}_{i} = -i\hbar\, \vb{d}/\vb{dr}_{i} $ --- оператор импульса нуклона с номером $i$,
% $V(\vb{r}_{1} \ldots \vb{r}_{A-1}, \vb{r}_{\Delta}, \vb{r}_{\rm p})$ --- общая форма записи потенциала взаимодействия между нуклонами ядра, $\Delta$-изобарой и протоном рассеяния,
% $\sigmabf$ --- матрицы Паули,
% $A_{i} = (\vb{A}_{i}, A_{i,0})$ --- векторный потенциал электромагнитного поля, который формируется движением нуклона с номером $i$ или $\Delta$-изобары,
% $A$ в суммировании --- массовое число ядра-мишени.
% Также мы ввели новый параметр $f_{\Delta}$, связанный с тем что уравнение Дирака введено для фермионов со спином 1/2, тогда как $\Delta$-изобара имеет спин 3/2.
% Считаем, что уравнение Паули --- это первое приближение уравнения Дирака, работает для фермионов со спином 1/2.
% А для $\Delta$-изобары мы возьмем аналог уравнения Паули лишь с измененным слагаемым спина
% (в этой работе примем $f_{\Delta}=3$, тогда спин изобары будет в три раза больше чем для нуклона).

We rewrite this hamiltonian (\ref{eq.pauli.2}) as
\begin{equation}
\begin{array}{lcl}
  \hat{H} = \hat{H}_{0} + \hat{H}_{\gamma},
\end{array}
\label{eq.pauli.4}
% \label{eq.2.1.1}
\end{equation}
where
\begin{equation}
\begin{array}{lllll}
\vspace{1mm}
  \hat{H}_{0} & = &
  \displaystyle\frac{1}{2m_{\rm p}}\, \vu{p}_{\rm p}^{2} +
  \displaystyle\sum_{j=1}^{A-1} \displaystyle\frac{1}{2m_{j}}\, \vu{p}_{j}^{2} +
  \displaystyle\frac{1}{2m_{\Delta}}\, \vu{p}_{\Delta}^{2} +
  V(\vb{r}_{1} \ldots \vb{r}_{A-1}, \vb{r}_{\Delta}, \vb{r}_{\rm p}), \\

\vspace{1mm}
  \hat{H}_{\gamma} & = &
  \biggl\{
    - \displaystyle\frac{z_{\rm p} e}{m_{\rm p}c}\; \vu{p}_{\rm p} \cdot \vb{A}_{\rm p} +
    \displaystyle\frac{z_{\rm p}^{2}e^{2}}{2m_{\rm p}c^{2}} \vb{A}_{\rm p}^{2} -
    \displaystyle\frac{z_{\rm p}e\hbar}{2m_{\rm p}c}\, \sigmabf \cdot \vb{rot A}_{\rm p} +
    z_{\rm p}e\, A_{\rm p,0}
  \biggr\}\; + \\
  % \biggl\{
    % - \displaystyle\frac{z_{K} e}{m_{K}c}\; \vu{p}_{K} \cdot \vb{A}_{K} + \displaystyle\frac{z_{K}^{2}e^{2}}{2m_{K}c^{2}} \vb{A}_{K}^{2}
    % - \displaystyle\frac{z_{K}e\hbar}{2m_{K}c}\, \sigmabf \cdot \vb{rot A}_{K}
    % + z_{K}e\, A_{K,0}
  % \biggr\}\; + \\

  & + &
  \displaystyle\sum_{j=1}^{A-1}
  \biggl\{
    - \displaystyle\frac{z_{j} e}{m_{j}c}\; \vu{p}_{j} \cdot \vb{A}_{j} +
    \displaystyle\frac{z_{j}^{2}e^{2}}{2m_{j}c^{2}} \vb{A}_{j}^{2} -
    \displaystyle\frac{z_{j}e\hbar}{2m_{j}c}\, \sigmabf \cdot \vb{rot A}_{j} +
    z_{j}e\, A_{j,0}
  \biggr\}\; + \\

  & + &
  \biggl\{
    - \displaystyle\frac{z_{\Delta} e}{m_{\Delta}c}\; \vu{p}_{\Delta} \cdot \vb{A}_{\Delta} +
    \displaystyle\frac{z_{\Delta}^{2}e^{2}}{2m_{\Delta}c^{2}} \vb{A}_{\Delta}^{2} -
    f_{\Delta} \cdot \displaystyle\frac{z_{\Delta}e\hbar}{2m_{\Delta}c}\, \sigmabf \cdot \vb{rot A}_{\Delta} +
    z_{\Delta}e\, A_{\Delta,0}
  \biggr\}.
\end{array}
\label{eq.pauli.5}
% \label{eq.2.1.2}
\end{equation}
%
% Здесь
% $\hat{H}_{0}$ --- гамильтониан, описывающий эволюцию нуклонов и $\Delta$-изобары ядра и протона рассеяния в изучаемом процессе рассеяния (без излучения фотонов),
% $\hat{H}_{\gamma}$ --- оператор, описывающий излучение тормозных фотонов в рассеянии.
Here,
$\hat{H}_{0}$ is hamiltonian describing evolution of nucleons of nucleus and $\Delta$-resonance in the scattering of proton (without emission of photons),
$\hat{H}_{\gamma}$ is operator describing emission of bremsstrahlung photons in this scattering.
%-----------------------------------------------------------------------------------------------------------------------

%-----------------------------------------------------------------------------------------------------------------------
% \subsection{Введение аномальных магнитных моментов нуклонов и изобары
% Introduction of magnetic moment of nucleons (and hyperons)
% \label{sec.2.2}}

To include magnetic moments of particles, we change Dirac's magnetic moment $\mu_{i}^{\rm (Dir)} = z_{i}\, e\hbar / (2m_{i}c)$ for each particle with number $i$ as
$\mu_{i}^{\rm (Dir)} \to \mu_{i}^{\rm (an)}\, \mu_{N}$,
where
$\mu_{N} = e\hbar / (2m_{\rm p}c)$ is nuclear magneton,
$\mu_{\rm p}^{\rm (an)} = 2.79284734462$ is anomalous magnetic moment for proton,
$\mu_{\rm n}^{\rm (an)} = -1.91304273$ is anomalous magnetic moment for neutron
$\mu_{j}$ are magnetic moments of protons or neutrons of nucleus
(measured in units of nuclear magneton $\mu_{N}$, see Ref.~\cite{RewPartPhys_PDG.2018}).
We neglect terms at $\vb{A}_{j}^{2}$ and $A_{j,0}$, and use Coulomb gauge.%
\footnote{%
% \begin{itemize}
% \vspace{2mm}
% \item
In QED one can write down gauge for potential of electromagnetic field as
$A_{\nu}^{\prime} (x) = A_{\nu} (x) + \partial_{\nu} f(x)$ ($\nu = 0, 1,2,3$).
Function $f(x)$ can be changed (but equations of motion are not changed),
and it can be found in any fixed frame as $A_{0}=0$.
In result, we obtain Coulomb gauge $\div \vb{A} = 0$ in QED (for example, see Ref.~\cite{Bogoliubov.1980}, p.~37),
which is used in the formalism.
%
% \vspace{2mm}
% \item
QED is perturbative theory. Here it is supposed that matrix elements based on terms with $\vb{A}^{2}$ are smaller than (non-zero) matrix elements with $\vb{A}$.
Similar situation exits in determination of different electromagnetic processes based on calculations of $S$-matrix in different orders in QED.
Following to such a logic, terms with $\vb{A}^{2}$ are neglected in formalism and calculations,
analysis of such terms with $\vb{A}^{2}$ are omitted.
% \end{itemize}
%
Way to neglect terms with $A_0$ and $\vb{A}^2$ was used by many authors in study of bremsstrahlung in the different nuclear processes
(for example, see Refs.~\cite{Papenbrock.1998.PRLTA,Tkalya.1999.PHRVA}).
% T.~Papenbrock and G.~F.~Bertsch, Phys. Rev. Lett. \textbf{80}, 4141 (1998);
% E.~V.~Tkalya, Phys.~Rev.~\textbf{C 60}, 0446 (1999)].
% We just continue this logic.
}

Operator of emission~(\ref{eq.pauli.5}) is transformed to
\begin{equation}
\begin{array}{llll}
  \hat{H}_{\gamma} & = &
    - \displaystyle\frac{z_{\rm p} e}{m_{\rm p}c}\; \vu{p}_{\rm p} \cdot \vb{A}_{\rm p} -
    \mu_{N}\, \mu_{\rm p}\, \sigmabf \cdot \vu{H}_{\rm p}
  + % \\

  \displaystyle\sum_{j=1}^{A}
  \biggl\{
    - \displaystyle\frac{z_{j} e}{m_{j}c}\; \vu{p}_{j} \cdot \vb{A}_{j} -
    \mu_{N}\, \mu_{j}\, \sigmabf \cdot \vu{H}_{j}
  \biggr\},
% +
%   \biggl\{
%     - \displaystyle\frac{z_{\Delta} e}{m_{\Delta}c}\; \vu{p}_{\Delta} \cdot \vb{A}_{\Delta} -
%     f_{\Delta} \cdot \mu_{\Delta}\, \sigmabf \cdot \vu{H}_{\Delta}
%   \biggr\}.
\end{array}
\label{eq.2.2.6}
\end{equation}
where
\begin{equation}
  \vu{H} = \vb{rot\: A} = \bigl[ \curl{\vb{A}} \bigr]
\label{eq.2.2.5}
\end{equation}
%
% где $\mu_{\Delta} = f_{\Delta} \cdot \mu_{p\,{\rm or}\, n}$ ($f_{\Delta}=3$).
% Это выражение --- многонуклонное обобщение оператора излучения $\hat{W}$ в~(4) в работе~\cite{Maydanyuk.2012.PRC} с
% включенными аномальными магнитными моментами нуклонов и изобары.
and $\mu_{\Delta} = f_{\Delta} \cdot \mu_{p,\, n}$ ($f_{\Delta}=3$).%
\footnote{One can see that formulation of $f_{\Delta}$ violates gauge invariance.
Note that in the shell model of nucleus the full magnetic moments for protons and neutrons in nucleus are depended on the shells of nucleus
(for example, see Eqs.~(118.12)--(118.14), p.~583--591 in Ref.~\cite{Landau.v3.1989}).
This formulation violates gauge invariance also.
However, level of accuracy of determination of some spectroscopic characteristics of nuclei in frameworks of the nuclear shell model is not bad.}
This expression is many-nucleon generalization of operator of emission $\hat{W}$ in Eq.~(4) in Ref.~\cite{Maydanyuk.2012.PRC} with
included magnetic moments for nucleons and $\Delta$-resonance.
%-----------------------------------------------------------------------------------------------------------------------

%-----------------------------------------------------------------------------------------------------------------------
\subsection{Formalism in space representation
\label{sec.2.3}}

Substituting the following definition for the potential of electromagnetic field:
\begin{equation}
\begin{array}{lcl}
  \vb{A} & = &
  \displaystyle\sum\limits_{\alpha=1,2}
    \sqrt{\displaystyle\frac{2\pi\hbar c^{2}}{w_{\rm ph}}}\; \vb{e}^{(\alpha),\,*}
    e^{-i\, \vb{k_{\rm ph}r}},
\end{array}
\label{eq.2.3.1}
\end{equation}
we obtain:
\begin{equation}
\begin{array}{lcl}
  \vu{H} & = &
  % \vb{rot\: A} =
  \bigl[ \curl{\vb{A}} \bigr] =
  \sqrt{\displaystyle\frac{2\pi\hbar c^{2}}{w_{\rm ph}}}\,
    \displaystyle\sum\limits_{\alpha=1,2}
    \Bigl\{ -i\, e^{-i\, \vb{k_{\rm ph}r}}\, \bigl[ \vb{k_{\rm ph}} \times \vb{e}^{(\alpha),\,*} \bigr] +
      e^{-i\, \vb{k_{\rm ph}r}}\, \bigl[ \curl{\vb{e}^{(\alpha),\,*}} \bigr] \Bigr\}.
\end{array}
\label{eq.2.3.2}
\end{equation}
Here,
$\vb{e}^{(1)}$ and $\vb{e}^{(2)}$ are two independent unit vectors of polarizations for photon %with momentum $\vb{k}_{\rm ph}$
% $\vb{e}^{(\alpha)}$ are unit vectors of polarization of the photon emitted
[$\vb{e}^{(\alpha), *} = \vb{e}^{(\alpha)}$, $\alpha=1,2$], $\vb{k}_{\rm ph}$ is wave vector of the photon and $w_{\rm ph} = k_{\rm ph} c = \bigl| \vb{k}_{\rm ph}\bigr|c$.
Vectors $\vb{e}^{(\alpha)}$ are perpendicular to $\vb{k}_{\rm ph}$ in Coulomb gauge, satisfy Eq.~(8) in Ref.~\cite{Liu_Maydanyuk_Zhang_Liu.2019.PRC.hypernuclei}.
% We have two independent polarizations $\vb{e}^{(1)}$ and $\vb{e}^{(2)}$ for the photon with impulse $\vb{k}_{\rm ph}$ ($\alpha=1,2$).
One can develop formalism simpler in the system of units where $\hbar = 1$ and $c = 1$, but we shall write constants $\hbar$ and $c$ explicitly.
Also we have properties:
\begin{equation}
\begin{array}{lclc}
  \Bigl[ \vb{k}_{\rm ph} \times \vb{e}^{(1)} \Bigr] = k_{\rm ph}\, \vb{e}^{(2)}, &
  \Bigl[ \vb{k}_{\rm ph} \times \vb{e}^{(2)} \Bigr] = -\, k_{\rm ph}\, \vb{e}^{(1)}, &
  \Bigl[ \vb{k}_{\rm ph} \times \vb{e}^{(3)} \Bigr] = 0, &
  \displaystyle\sum\limits_{\alpha=1,2,3} \Bigl[ \vb{k}_{\rm ph} \times \vb{e}^{(\alpha)} \Bigr] = k_{\rm ph}\, (\vb{e}^{(2)} - \vb{e}^{(1)}).
\end{array}
\label{eq.2.3.3}
\end{equation}
We substitute formulas (\ref{eq.2.3.1}) and (\ref{eq.2.3.2}) to formula~(\ref{eq.2.2.6}) for operator of emission and obtain:
\begin{equation}
\begin{array}{lcl}
  \hat{H}_{\gamma} & = &
  \sqrt{\displaystyle\frac{2\pi\hbar c^{2}}{w_{\rm ph}}}\; \mu_{N}\,
  \displaystyle\sum\limits_{\alpha=1,2}
    e^{-i\, \vb{k_{\rm ph}r}_{K}}\,
  \biggl\{
    i\, 2 z_{\rm p}\, \vb{e}^{(\alpha)} \cdot \grad_{\rm p} +
    \mu_{\rm p}\, \sigmabf \cdot \Bigl( i\, \bigl[ \vb{k_{\rm ph}} \times \vb{e}^{(\alpha)} \bigr] - \bigl[ \grad_{\rm p} \times \vb{e}^{(\alpha)} \bigr] \Bigr)
  \biggr\}\; + \\
  % \biggl\{ i\, \mu_{N}\, \displaystyle\frac{2 z_{K} m_{\rm p}}{m_{K}}\: \vb{e}^{(\alpha)} \cdot \grad_{K} \biggr\}\; + \\

  & + &
  \sqrt{\displaystyle\frac{2\pi\hbar c^{2}}{w_{\rm ph}}}\; \mu_{N}\,
  \displaystyle\sum_{j=1}^{A}
  \displaystyle\sum\limits_{\alpha=1,2}
    e^{-i\, \vb{k_{\rm ph}r}_{j}}\;
    \biggl\{
      i\, \displaystyle\frac{2 z_{j} m_{\rm p}}{m_{Aj}}\: \vb{e}^{(\alpha)} \cdot \grad_{j} +
      \mu_{j}\, \sigmabf \cdot \Bigl( i\, \bigl[ \vb{k_{\rm ph}} \times \vb{e}^{(\alpha)} \bigr] - \bigl[ \grad_{j} \times \vb{e}^{(\alpha)} \bigr] \Bigr)
    \biggr\}.
\end{array}
\label{eq.2.3.4}
\end{equation}
%
% where $\mu_{N}$ is nuclear magneton defined after Eqs.~(\ref{eq.2.2.3}).
This expression coincides with operator of emission $\hat{W}$ in form (6) in Ref.~\cite{Maydanyuk.2012.PRC} in the limit case of problem of one nucleon with charge $Z_{\rm eff}$ in the external field
[taking Eqs.~(\ref{eq.2.3.3}),
% $\vb{e}^{(\alpha), *} = \vb{e}^{(\alpha)}$ and
$\hbar = 1$ into account].
We have included terms for $\Delta$-resonance to summation in Eq.~(\ref{eq.2.3.4}) and further in this paper, for convenience.
%-----------------------------------------------------------------------------------------------------------------------

%-----------------------------------------------------------------------------------------------------------------------
% \section{Transition to coordinates of relative distances
% \label{sec.2.4}}
\subsection{Operator of emission with relative coordinates
\label{sec.2.5}}

We define coordinates of center-of-mass for the nucleus as $\vb{R}_{A}$, for the complete system as $\vb{R}$,
relative coordinate $\vb{r}$ between the scattered proton and center-of-mass of nucleus-target,
relative coordinates $\rhobf_{A j}$ of nucleons (and $\Delta$-resonance) of the nucleus concerning to its center-of-mass.
% adapting formalism in Eqs.~(11)--(14) and Appendix~A in Ref.~\cite{Liu_Maydanyuk_Zhang_Liu.2019.PRC.hypernuclei} for proton-nucleus scattering.
Following to formalism in Ref.~\cite{Maydanyuk.2012.PRC} [see Eqs.~(3), (4) and explanation in the text of that paper;
also see Appendix~A in Ref.~\cite{Liu_Maydanyuk_Zhang_Liu.2019.PRC.hypernuclei}],
we find
new relative coordinate $\vb{r} = \vb{r}_{\rm p} - \vb{R}_{A}$,
new relative coordinates $\rhobf_{Aj} = \vb{r}_{Aj} - \vb{R}_{A}$ for nucleons for the nucleus.
We calculate momentums $\vu{P}$, $\vu{p}$, $\vb{\tilde{p}}_{Aj}$
% $\vu{p}_{\rm p}$, $\vu{p}_{Aj}$ %, $\vu{p}_{AA}$
corresponding to independent variables $\vb{R}$, $\vb{r}$, $\rhobf_{A j}$ at $j = 1 \ldots A-1$ (defined like $\vu{p}_{i} = -i\hbar\, \vb{d}/\vb{dr}_{i}$),
and rewrite formalism above.%
% we find coordinates of relative distances and corresponding moments (see Appendix~\ref{sec.app.relativecoord}), and rewrite formalism above.
%-----------------------------------------------------------------------------------------------------------------------
%
%-----------------------------------------------------------------------------------------------------------------------
% Подставляя формулы (\ref{eq.2.4.4.5}) в это выражение, мы найдем
% (см. вычисления в Прил.~\ref{sec.app.OpEmission}, конечные формулы (\ref{eq.app.OpEmission.4.1})--(\ref{eq.app.OpEmission.4.6});
% также см. решения для операторов $\hat{H}_{P}$, $\hat{H}_{p}$, $\hat{H}_{k}$ в~(16)--(18) в Прил.~B в статье~\cite{Liu_Maydanyuk_Zhang_Liu.2019.PRC.hypernuclei},
% решения для $\Delta \hat{H}_{\gamma E}$, $\Delta \hat{H}_{\gamma M}$ в Прил. A в статье~\cite{Maydanyuk_Zhang_Zou.2019.brem_alpha_nucleus.arxiv},
% и мы усовершенствовали новое решение для последнего матричного элемента $\hat{H}_{k}$):
%
\footnote{One can write useful formulas for nucleon with number $A$ of the nucleus-target as
% (such variables are dependent on $\rhobf_{Aj}$ and $\vu{p}_{Aj}$ at $j=1 \ldots A-1$):
%
\begin{equation}
\begin{array}{lllll}
  \rhobf_{AA} = -\, \displaystyle\frac{1}{m_{AA}} \displaystyle\sum_{k=1}^{A-1} m_{Ak}\, \rhobf_{A k}, &
  \vu{p}_{AA} =
    \displaystyle\frac{m_{AA}}{m_{A} + m_{p}}\, \vu{P} -
    \displaystyle\frac{m_{AA}}{m_{A}}\,\vu{p} -
    \displaystyle\frac{m_{AA}}{m_{A}}\, \displaystyle\sum_{k=1}^{A-1} \vb{\tilde{p}}_{Ak}.
\end{array}
\label{eq.footnote.1}
\end{equation}
}

% Substituting formulas (\ref{eq.2.4.3.3}) to these expressions, we find (see calculations in Appendix~\ref{sec.app.1}):
Following to formalism in Ref.~\cite{Liu_Maydanyuk_Zhang_Liu.2019.PRC.hypernuclei},
we calculate operator of emission of bremsstrahlung photons in the scattering of proton off nucleus in the laboratory frame as
\begin{equation}
  \hat{H}_{\gamma} =
  % \displaystyle\sum\limits_{i=1}^{A+1} \bar{h}_{\gamma 0} (\vb{r}_{i}) =
%   \displaystyle\sum\limits_{i=1}^{A+1}
%   \Bigl[
%     - \displaystyle\frac{z_{i}e}{2\,m_{i}c} (-i\hbar\, \vb{div A}_{i} + 2\,\vb{A_{i}p_{i}}) +
%     \displaystyle\frac{z_{i}^{2}e^{2}}{2\,m_{i}c^{2}}\, \vb{A}_{i}^{2} -
%     \displaystyle\frac{z_{i}e}{2\,m_{i}c}\, \sigmabf \vb{H}_{i}
%   \Bigr], \\
%
%   \hat{H}_{\gamma 0} =
  \hat{H}_{P} + \hat{H}_{p} + \Delta \hat{H}_{\gamma E} + \Delta \hat{H}_{\gamma M} + \hat{H}_{k},
\label{eq.2.5.2}
\end{equation}
where
operators $\hat{H}_{P}$, $\hat{H}_{p}$, $\hat{H}_{k}$ are calculated in Eqs.~(16)--(19) in Ref.~\cite{Liu_Maydanyuk_Zhang_Liu.2019.PRC.hypernuclei} (at $m_{\alpha} \to m_{p}$ replacement) and
(see solutions for $\Delta \hat{H}_{\gamma E}$, $\Delta \hat{H}_{\gamma M}$ in Appendix A in Ref.~\cite{Maydanyuk_Zhang_Zou.2019.brem_alpha_nucleus.arxiv})
\begin{equation}
\begin{array}{lcl}
\vspace{0.4mm}
  \Delta \hat{H}_{\gamma E} & = &
  -\, \sqrt{\displaystyle\frac{2\pi c^{2}}{\hbar w_{\rm ph}}}\:
    2\, \mu_{N}\, e^{-i\, \vb{k_{\rm ph}}\vb{R}}\,
    \displaystyle\sum\limits_{\alpha=1,2} \vb{e}^{(\alpha)}\; \times \\
  & \times &
  \biggl\{
    e^{i\, c_{\rm p}\, \vb{k_{\rm ph}} \vb{r}}\,
    \displaystyle\sum_{j=1}^{A-1}
      \displaystyle\frac{z_{j}\,m_{\rm p}}{m_{Aj}}\, e^{-i\, \vb{k_{\rm ph}} \rhobf_{Aj}}\, \vb{\tilde{p}}_{Aj} -
    \displaystyle\frac{m_{\rm p}}{m_{A}}\, e^{i\, c_{\rm p}\, \vb{k_{\rm ph}} \vb{r}}\,
      \displaystyle\sum_{j=1}^{A} z_{j}\, e^{-i\, \vb{k_{\rm ph}} \rhobf_{Aj}}\, \displaystyle\sum_{k=1}^{A-1} \vb{\tilde{p}}_{Ak}
  \biggr\},
\end{array}
\label{eq.2.5.6}
% \label{eq.app.OpEmission.3.3}
% \label{eq.10.1.5}
\end{equation}
\begin{equation}
\begin{array}{lll}
\vspace{0.4mm}
  & \Delta \hat{H}_{\gamma M} =
  -\, i\, \sqrt{\displaystyle\frac{2\pi c^{2}}{\hbar w_{\rm ph}}}\:
   \mu_{N}\:
    e^{-i\, \vb{k_{\rm ph}} \vb{R}}\,
  \displaystyle\sum\limits_{\alpha=1,2} \; \times \\

% \vspace{0.4mm}
  \times &
  \biggl\{
    e^{i\, \vb{k_{\rm ph}} c_{\rm p}\, \vb{r}}\,
    \displaystyle\sum_{j=1}^{A-1}
      \mu_{j}\, e^{-i\, \vb{k_{\rm ph}} \rhobf_{Aj}}\, \sigmabf \cdot \bigl[ \vb{\tilde{p}}_{Aj} \times \vb{e}^{(\alpha)} \bigr] -

    e^{i\, \vb{k_{\rm ph}} c_{\rm p}\, \vb{r}}\,
    \displaystyle\sum_{j=1}^{A}
      \mu_{j}\, \displaystyle\frac{m_{Aj}}{m_{A}}\,
      e^{-i\, \vb{k_{\rm ph}} \rhobf_{Aj}}\,
      \displaystyle\sum_{k=1}^{A-1} \sigmabf \cdot \bigl[ \vb{\tilde{p}}_{Ak} \times \vb{e}^{(\alpha)} \bigr]
  \biggr\}.
\end{array}
\label{eq.2.5.7}
% \label{eq.app.OpEmission.3.4}
% \label{eq.10.1.6}
\end{equation}
Here,
% $\mu_{N} = e\hbar / (2m_{\rm p}c)$ is nuclear magneton, % $ = 3.152 451 2550\; 10^{-14}$~MeV T$^{-1}$ is nuclear magneton,
% $m_{i}$ and $z_{i}$ are mass and electric charge of nucleon with number $i$,
% $m_{\rm p}$ and $m_{A}$ are masses of the scattered proton and nucleus-target,
% $c_{A} = \frac{m_{A}}{m_{A}+m_{K}}$ and $c_{\rm p} = \frac{m_{\rm p}}{m_{A}+m_{\rm p}}$,
% $\mu_{j}$ are magnetic moments of protons or neutrons of nucleus
% [measured in units of nuclear magneton $\mu_{N}$].
%
% $\vb{r}_{p}$, $\vb{R}_{A}$ and $\vb{R}$ are coordinates of the scattered proton, center of nucleus and center of masses of complete system in laboratory frame
% [adapting formalism in Eqs.~(11)--(14) and Appendix~A in Ref.~\cite{Liu_Maydanyuk_Zhang_Liu.2019.PRC.hypernuclei} for proton-nucleus scattering].
% $\vb{r} = \vb{r}_{p} - \vb{R}_{A}$,
% $\rhobf_{A j} = \vb{r}_{j} - \vb{R}_{A}$ are relative coordinates for nucleons ($j=1 \ldots A-1$),
% $\vb{\hat{P}}$, $\vb{\hat{p}}$ and $\vb{\tilde{p}}_{Aj}$ are momentums corresponding to $\vb{R}$, $\vb{r}$, $\rhobf_{Aj}$ at $j=1 \ldots A-1$
% (defined like $\vu{p}_{i} = -i\hbar\, \vb{d}/\vb{dr}_{i}$).
% $\vb{e}^{(1)}$ and $\vb{e}^{(2)}$ are unit vectors of polarizations for photon with momentum $\vb{k}_{\rm ph}$.
%
new coefficients $c_{A} = \frac{m_{A}}{m_{A}+m_{\rm p}}$ and $c_{\rm p} = \frac{m_{\rm p}}{m_{A} + m_{\rm p}}$ are introduced,
where
$m_{\rm p}$ and $m_{A}$ are masses of proton scattered and nucleus-target,
$m_{Aj}$ is mass of nucleon of nucleus with number $j$, and
(%$\alpha=1,2$,
% $\vb{e}^{(\alpha)}$ are unit vectors of polarization of the photon emitted ($\vb{e}^{(\alpha),*} = \vb{e}^{(\alpha)}$),
% $\vb{k}_{\rm ph}$ is wave vector of the photon and $w_{\rm ph} = k_{\rm ph} c = \bigl| \vb{k}_{\rm ph}\bigr|c$.
$\vb{e}^{(\alpha)}$ are perpendicular to $\vb{k}_{\rm ph}$ in Coulomb gauge, satisfy Eq.~(8) in Ref.~\cite{Liu_Maydanyuk_Zhang_Liu.2019.PRC.hypernuclei}.
% (we neglect terms at $\vb{A}_{j}^{2}$ and $A_{j,0}$, and use Coulomb gauge),
% $A_{i} = (\vb{A}_{i}, A_{i,0})$ is a potential of electromagnetic field formed by moving nucleon with number $i$ or $K$-meson,
% $\sigmabf$ are Pauli matrixes,
$A$ in summation is mass number of the nucleus-target.
%-----------------------------------------------------------------------------------------------------------------------
%
%-----------------------------------------------------------------------------------------------------------------------
% A summation of expression (\ref{eq.2.5.4}) and $\hat{H}_{k}$ is many-nucleon generalization of operator of emission $\hat{W}$ in Eq.~(6) in Ref.~\cite{Maydanyuk.2012.PRC} with
% included magnetic moments for nucleons.
%-----------------------------------------------------------------------------------------------------------------------

%-----------------------------------------------------------------------------------------------------------------------
\subsection{Matrix elements of emission of bremsstrahlung photons
\label{sec.13}}

% \section{Wave function of scattering of proton on nucleus with $\Delta$-resonance
% \label{sec.2.6}}

We define the wave function of the full nuclear system as
$\Psi = \Phi (\vb{R}) \cdot \Phi_{\rm p - nucl} (\vb{r}) \cdot \psi_{\rm nucl} (\beta_{A})$
following the formalism in Ref.~\cite{Maydanyuk_Zhang.2015.PRC} for the proton-nucleus scattering [see Sect.~II.B, Eqs.~(10)--(13)],
and we add description of many-nucleon structure of the nucleus as in Ref.~\cite{Maydanyuk_Zhang_Zou.2016.PRC}.
Here,
$\beta_{A}$ is the set of numbers $1 \cdots A$ of nucleons of the nucleus,
$\Phi (\vb{R})$ is the function describing motion of center-of-mass of the full nuclear system in laboratory frame,
$\Phi_{\rm p - nucl} (\vb{r})$ is the function describing relative motion of the scattered proton concerning to nucleus (without description of internal relative motions of nucleons in the nucleus),
% $\psi_{\rm p} (\beta_{p})$ is the wave function of the scattered proton,
$\psi_{\rm nucl} (\beta_{A})$ is the many-nucleon function of the nucleus,
defined in Eq.~(12) Ref.~\cite{Maydanyuk_Zhang.2015.PRC} on the basis of one-nucleon functions $\psi_{\lambda_{s}}(s)$,
$\beta_{A}$ is the set of numbers $1 \cdots A$ of nucleons of the nucleus.
One-nucleon functions $\psi_{\lambda_{s}}(s)$ represent the multiplication of space and spin-isospin
functions as $\psi_{\lambda_{s}} (s) = \varphi_{n_{s}} (\vb{r}_{s})\, \bigl|\, \sigma^{(s)} \tau^{(s)} \bigr\rangle$,
where
$\varphi_{n_{s}}$ is the space function of the nucleon with number $s$,
$n_{s}$ is the number of state of the space function of the nucleon with number $s$,
$\bigl|\, \sigma^{(s)} \tau^{(s)} \bigr\rangle$ is the spin-isospin function of the nucleon with number $s$.

We define the matrix element of emission, using the wave functions $\Psi_{i}$ and $\Psi_{f}$ of the full nuclear system in states before emission of photons ($i$-state) and after such emission ($f$-state),
as
\begin{equation}
  F = \langle \Psi_{f} |\, \hat{H}_{\gamma} |\, \Psi_{i} \rangle.
\label{eq.13.1.1}
\end{equation}
In this matrix element we should integrate over all independent variables i.e.
% These variables are
space variables $\vb{R}$, $\vb{r}$, $\rhobf_{Am}$.
We should take into account space representation of all used moments $\vu{P}$, $\vu{p}$, $\vb{\tilde{p}}_{A m}$
(as
$\vu{P} = -i\hbar\, \vb{d/dR}$,
$\vu{p} = -i\hbar\, \vb{d/dr}$,
$\vb{\tilde{p}}_{A m} = -i\hbar\, \vb{d/d} \rhobf_{Am}$).
Using formulas (\ref{eq.2.5.2})--(\ref{eq.2.5.7}) for the operator of emission, we calculate
[see Appendix~\ref{sec.app.short.2.7} for details, Eqs.~(\ref{eq.app.short.2.7.2}), (\ref{eq.app.short.2.12.3})--(\ref{eq.app.short.2.12.8})]
%
% [see Eqs.~(23), (34) and Appendixes B--D in Ref.~\cite{Maydanyuk_Zhang_Zou.2019.brem_alpha_nucleus.arxiv} adapting calculations for proton-nucleus scattering]:
%
\begin{equation}
\begin{array}{lll}
  \langle \Psi_{f} |\, \hat{H}_{\gamma} |\, \Psi_{i} \rangle \;\; = \;\;
  \sqrt{\displaystyle\frac{2\pi\, c^{2}}{\hbar w_{\rm ph}}}\,
  M_{\rm full}, &
  % \Bigl\{ M_{P} + M_{p}^{(E)} + M_{p}^{(M)} + M_{k} + M_{\Delta E} + M_{\Delta M} \Bigr\}, &

  M_{\rm full} = M_{P} + M_{p}^{(E)} + M_{p}^{(M)} + M_{k} + M_{\Delta E} + M_{\Delta M},
\end{array}
\label{eq.13.1.2}
\end{equation}
where
\begin{equation}
\begin{array}{lll}
\vspace{-0.2mm}
  M_{p}^{(E)} & = &
  2\,i \hbar\, (2\pi)^{3} \displaystyle\frac{m_{\rm p}}{\mu}\: \mu_{N}
  \displaystyle\sum\limits_{\alpha=1,2}
  \displaystyle\int\limits_{}^{}
    \Phi_{\rm p - nucl, f}^{*} (\vb{r})\;
    e^{-i\, \vb{k}_{\rm ph} \vb{r}} \cdot
    Z_{\rm eff} (\vb{k}_{\rm ph}, \vb{r}) \cdot \vb{e}^{(\alpha)}\, \vb{\displaystyle\frac{d}{dr}} \cdot
    \Phi_{\rm p - nucl, i} (\vb{r})\; \vb{dr}, \\

  M_{p}^{(M)} & = &
  -\, \hbar\, (2\pi)^{3} \displaystyle\frac{m_{\rm p}}{\mu}\: \mu_{N}
  \displaystyle\sum\limits_{\alpha=1,2}
  \displaystyle\int\limits_{}^{}
    \Phi_{\rm p - nucl, f}^{*} (\vb{r})\;
    e^{-i\, \vb{k}_{\rm ph} \vb{r}} \cdot
    \vb{M}_{\rm eff} (\vb{k}_{\rm ph}, \vb{r}) \cdot \Bigl[ \vb{\displaystyle\frac{d}{dr}} \times \vb{e}^{(\alpha)} \Bigr] \cdot
    \Phi_{\rm p - nucl, i} (\vb{r})\; \vb{dr},
\end{array}
\label{eq.13.1.3}
\end{equation}
\begin{equation}
\begin{array}{lllll}
\vspace{-0.1mm}
  M_{P} & = &
  \displaystyle\frac{\hbar\, (2\pi)^{3}}{m_{A} + m_{p}}\, \mu_{N}\,
  \displaystyle\sum\limits_{\alpha=1,2}
  \displaystyle\int\limits_{}^{}
    \Phi_{\rm p - nucl, f}^{*} (\vb{r})\;
  \biggl\{
    2\, m_{\rm p}\;
    \Bigl[
      e^{-i\, c_{A}\, \vb{k_{\rm ph}} \vb{r}} F_{p,\, {\rm el}} + e^{i\, c_{p}\, \vb{k_{\rm ph}} \vb{r}} F_{A,\, {\rm el}}
    \Bigr]\, \vb{e}^{(\alpha)} \cdot \vb{K}_{i}\; + \\

  & + &
    i\: \Bigl[
      e^{-i\, c_{A}\, \vb{k_{\rm ph}} \vb{r}}\, \vb{F}_{p,\, {\rm mag}} + e^{i\, c_{p}\, \vb{k_{\rm ph}} \vb{r}}\, \vb{F}_{A,\, {\rm mag}}
    \Bigr] \cdot
    \bigl[ \vb{K}_{i} \cp \vb{e}^{(\alpha)} \bigr]
  \biggr\} \cdot
  \Phi_{\rm p - nucl, i} (\vb{r})\; \vb{dr},
\end{array}
\label{eq.13.1.4}
\end{equation}
\begin{equation}
\begin{array}{lcl}
% \vspace{-0.1mm}
  M_{k} & = &
  i\, \hbar\, (2\pi)^{3}  \mu_{N}\,
  \displaystyle\sum\limits_{\alpha=1,2}
    \bigl[ \vb{k_{\rm ph}} \cp \vb{e}^{(\alpha)} \bigr]
  \displaystyle\int\limits_{}^{}
    \Phi_{\rm p - nucl, f}^{*} (\vb{r}) \cdot
    \Bigl\{ e^{-i\, c_{A}\, \vb{k_{\rm ph}} \vb{r}}\, \vb{D}_{p,\, {\rm k}} + e^{i\, c_{p}\, \vb{k_{\rm ph}} \vb{r}}\, \vb{D}_{A,\, {\rm k}} \Bigr\} \cdot
    \Phi_{\rm p - nucl, i} (\vb{r})\; \vb{dr},
\end{array}
\label{eq.13.1.5}
\end{equation}
%
% \begin{equation}
% \begin{array}{lll}
% \vspace{-0.1mm}
%   M_{\Delta E} & = &
%   -\, (2\pi)^{3}\, 2\, \mu_{N}
%   \displaystyle\sum\limits_{\alpha=1,2} \vb{e}^{(\alpha)}
%   \displaystyle\int\limits_{}^{}
%     \Phi_{\rm p - nucl, f}^{*} (\vb{r})\;
%   \biggl\{
%     \Bigl[ e^{-i\, c_{A}\, \vb{k_{\rm ph}} \vb{r}}\, \vb{D}_{p 1,\, {\rm el}} + e^{i\, c_{p}\, \vb{k_{\rm ph}} \vb{r}}\, \vb{D}_{A 1,\, {\rm el}} \Bigr]\; - \\
%   &- &
%     \Bigl[ e^{-i\, c_{A}\, \vb{k_{\rm ph}} \vb{r}}\, \vb{D}_{p 2,\, {\rm el}} +
%     \displaystyle\frac{m_{\rm p}}{m_{A}}\, e^{i\, c_{p}\, \vb{k_{\rm ph}} \vb{r}}\, \vb{D}_{A 2,\, {\rm el}} \Bigr]
%   \biggr\} \cdot
%   \Phi_{\rm p - nucl, i} (\vb{r})\; \vb{dr},
% \end{array}
% \label{eq.13.1.6}
% \end{equation}
%
% \begin{equation}
% \begin{array}{lll}
% \vspace{-0.1mm}
%   M_{\Delta M} & = &
%   -\, i\, (2\pi)^{3}\,  \mu_{N}\,
%   \displaystyle\sum\limits_{\alpha=1,2}
%   \displaystyle\int\limits_{}^{}
%     \Phi_{\rm p - nucl, f}^{*} (\vb{r})\;
%   \biggl\{
%     \Bigl[ e^{-i\, c_{A}\, \vb{k_{\rm ph}} \vb{r}}\; D_{p 1,\, {\rm mag}} (\vb{e}^{(\alpha)}) + e^{i\, c_{p}\, \vb{k_{\rm ph}} \vb{r}}\; D_{A 1,\, {\rm mag}} (\vb{e}^{(\alpha)}) \Bigr]\; - \\
%   & - &
%     \Bigl[ e^{-i\, c_{A}\, \vb{k_{\rm ph}} \vb{r}}\; D_{p 2,\, {\rm mag}} (\vb{e}^{(\alpha)}) + e^{i\, c_{p}\, \vb{k_{\rm ph}} \vb{r}}\; D_{A 2,\, {\rm mag}} (\vb{e}^{(\alpha)}) \Bigr]
%   \biggr\} \cdot
%   \Phi_{\rm p - nucl, i} (\vb{r})\; \vb{dr}
% \end{array}
% \label{eq.13.1.7}
% \end{equation}
%
\begin{equation}
\begin{array}{lll}
  M_{\Delta E} & = &
  -\, (2\pi)^{3}\, 2\, \mu_{N}
  \displaystyle\sum\limits_{\alpha=1,2} \vb{e}^{(\alpha)}
  \displaystyle\int\limits_{}^{}
    \Phi_{\rm p - nucl, f}^{*} (\vb{r})\;
  \biggl\{
    e^{i\, c_{p}\, \vb{k_{\rm ph}} \vb{r}}\, \vb{D}_{A 1,\, {\rm el}} -
    \displaystyle\frac{m_{\rm p}}{m_{A}}\, e^{i\, c_{p}\, \vb{k_{\rm ph}} \vb{r}}\, \vb{D}_{A 2,\, {\rm el}}
  \biggr\} \cdot
  \Phi_{\rm p - nucl, i} (\vb{r})\; \vb{dr},
\end{array}
\label{eq.13.1.10}
\end{equation}
\begin{equation}
\begin{array}{lll}
  M_{\Delta M} & = &
  -\, i\, (2\pi)^{3}\,  \mu_{N}\,
  \displaystyle\sum\limits_{\alpha=1,2}
  \displaystyle\int\limits_{}^{}
    \Phi_{\rm p - nucl, f}^{*} (\vb{r})\;
  \biggl\{
    e^{i\, c_{p}\, \vb{k_{\rm ph}} \vb{r}}\; D_{A 1,\, {\rm mag}} (\vb{e}^{(\alpha)}) -
    e^{i\, c_{p}\, \vb{k_{\rm ph}} \vb{r}}\; D_{A 2,\, {\rm mag}} (\vb{e}^{(\alpha)})
  \biggr\} \cdot
  \Phi_{\rm p - nucl, i} (\vb{r})\; \vb{dr}
\end{array}
\label{eq.13.1.11}
\end{equation}
and $\vb{K}_{i} = \vb{K}_{f} + \vb{k}_{\rm ph}$.
Here, $\mu = m_{\rm p} m_{A} / (m_{\rm p} + m_{A})$ is reduced mass and
the effective electric charge and magnetic moment are [see Eqs.~(\ref{eq.app.short.2.10.3}), (\ref{eq.app.short.2.10.4})]
\begin{equation}
\begin{array}{lll}
\vspace{1.0mm}
  Z_{\rm eff} (\vb{k}_{\rm ph}, \vb{r}) =
  e^{i\, \vb{k_{\rm ph}} \vb{r}}\,
  \Bigl[
    e^{-i\, c_{A} \vb{k_{\rm ph}} \vb{r}}\, \displaystyle\frac{m_{A}}{m_{p} + m_{A}}\, F_{p,\, {\rm el}} -
    e^{i\, c_{p} \vb{k_{\rm ph}} \vb{r}}\, \displaystyle\frac{m_{p}}{m_{p} + m_{A}}\, F_{A,\, {\rm el}}
  \Bigr], \\

  \vb{M}_{\rm eff} (\vb{k}_{\rm ph}, \vb{r}) =
  e^{i\, \vb{k_{\rm ph}} \vb{r}}\,
  \Bigl[
    e^{-i\, c_{A} \vb{k_{\rm ph}} \vb{r}}\,  \displaystyle\frac{m_{A}}{m_{p} + m_{A}}\, \vb{F}_{p,\, {\rm mag}} -
    e^{i\, c_{p} \vb{k_{\rm ph}} \vb{r}}\,  \displaystyle\frac{m_{p}}{m_{p} + m_{A}}\, \vb{F}_{A,\, {\rm mag}}
  \Bigr].
\end{array}
\label{eq.13.1.8}
% \label{eq.2.13.1.3}
\end{equation}
Here,
$F_{{\rm p},\, {\rm el}}$,
$F_{A,\, {\rm el}}$,
$\vb{F}_{{\rm p},\, {\rm mag}}$,
$\vb{F}_{A,\, {\rm mag}}$,
$\vb{D}_{A 1,\, {\rm el}}$,
$\vb{D}_{A 2,\, {\rm el}}$,
$D_{A 1,\, {\rm mag}}$,
$D_{A 2,\, {\rm mag}}$,
$\vb{D}_{{\rm p},\, {\rm k}}$,
$\vb{D}_{A,\, {\rm k}}$,
$D_{{\rm p}, P\, {\rm el}}$,
$D_{A,P\, {\rm el}}$,
$\vb{D}_{{\rm p}, P\, {\rm mag}}$,
$\vb{D}_{A,P\, {\rm mag}}$
are electric and magnetic form factors defined in
Appendix~\ref{sec.app.short.2.9} [see Eqs.~(\ref{eq.app.short.2.9.7}), (\ref{eq.app.short.2.9.b.4}), (\ref{eq.app.short.2.9.b.4}), (\ref{eq.app.short.2.9.c.2}), (\ref{eq.app.short.2.9.d.2})].
% -----------------------------------------------------------------------------------------------------------------------

%-----------------------------------------------------------------------------------------------------------------------
\subsection{Dipole approximation of effective electric charge and magnetic moment of nuclear system
\label{sec.13}}

After calculations, we obtain the matrix elements for the coherent bremsstrahlung
[see Appendix~\ref{sec.2.13}, Eqs.%(\ref{eq.2.13.2.6}),
(\ref{eq.2.13.5.5})]
\begin{equation}
\begin{array}{lll}
\vspace{1.5mm}
  M_{p}^{(E,\, {\rm dip})} =
  i \hbar\, (2\pi)^{3}
  \displaystyle\frac{2\, \mu_{N}\,  m_{\rm p}}{\mu}\;
  % \displaystyle\frac{e}{\mu c}\;
  Z_{\rm eff}^{\rm (dip)}\;
  \displaystyle\sum\limits_{\alpha=1,2} \vb{e}^{(\alpha)} \cdot \vb{I}_{1}, \\

  M_{p}^{(M,\, {\rm dip})} =
  \hbar\, (2\pi)^{3}\, \displaystyle\frac{\mu_{N}}{\mu} \cdot \alpha_{M} \cdot
  (\vb{e}_{\rm x} + \vb{e}_{\rm z})\,
  \displaystyle\sum\limits_{\alpha=1,2}
  \Bigl[ \vb{I}_{1} \times \vb{e}^{(\alpha)} \Bigr]
\end{array}
\label{eq.resultingformulas.1}
\end{equation}
and for incoherent bremsstrahlung [Appendixes~\ref{sec.2.15}--\ref{sec.2.16}, Eqs.~(\ref{eq.2.14.7}), (\ref{eq.2.15.6}), (\ref{eq.2.16.5})]
\begin{equation}
\begin{array}{lll}
\vspace{1.4mm}
  M_{\Delta E} = 0, \\
\vspace{1.4mm}
  M_{\Delta M} = i\, \hbar\, (2\pi)^{3}\, \mu_{N}\, f_{1} \cdot |\vb{k}_{\rm ph}| \cdot Z_{\rm A} (\vb{k}_{\rm ph}) \cdot I_{2}, \\

  M_{k} =
    -\, i\, \hbar\, (2\pi)^{3}\, \mu_{N} \cdot k_{\rm ph}\, z_{\rm p}\: \mu_{\rm p} \cdot I_{3} -
    \displaystyle\frac{\bar{\mu}_{\rm pn}}{f_{1}} \cdot M_{\Delta M}.
\end{array}
\label{eq.resultingformulas.2}
\end{equation}
Integrals are
\begin{equation}
\begin{array}{lllll}
\vspace{0.5mm}
  \vb{I}_{1} = \biggl\langle\: \Phi_{\rm p - nucl, f} (\vb{r})\; \biggl|\, e^{-i\, \vb{k}_{\rm ph} \vb{r}}\; \vb{\displaystyle\frac{d}{dr}} \biggr|\: \Phi_{\rm p - nucl, i} (\vb{r})\: \biggr\rangle_\mathbf{r}, \\
\vspace{0.5mm}
  I_{2} = \Bigl\langle \Phi_{\rm p - nucl, f} (\vb{r})\; \Bigl|\, e^{i\, c_{p}\, \vb{k_{\rm ph}} \vb{r}}\, \Bigr|\, \Phi_{\rm p - nucl, i} (\vb{r})\: \Bigr\rangle_\mathbf{r}, \\
  I_{3} = \Bigl\langle \Phi_{\rm p - nucl, f} (\vb{r})\; \Bigl|\, e^{-i\, c_{A}\, \vb{k_{\rm ph}} \vb{r}}\, \Bigr|\, \Phi_{\rm p - nucl, i} (\vb{r})\: \Bigr\rangle_\mathbf{r}.
\end{array}
\label{eq.resultingformulas.3}
\end{equation}
The effective electric charge and magnetic moment (\ref{eq.13.1.8}) are [see Eqs.~(\ref{eq.2.13.2.1}), (\ref{eq.app.matr_el.coh_mag.7})]
\begin{equation}
\begin{array}{llll}
  Z_{\rm eff}^{\rm (dip)} (\vb{k}_{\rm ph}) = \displaystyle\frac{m_{A}\, z_{\rm p} - m_{\rm p}\, Z_{\rm A}(\vb{k}_{\rm ph})}{m_{\rm p} + m_{A}}, \\

  \vb{M}_{\rm eff}^{\rm (dip)} (\vb{k}_{\rm ph}) =
  \Bigl[
    z_{\rm p}\, m_{A}\, \mu_{\rm p} -
    Z_{\rm A} (\vb{k}_{\rm ph})\: m_{p}\, \bar{\mu}_{\rm pn}
  \Bigr] \cdot
  \displaystyle\frac{m_{p}}{m_{p} + m_{A}}\; (\vb{e}_{\rm x} + \vb{e}_{\rm z}),
\end{array}
\label{eq.resultingformulas.4}
\end{equation}
and we obtain
\begin{equation}
\begin{array}{lll}
  \alpha_{M} =
    \Bigl[ Z_{\rm A} (\vb{k}_{\rm ph})\: m_{p}\, \bar{\mu}_{\rm pn} - z_{\rm p}\, m_{A}\, \mu_{\rm p} \Bigr] \cdot \displaystyle\frac{m_{p}}{m_{p} + m_{A}}, &
  f_{1} = \displaystyle\frac{A-1}{2A}\: \bar{\mu}_{\rm pn}.
\end{array}
\label{eq.resultingformulas.5}
\end{equation}
Here,
$\bar{\mu}_{\rm pn} = \mu_{\rm p} + \kappa\,\mu_{\rm n}$,
$\kappa = (A-N)/N$, $A$ and $N$ are numbers of nucleons and neutrons in nucleus,
$\mu_{\rm p}$ and $\mu_{\rm n}$ are magnetic moments of proton and neutron.
% где $\kappa = (A-N)/N$, $A$ и $N$ --- числа нуклонов и нейтронов в ядре.

Calculation of integrals (\ref{eq.resultingformulas.3}) is straightforward. In result, we obtain:
% We rewrite the found solutions [see Eqs.~(\ref{eq.simplestcase.4}), (\ref{eq.simplestcase.7})]:
%
\begin{equation}
\begin{array}{lll}
\vspace{1.5mm}
  M_{p}^{(E,\, {\rm dip})} & \simeq &
  -\, \hbar\, (2\pi)^{3}\, \displaystyle\frac{2\, \mu_{N}\,  m_{\rm p}}{\mu}\; Z_{\rm eff}^{\rm (dip)} \cdot \displaystyle\frac{\sqrt{3}}{6} \cdot J_{1}(0,1,0), \\

\vspace{1.5mm}
  M_{p}^{(M,\, {\rm dip})} & \simeq &
  -\, i\, \hbar\, (2\pi)^{3}\, \displaystyle\frac{\mu_{N}\, \alpha_{M}}{\mu} \cdot \displaystyle\frac{\sqrt{3}}{6} \cdot J_{1}(0,1,0), \\

\vspace{1.5mm}
  M_{\Delta M} & \simeq  &
   \hbar\, (2\pi)^{3}\, \mu_{N}\, f_{1}\, k_{\rm ph}\, Z_{\rm A} (\vb{k}_{\rm ph}) \cdot \displaystyle\frac{\sqrt{3}}{2} \cdot \tilde{J}\, (-c_{p}, 0,1,1), \\

  M_{k} & \simeq &
    -\, \hbar\, (2\pi)^{3}\, \mu_{N} \cdot k_{\rm ph}\, z_{\rm p}\: \mu_{\rm p}^{\rm (an)} \cdot \displaystyle\frac{\sqrt{3}}{2} \cdot \tilde{J}\, (c_{A}, 0,1,1) -
    \displaystyle\frac{\bar{\mu}_{\rm pn}^{\rm (an)}}{f_{1}} \cdot M_{\Delta M},
\end{array}
\label{eq.resultingformulas.6}
% \label{eq.simplestcase.14}
\end{equation}
where
\begin{equation}
\begin{array}{llllll}
  J_{1}(l_{i},l_{f},n) & = & \displaystyle\int\limits^{+\infty}_{0} \displaystyle\frac{dR_{i}(r, l_{i})}{dr}\: R^{*}_{f}(l_{f},r)\, j_{n}(k_{\rm ph}r)\; r^{2} dr, &
  \tilde{J}\,(c, l_{i},l_{f},n) & = & \displaystyle\int\limits^{+\infty}_{0} R_{i}(l_{i}, r)\, R^{*}_{f}(l_{f},r)\, j_{n}(c\, k_{\rm ph}r)\; r^{2} dr.
%  \breve{J}\,(c_{A}, l_{i}, l_{f},n) & = & \displaystyle\int\limits^{+\infty}_{0} R_{i}(r)\, R^{*}_{l,f}(r)\, V(\mathbf{r})\, j_{n}(c_{A}\,kr)\; r^{2} dr.
\end{array}
\label{eq.resultingformulas.7}
\end{equation}
Here, $R_{i,f}$ is radial part of wave function $\Phi_{\rm p - nucl} (\vb{r})$ in $i$-state or $f$-state,
$j_{n}(k_{\rm ph}r)$ is spherical Bessel function of order $n$.
% *******************************************************************************************************************

% *******************************************************************************************************************
We define cross-sections of the emitted bremsstrahlung photons on the basis of the full matrix element $p_{fi}$ % (\ref{eq.2.5.6.2})
in frameworks of formalism given in Refs.~\cite{Maydanyuk_Zhang_Zou.2016.PRC,Maydanyuk.2012.PRC,Maydanyuk_Zhang.2015.PRC}
(see Eq.~(22) in Ref.~\cite{Maydanyuk_Zhang_Zou.2016.PRC}, reference therein) and we do not repeat it in this paper.
Finally, we obtain the bremsstrahlung cross-sections as%
%
% \footnote{We obtain the formula (\ref{eq.2.6.1}) in dependence on mass of proton $m_{\rm p}$ while in Ref.~\cite{Maydanyuk.2012.PRC} we had the bremsstrahlung probability (49) in dependence on the reduced mass $\mu$.
% Such a difference is explained by that in the current paper we develop formalism on the basis of the emission operator of the many-nucleon system (\ref{eq.2.2.3})
% while in Ref.~\cite{Maydanyuk.2012.PRC} we started the formalism on the basis of the operator of emission (4) of the proton-nucleus system defined via the reduced mass of proton and nucleus.}
%
\begin{equation}
\begin{array}{llll}
  \displaystyle\frac{d \sigma}{dw_{\rm ph}} =
    \displaystyle\frac{e^{2}}{2\pi\,c^{5}}\: \displaystyle\frac{w_{\rm ph}\,E_{i}}{m_{\rm p}^{2}\,k_{i}}\:
    \bigl| p_{fi} \bigr|^{2}, &

  \displaystyle\frac{d^{2} \sigma}{dw_{\rm ph}\, d \cos \theta} =
    \displaystyle\frac{e^{2}}{2\pi\,c^{5}}\: \displaystyle\frac{w_{\rm ph}\,E_{i}}{m_{\rm p}^{2}\,k_{i}}\:
    \bigl\{ p_{fi} \displaystyle\frac{d\,p_{fi}^{*}}{d\, \cos \theta}  + c.\,c. \bigr\}, &
  M_{\rm full} = - \displaystyle\frac{e}{m_{\rm p}}\, p_{fi},
\end{array}
\label{eq.model.bremprobability.1}
\end{equation}
%
% where $p_{fi}$ is proportional to the electrical component $p_{\rm el}$ in Eqs.~(10) in \cite{Maydanyuk.2012.PRC}
% [with the additional factor of $2\, e^{-\, (a^{2} k_{x}^{2} + b^{2} k_{y}^{2} + c^{2} k_{z}^{2})\,/4}$ and the included effective charge $\tilde{Z}_{\rm eff}^{\rm (dip)}$] and
% $d\,p_{fi} (\theta_{f})\, / d\,\cos{\theta_{f}}$ is defined by the same way as $d\,p\, (k_{i}, k_{f}, \theta_{f})\, / d\,\cos{\theta_{f}}$ in Ref.~\cite{Maydanyuk.2012.PRC}.
%
where c.\,c. is complex conjugation.
We calculate the different contributions of the emitted photons to the full bremsstrahlung spectrum.
For estimation of the interesting contribution, we use the corresponding matrix element of emission.
In this paper we calculate the matrix elements on the basis of wave functions with quantum numbers $l_{i}=0$, $l_{f}=1$ and $l_{\rm ph}=1$
[here, $l_{i}$ and $l_{f}$ are orbital quantum numbers of wave function $\Phi_{\rm p - nucl} (\vb{r})$ % defined in Eq.~(\ref{eq.18.1.2})
for states before emission of photon and after this emission,
$l_{\rm ph}$ is orbital quantum number of photon in the multipole approach].
% defined in Eq.~(\ref{eq.app.integrals.3})]
% *******************************************************************************************************************

% *******************************************************************************************************************
% \newpage
\section{Analysis
\label{sec.analysis}}

\subsection{The spectra for $\Delta$-nuclei, coherent contributions VS incoherent ones
\label{sec.analysis.1}}

Previously, processes with emission of photons in the scattering of protons on the \isotope[12]{C}, \isotope[40]{Ca}, \isotope[208]{Pb} nuclei in the $\Delta$-resonance energy region were analyzed~\cite{Gil_Oset.1998.PLB.v416}.
So, we use these nuclei for analysis.
To test our calculations of emission of bremsstrahlung photons,
we choose the scattering of $p + \isotope[197]{Au}$ at proton beam energy of 190~MeV.
%
% where experimental bremsstrahlung data \cite{Goethem.2002.PRL} are obtained with high accuracy
% (note that calculations in Ref.~\cite{Gil_Oset.1998.PLB.v416} were predictions, not tested on the bremsstrahlung experimental data).
%
Such a choice is explained by the following.
We estimate that accuracy of measurements of bremsstrahlung photons and presented data are the highest for the nucleus \isotope[197]{Au} (see Ref.~\cite{Goethem.2002.PRL})
in comparison with other experiments in measurements of bremsstrahlung in the proton nucleus scattering.
Even, if to compare those data with other measurements of bremsstrahlung photons for other nuclear reactions
($\alpha$-decay, fission, $\alpha$-nucleus scattering, nucleus-nucleus scattering), data in Ref.~\cite{Goethem.2002.PRL} were obtained with the highest accuracy
(those data provide reach information).
So, those data is a good basis for analysis and tests of different models.
By such a reason, calculations in this paper are compared with experimental data in Ref.~\cite{Goethem.2002.PRL} for the nucleus \isotope[197]{Au} (at energies of proton beam used in experiments).
Those calculations are performed for test the model, before its next use in the paper.

One can note other experimental bremsstrahlung data for $p + \isotope[208]{Pb}$ at energy of proton beam of 140~MeV obtained by Edington and Rose in Ref.~\cite{Edington.1966.NP},
and for $p + \isotope[64]{Cu}$, $p + \isotope[107]{Ag}$ at energy of proton beam of 72~MeV obtained by Kwato~Njock et al. in Ref.~\cite{Kwato_Njock.1988.PLB}.
% $p + \isotope[9]{Be}$, $p + \isotope[12]{C}$
One can find some contradiction between those data and data in Ref.~\cite{Goethem.2002.PRL}.
% (I omit analysis of those results, that was in the previous papers partially).
But, once again, accuracy of the data in Ref.~\cite{Goethem.2002.PRL} is higher, so we choose data in Ref.~\cite{Goethem.2002.PRL} for tests in the manuscript.

We calculate
wave function of relative motion between proton and center-of-mass of nucleus numerically
concerning to the proton-nucleus potential in form of $V (r) = v_{c}(r) + v_{N}(r) + v_{\rm so}(r) + v_{l} (r)$,
where $v_{c}(r)$, $v_{N}(r)$, $v_{\rm so}(r)$, and $v_{l} (r)$ are Coulomb, nuclear, spin-orbital, and centrifugal components, respectively.
Parameters of this potential are defined in Eqs.~(46)--(47) in Ref.~\cite{Maydanyuk_Zhang.2015.PRC}.
We calculate the bremsstrahlung cross-section by Eq.~(\ref{eq.model.bremprobability.1}),
where we include
matrix elements of coherent emission $M_{p}^{(E,\, {\rm dip})}$, $M_{p}^{(M,\, {\rm dip})}$ in Eqs.~(\ref{eq.resultingformulas.1}),
and matrix elements of incoherent emission $ M_{\Delta M}$, $M_{k}$ in Eqs.~(\ref{eq.resultingformulas.2}).
% The boundary conditions and normalization are used in form of~(B.1)--(B.9) in  \cite{Maydanyuk.2011.JPG}.
%
% \footnote{One proton from beam can transfer energy to one nucleon of nucleus for transition $NN \to \Delta N$ with formation of $\Delta$-resonance in nucleus.
% Other protons of beam with nucleons of nucleus-target and $\Delta$-resonance can emit bremsstrahlung photons.
% We analyze such a possibility in this paper.}

\begin{figure}[htbp]
\centerline{\includegraphics[width=90mm]{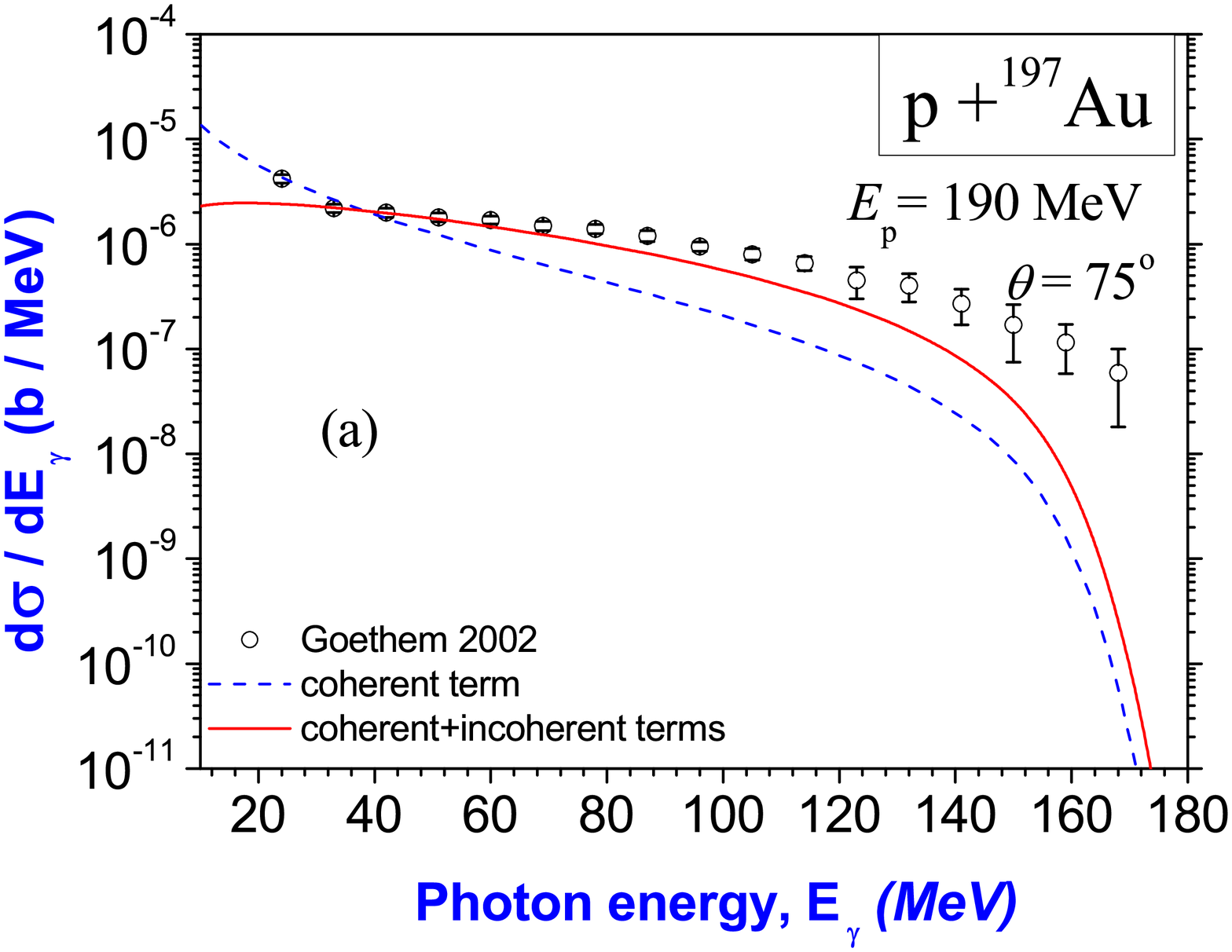}
\hspace{-1mm}\includegraphics[width=90mm]{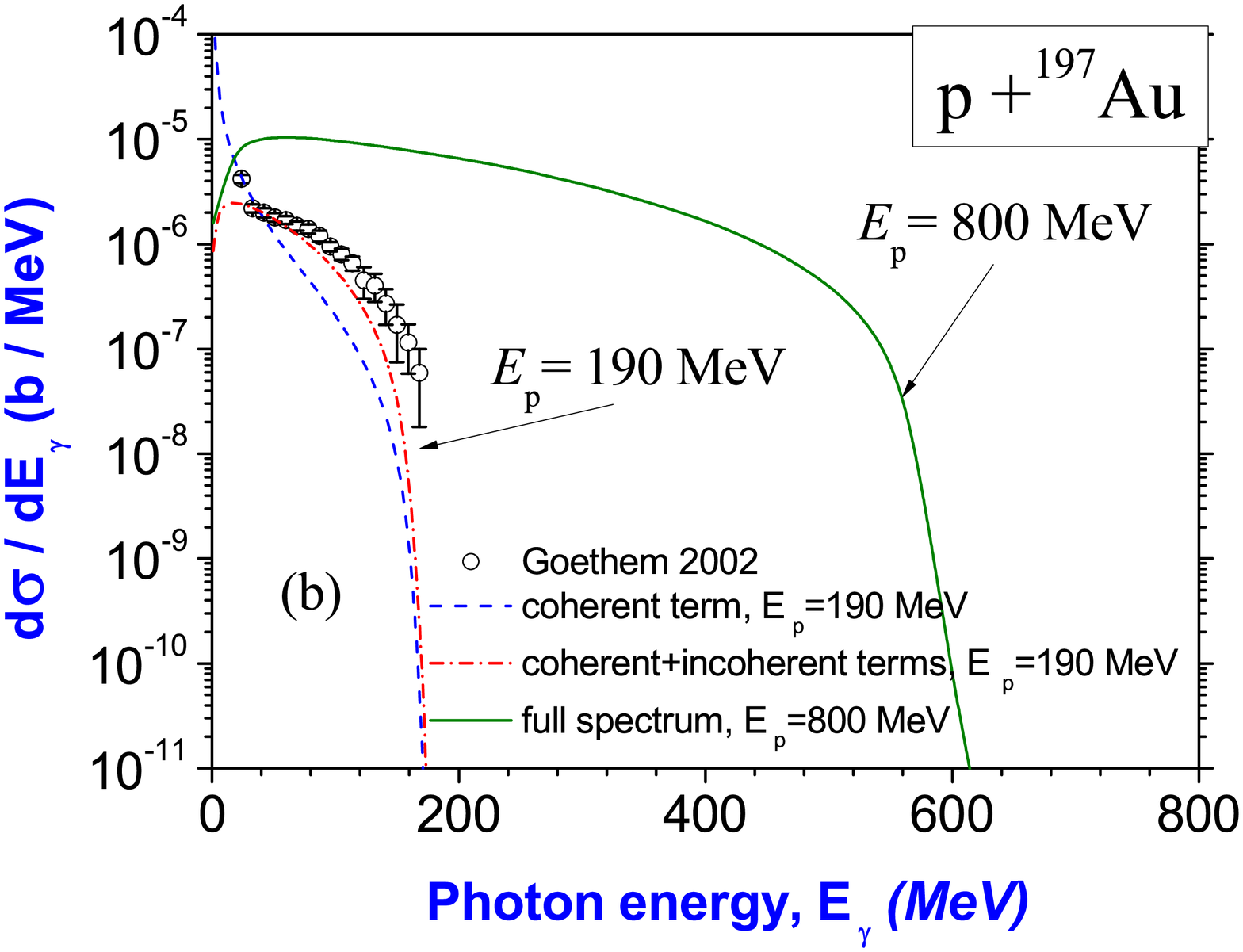}}
% \hspace{-1mm}\includegraphics[width=90mm]{Figures/fig.1b_p197Au_190MeV_magnetic-momentums_error.eps}}
\vspace{-4mm}
\caption{\small (Color online)
Panel (a):
The calculated bremsstrahlung spectra (with coherent and incoherent terms)
in the scattering of protons off the \isotope[197]{Au} nuclei at energy of proton beam of $E_{\rm p}=190$~MeV
in comparison with experimental data~\cite{Goethem.2002.PRL}
[matrix elements are defined in Eqs.~(\ref{eq.resultingformulas.1})--(\ref{eq.resultingformulas.2}),
$Z_{A} (k_{\rm ph})\simeq Z_{A}$ is electric charge of nucleus,
we normalize each calculated spectrum on the second point of experimental data].
Here,
experimental data given by open circles (Goethem 2002) are extracted from Ref.~\cite{Goethem.2002.PRL},
blue dashed line is coherent contribution defined by $M_{p}^{(E,\, {\rm dip})}$,
red solid line is full spectrum with coherent and incoherent contributions defined by $M_{p}^{(E,\, {\rm dip})}$, $M_{p}^{(M,\, {\rm dip})}$, $M_{\Delta M}$ and $M_{k}$.
%
% \vspace{1.5mm}
% \newline
Panel (b):
New calculated spectrum of full bremsstrahlung for \isotope[197]{Au} at energy of proton beam of $E_{\rm p}=800$~MeV
in comparison with the bremsstrahlung spectrum for the same reaction at energy of proton beam of $E_{\rm p}=190$~MeV shown in Fig.~(a).
%
% Ration between incoherent and coherent contributions in dependence on energy of emitted bremsstrahlung photon
% [we define ratio as $\varepsilon = \sigma_{\rm incoh}/\sigma_{\rm coh}$, in calculations we use factor of incoherence of $f_{\rm incoh}=1.0$]
% related to full spectrum shown by dashed brown line in figure (a)].
% One can see that
% (a) incoherent emission is essentially more intensive than the coherent emission,
% (b) role of incoherent processes is increased at increasing of energy of photon,
% (c) better agreement with experimental data at $f_{\rm incoh}=0.001$ (than at $f_{\rm incoh}=1$) indicates on presence of some unknown effect,
% which highly suppresses the incoherent processes.
%
% factor $f_{\rm incoh}$ which suppresses the intensity of incoherent processes.
% One can see clear changes of the full spectrum in dependence on factor $f_{\rm incoh}$,
% red dash-dotted solid line (at $f=0.001$) corresponds to the best agreement with experimental data.
% This result confirms the important (and not small) role of incoherent emission in bremsstrahlung.
% (b) Contributions of the bremsstrahlung emission given by term $p_{\rm q,1}$ with different $Q$ in comparison with electric and magnetic emissions.
% Можно видеть, что магнитное излучение вносит вклад около 28 процентов в диапазоне энергий 50--300~кэВ.
\label{fig.1}}
\end{figure}

Results of previous study of bremsstrahlung emission
\cite{Maydanyuk_Zhang.2015.PRC,Maydanyuk.2012.PRC,Maydanyuk_Zhang_Zou.2016.PRC,Liu_Maydanyuk_Zhang_Liu.2019.PRC.hypernuclei}
% \cite{Maydanyuk_Zhang.2015.PRC,Maydanyuk.2012.PRC,Maydanyuk_Zhang_Zou.2016.PRC,Liu_Maydanyuk_Zhang_Liu.2019.PRC.hypernuclei,Maydanyuk_Zhang_Zou.2019.PRC.microscopy}
show that incoherent emission is essentially larger than coherent one.
By such a reason, we start calculations for \isotope[197]{Au}, where experimental data exist.
Results of such calculations with inclusion of coherent and incoherent contributions in comparison with experimental data are presented in Fig.~\ref{fig.1}~(a).
The calculated spectra are normalized on one point of these experimental data.
% These calculations are shown in Fig.~1~(a).
% All other calculated spectra for different nuclei and energies of proton beam use the same coefficient of normalization (this gives possibility to predict new spectra).
%
This shows that bremsstrahlung model with the coherent and incoherent contributions provides the spectrum in good agreement with experimental data.
Note that the first point in experimental data is not explained in satisfactory way by model with included incoherent bremsstrahlung contribution.
Without the incoherent contribution, the calculated renormalized spectrum is in worse agreement with experimental data.
Presence of such a point could indicate on existence of some unknown processes forming more intensive coherent bremsstrahlung emission at low energies of photons.
That problem can motivate on next investigations and developments of the model or more precise measurements of emission of photons at low energies in possible future experiments.
But, in current research we omit analysis of this problem.

We obtain
\begin{equation}
\begin{array}{lll}
  M_{p}^{(E,\, {\rm dip})} + M_{p}^{(M,\, {\rm dip})} =
  M_{p}^{(E,\, {\rm dip})} \cdot \Bigl\{ 1 + i \displaystyle\frac{\alpha_{M}}{2\, m_{\rm p}\, Z_{\rm eff}^{\rm (dip)}} \Bigr\}.
\end{array}
\label{eq.analysis.1.1}
% \label{eq.simplestcase.16}
\end{equation}
This formula shows role of magnetic emission on the basis of electric one in the full coherent emission of photons.
% Here, factor $i \displaystyle\frac{\alpha_{M}}{2\, m_{\rm p}\, Z_{\rm eff}^{\rm (dip)}}$ describes suppression or reinforcement of electric emission of the bremsstrahlung photons
% (which is mainly studied in many papers of other authors)
% via additional inclusion of magnetic emission [that factor is different for different nuclei].
%
From previous research we know, that increasing of energy of protons beam in nuclear scattering increases intensity of bremsstrahlung emission.
Indeed, new calculations for the same scattering $p + \isotope[197]{Au}$ but at energy of protons beam of $E_{\rm p}=800$~MeV are presented in Fig.~\ref{fig.1}~(b), confirming this property of bremsstrahlung.
Also from this figure one can see that difference in intensity of bremsstrahlung for these two cases is essential.

\begin{figure}[htbp]
\centerline{\includegraphics[width=90mm]{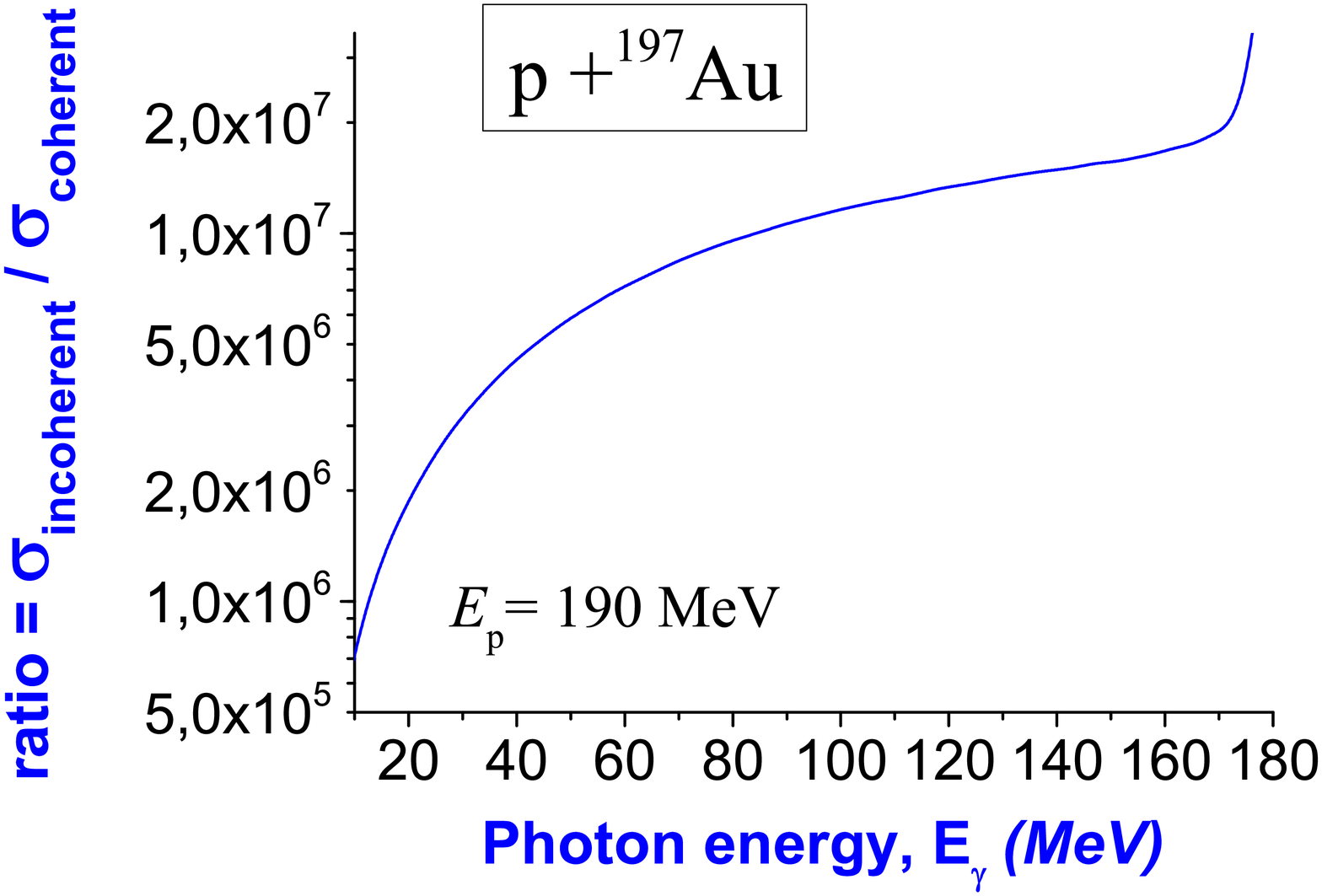}
% \hspace{-1mm}\includegraphics[width=90mm]{Figures/fig.1b_isobara_p197Au_800MeV_spectra.eps}
}
% \hspace{-1mm}\includegraphics[width=90mm]{Figures/fig.1b_p197Au_190MeV_magnetic-momentums_error.eps}}
\vspace{-4mm}
\caption{\small (Color online)
% Panel (a):
Ratio between incoherent and coherent bremsstrahlung contributions
% The calculated bremsstrahlung spectra (with coherent and incoherent terms)
in the scattering of protons off the \isotope[197]{Au} nuclei at energy of proton beam of $E_{\rm p}=190$~MeV
[matrix elements are defined in Eqs.~(\ref{eq.resultingformulas.1})--(\ref{eq.resultingformulas.2}),
$Z_{A} (k_{\rm ph})\simeq Z_{A}$ is electric charge of nucleus
% we normalize each calculated spectrum on the second point of experimental data
].
In the coherent bremsstrahlung, the magnetic emission based on $M_{p}^{(M,\, {\rm dip})}$ is almost the same as electric emission based on $M_{p}^{(E,\, {\rm dip})}$:
we obtain $\frac{\sigma_{\rm mag}^{\rm (coh)}}{\sigma_{\rm el}^{\rm (coh)}} = 3.3213$ in the full energy region of photons.
In the incoherent bremsstrahlung, role of background emission based on $M_{k}$ is a little larger than magnetic contribution based on $M_{\Delta M}$:
we obtain $\frac{\sigma_{\rm background}^{\rm (incoh)}}{\sigma_{\rm mag}^{\rm (incoh)}} = 4.04$ in the full energy region of photons.
\label{fig.2}}
\end{figure}

In Fig.~\ref{fig.2} we show ratio between contributions of the full incoherent contribution to the full coherent contribution in the bremsstrahlung emission during this scattering process
considered above at $E_{\rm p}=190$~MeV.
From this result one can see that role of incoherent emission is essentially larger than coherent emission.

\begin{figure}[htbp]
\centerline{\includegraphics[width=90mm]{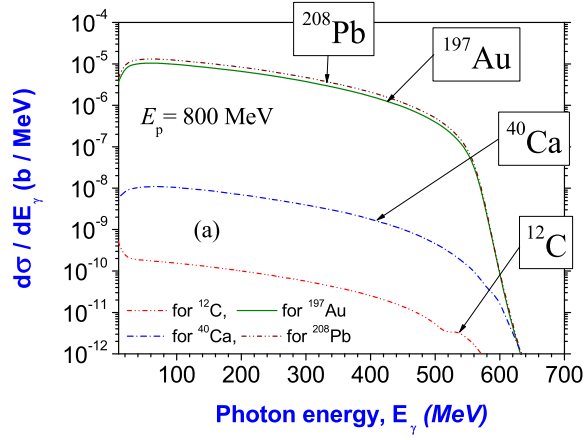}
% \hspace{-1mm}\includegraphics[width=90mm]{Figures/fig.1b_isobara_p197Au_800MeV_spectra.eps}
}
\vspace{-4mm}
\caption{\small (Color online)
The calculated full bremsstrahlung spectra % (with coherent and incoherent terms)
in the scattering of protons off the \isotope[12]{C}, \isotope[40]{Ca}, \isotope[197]{Au}, \isotope[208]{Pb} nuclei at energy of proton beam of $E_{\rm p}=800$~MeV
\label{fig.3}}
\end{figure}

Let us analyze how much the bremsstrahlung spectra are different for different nuclei.
Such calculations for the \isotope[12]{C}, \isotope[40]{Ca}, \isotope[197]{Au}, \isotope[208]{Pb} nuclei at energy of proton beam of $E_{\rm p}=800$~MeV are presented in Fig.~\ref{fig.3}.
From this figure one can conclude that probability of emission is larger for heavier nuclei, and difference between the spectra for light and heavy nuclei is essential.

% Now we will perform the following analysis.
Now let's suppose possibility of formation of $\Delta$-resonance in the nucleus-target and emission of photons in the scattering, following to Ref.~\cite{Gil_Oset.1998.PLB.v416}.%
%
% Here, emission of photons is studied from nuclear matter.
% That paper helps me to understand how to implement that process inside nuclear matter in the bremsstrahlung formalism in fast way, idea was clear and enough simple.
\footnote{However, formalism in Ref.~\cite{Gil_Oset.1998.PLB.v416} does not include quantum fluxes, which can be useful in analysis of scattering.
As example, one can note study of fusion in capture of $\alpha$-particles by nuclei (see Refs.~\cite{Maydanyuk.2015.NPA,Maydanyuk_Zhang_Zou.2017.PRC}, reference therein on that method, its applications on other reactions).
In particular, cross-section of capture can be essentially changed in dependence on different variants of fusion and quantum fluxes inside nucleus-target.
Those processes and effects are not taken into account in the nuclear formalism in Ref.~\cite{Gil_Oset.1998.PLB.v416}.}
One can estimate how much the bremsstrahlung emission is changed after inclusion of $\Delta$-resonance to the formalism.
For that, one can suppose that such photons are emitted by protons in beam and the modified nucleus, where transition of one nucleon to $\Delta$-resonance takes place
(we take transition $p N \to \Delta^{+} N$ in nucleus for analysis in this paper, while another case $n N \to \Delta^{0} N$ can be studied in analogous way).

\begin{figure}[htbp]
\centerline{\includegraphics[width=90mm]{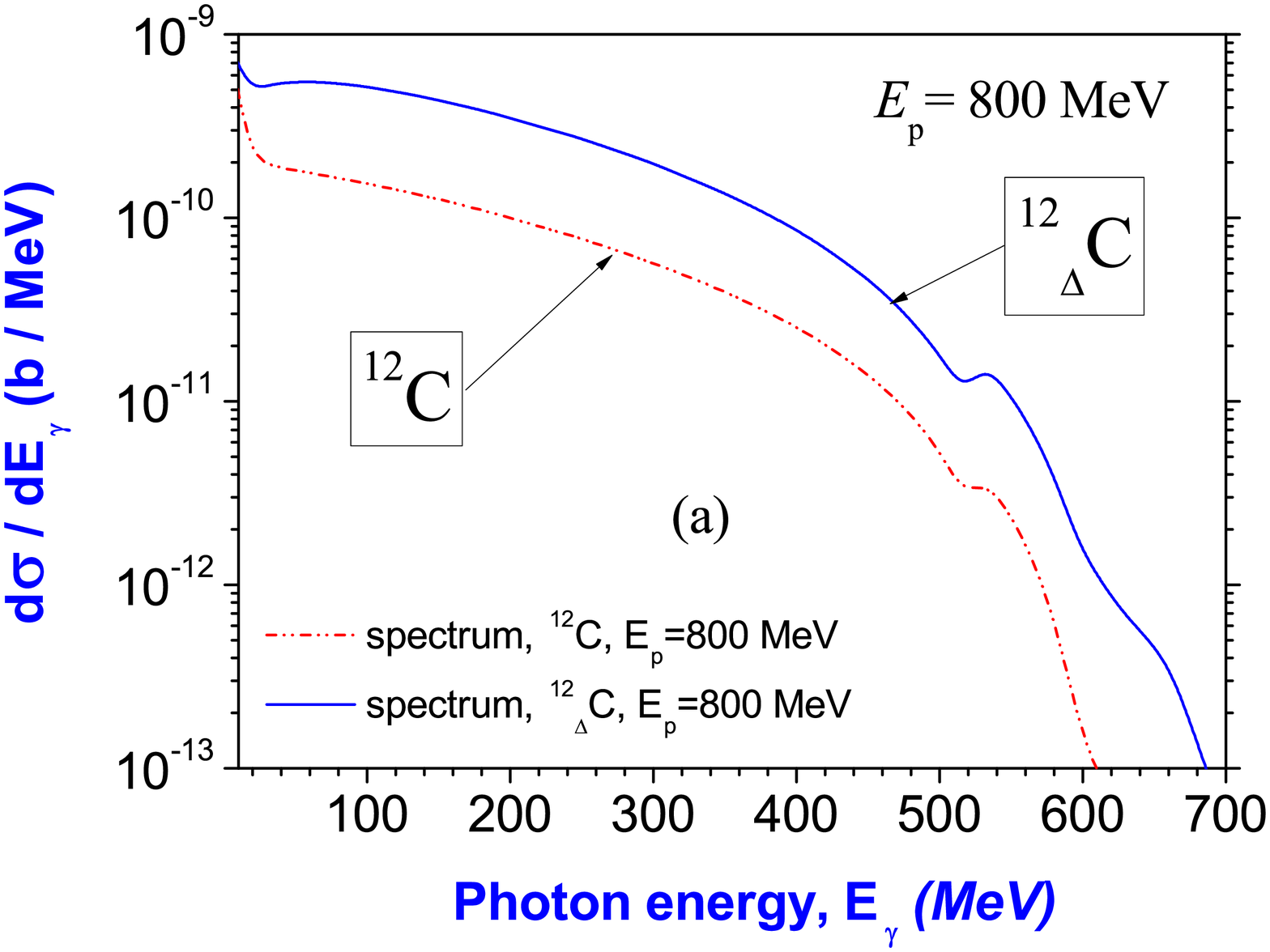}
\hspace{-1mm}\includegraphics[width=90mm]{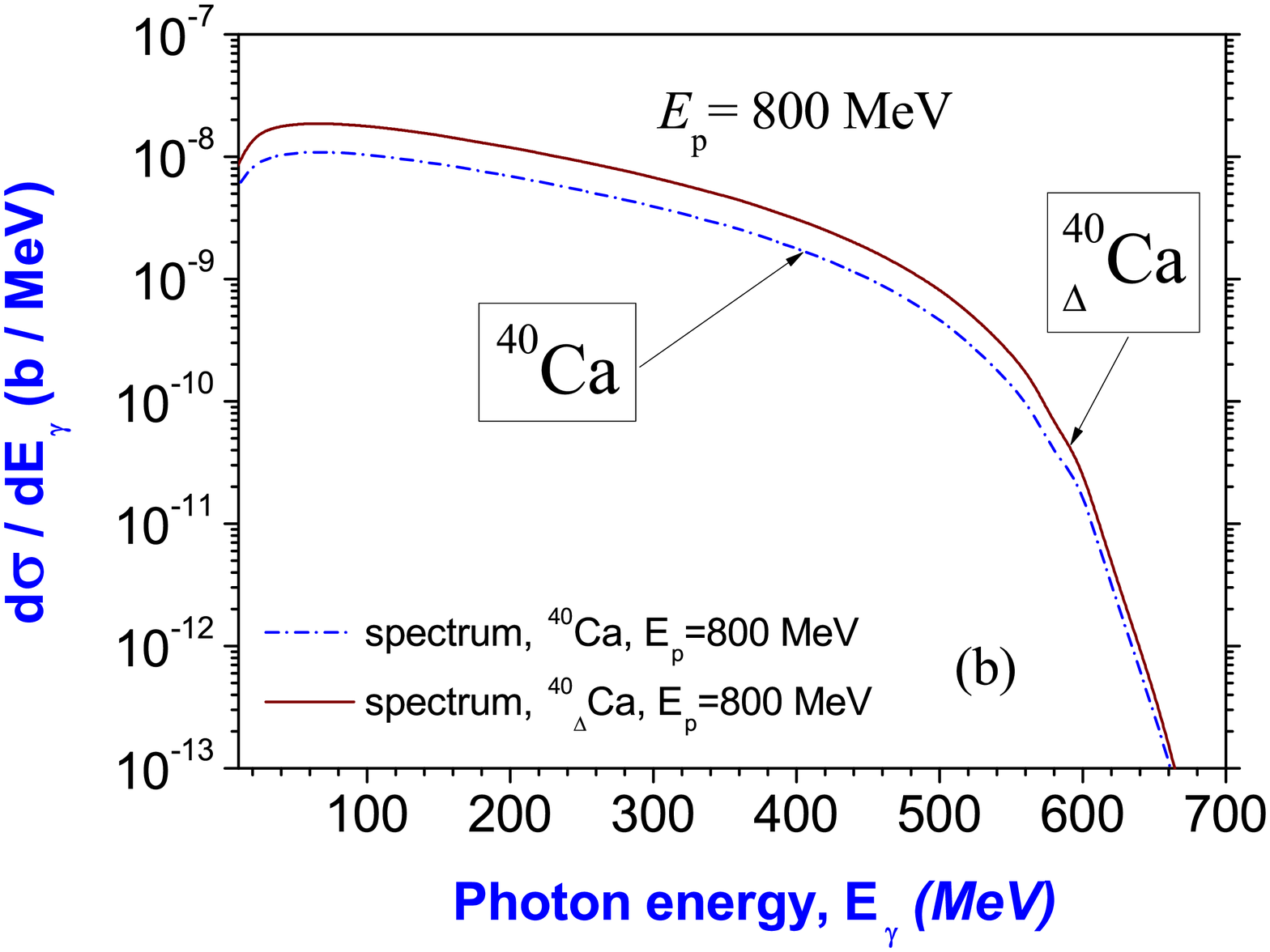}}
\vspace{-4mm}
\caption{\small (Color online)
The calculated full bremsstrahlung spectra % (with coherent and incoherent terms)
in the scattering of protons off
the \isotope[12]{C}, \isotope[12][\Delta]{C} nuclei (a) and
the \isotope[40]{Ca}, \isotope[40][\Delta]{Ca} nuclei (b) at energy of proton beam of $E_{\rm p}=800$~MeV
\label{fig.4}}
\end{figure}

The first question is in which nuclei are the most convenient to find larger difference between the spectra for normal nuclei and nuclei with included $\Delta$-resonance.
Analyzing formalism, one can find that such difference can be from effective electric charge and magnetic moment of proton-nucleus system.
Difference between such characteristics is larger for more light nuclei.
However, probability of photons for more light nuclei is smaller (see Fig.~\ref{fig.3}).
So, we have unclear situation in finding proper direction.
In Fig.~\ref{fig.4} we show calculations of the spectra for the \isotope[12]{C}, \isotope[40]{Ca} nuclei in comparison with
the \isotope[12][\Delta]{C}, \isotope[40][\Delta]{Ca} nuclei.
From such results one can conclude that the bremsstrahlung emission after formation of $\Delta$-resonance from one of nucleons of nucleus is changed not much for light and heavy nuclei, in general.
%-----------------------------------------------------------------------------------------------------------------------

%-----------------------------------------------------------------------------------------------------------------------
\subsection{Nuclei with highest enhancement of bremsstrahlung due to formation of $\Delta$-resonance
\label{sec.analysis.2}}

Now we will find nuclei, when transition from nucleon of nucleus to $\Delta$-resonance maximally reinforces the bremsstrahlung emission in the proton nucleus scattering.
% This is a case, when for normal nucleus.
%
Parameters important in this task are mass of $\Delta$-resonance, its spin, magnetic moment.
Change of mass of baryon in transition from nucleon to $\Delta$-resonances is not much sensitive in calculations of the bremsstrahlung probability,
so we will ignore it for more clear understanding of the model.
From analysis of the model above we find that the matrix elements $M_{\Delta M}$ and $M_{k}$ defined in  Eqs.~(\ref{eq.resultingformulas.2}) give the largest contributions to the bremsstrahlung emission.
On such a basis, from Eqs.~(\ref{eq.resultingformulas.2}) one can find that the most important parameter in this task is $\bar{\mu}_{\rm pn}$.
One can see that the highest change of the probability of bremsstrahlung at transition from nucleon to $\Delta$-resonance is in case when $\bar{\mu}_{\rm pn}$ is minimal for normal nucleus.
In particular, this is a condition when $\bar{\mu}_{\rm pn}$ is equal to zero for normal nucleus,
and we find
\begin{equation}
\begin{array}{lll}
  \displaystyle\frac{Z}{N} =
  \displaystyle\frac{\mu_{\rm p}} {|\mu_{\rm n}|} =
  \displaystyle\frac{2.79284734} {1.91304273} =
  1.459898,
\end{array}
\label{eq.analysis.2.1}
\end{equation}
where $Z$ and $N$ are numbers of protons and neutrons for the normal nucleus-target.
From Eq.~(\ref{eq.analysis.2.1}) one can find more simple condition $Z > N$.
Note that unstable nuclei satisfy to this condition mainly which cannot be used as target.
However, condition (\ref{eq.analysis.2.1}) indicates limit, which can be used in analysis to find the proper nucleus from stable isotopes.
% We conclude that the most suitable nucleus in this research can be found on the basis of analysis of different isotopes.
One can consider Eq.~(\ref{eq.analysis.2.1}) as condition of the highest enhancement of the bremsstrahlung emission due to creation of $\Delta$-resonance in the nucleus-target.

\begin{figure}[htbp]
\centerline{\includegraphics[width=90mm]{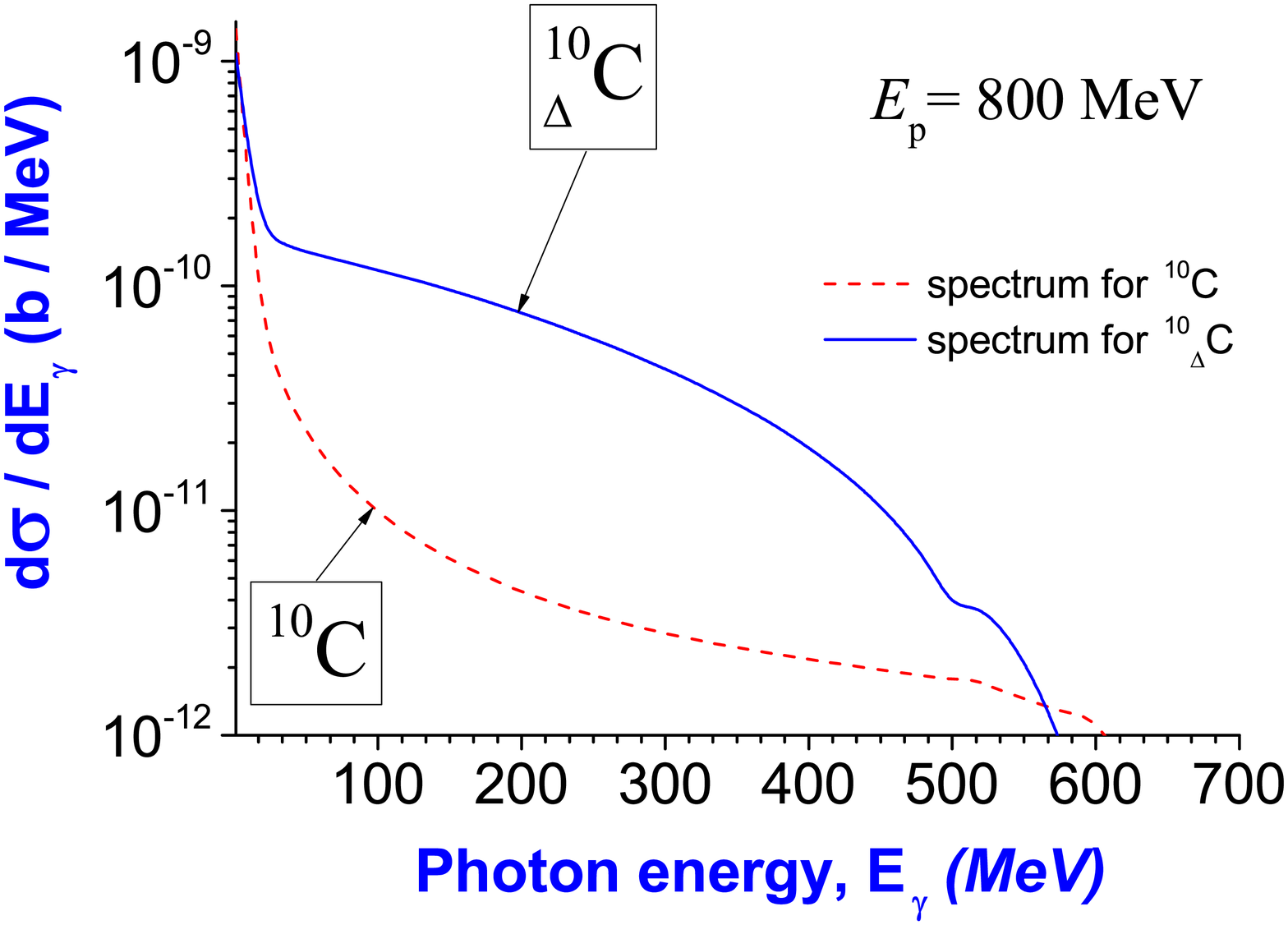}}
% \centerline{\includegraphics[width=90mm]{Figures/fig.5_isobara_p-10C-delta10C_800MeV_spectra.eps}}
% \hspace{-1mm}\includegraphics[width=90mm]{Figures/fig.3b_isobara_p-40Ca-isobara_800MeV_spectra.eps}}
\vspace{-4mm}
\caption{\small (Color online)
The calculated bremsstrahlung spectra in the scattering of protons on
\isotope[10]{C} and \isotope[10][\Delta]{C} at energy of proton beam of $E_{\rm p}=800$~MeV.
One can see that difference between the spectra for \isotope[10]{C} and \isotope[10][\Delta]{C} is larger
than difference between the spectra for \isotope[12]{C} and \isotope[12][\Delta]{C} shown in Fig.~\ref{fig.4}~(a).
\label{fig.5}}
\end{figure}

For example, let us look at isotopes of Carbon.
Nucleus \isotope[12]{C} % is not convenient for such a research as this nucleus
does not satisfy to more simple condition as $Z/N = 1$.
But, unstable nuclear system \isotope[10]{C} satisfies to that condition, and even to condition~(\ref{eq.analysis.2.1})
($Z/N = 1.5$ for \isotope[10]{C}).
% We find - for normal nucleus, and for $\Delta$-nucleus ...
Results of calculations of the bremsstrahlung spectra for \isotope[10]{C} and \isotope[10][\Delta]{C} are shown in Fig.~\ref{fig.5}.
In this figure one can see that the spectrum for \isotope[10][\Delta]{C} is larger than the spectrum for \isotope[10]{C}.
% This confirms correctness of our logic above.
Comparing with results in Fig.~\ref{fig.4}~(a), we see that
the difference between the spectra for normal nucleus \isotope[10]{C} and $\Delta$-nucleus \isotope[10][\Delta]{C} is larger than
difference between the spectra for \isotope[12]{C} and \isotope[12][\Delta]{C}.
% So, proper way in this research is to find more effective isotope.
%-----------------------------------------------------------------------------------------------------------------------

%-----------------------------------------------------------------------------------------------------------------------
\subsection{Bremsstrahlung emission for shortly living $\Delta$-resonance in nucleus
\label{sec.analysis.3}}

\begin{figure}[htbp]
\centerline{\includegraphics[width=90mm]{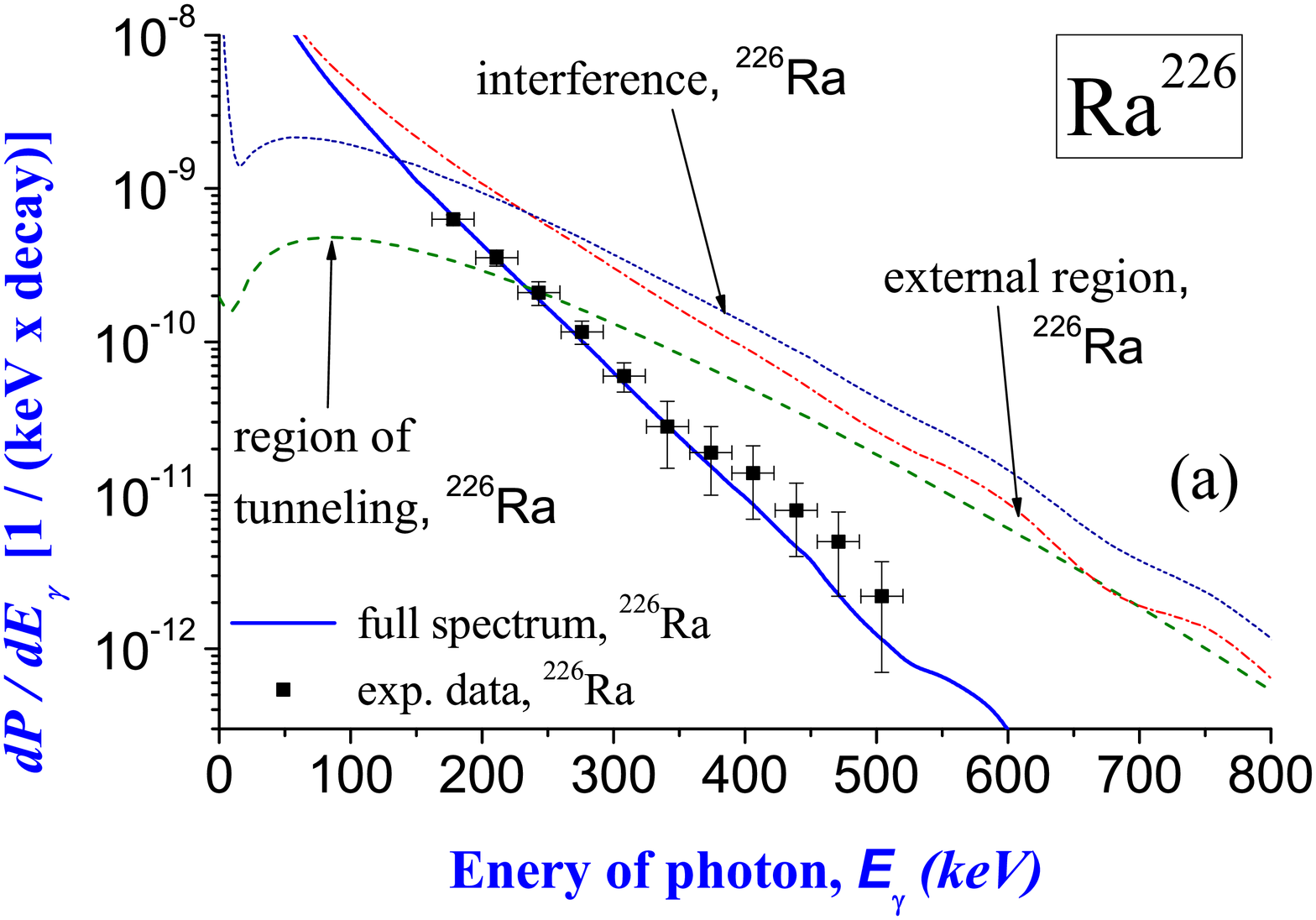}
\hspace{-1mm}\includegraphics[width=90mm]{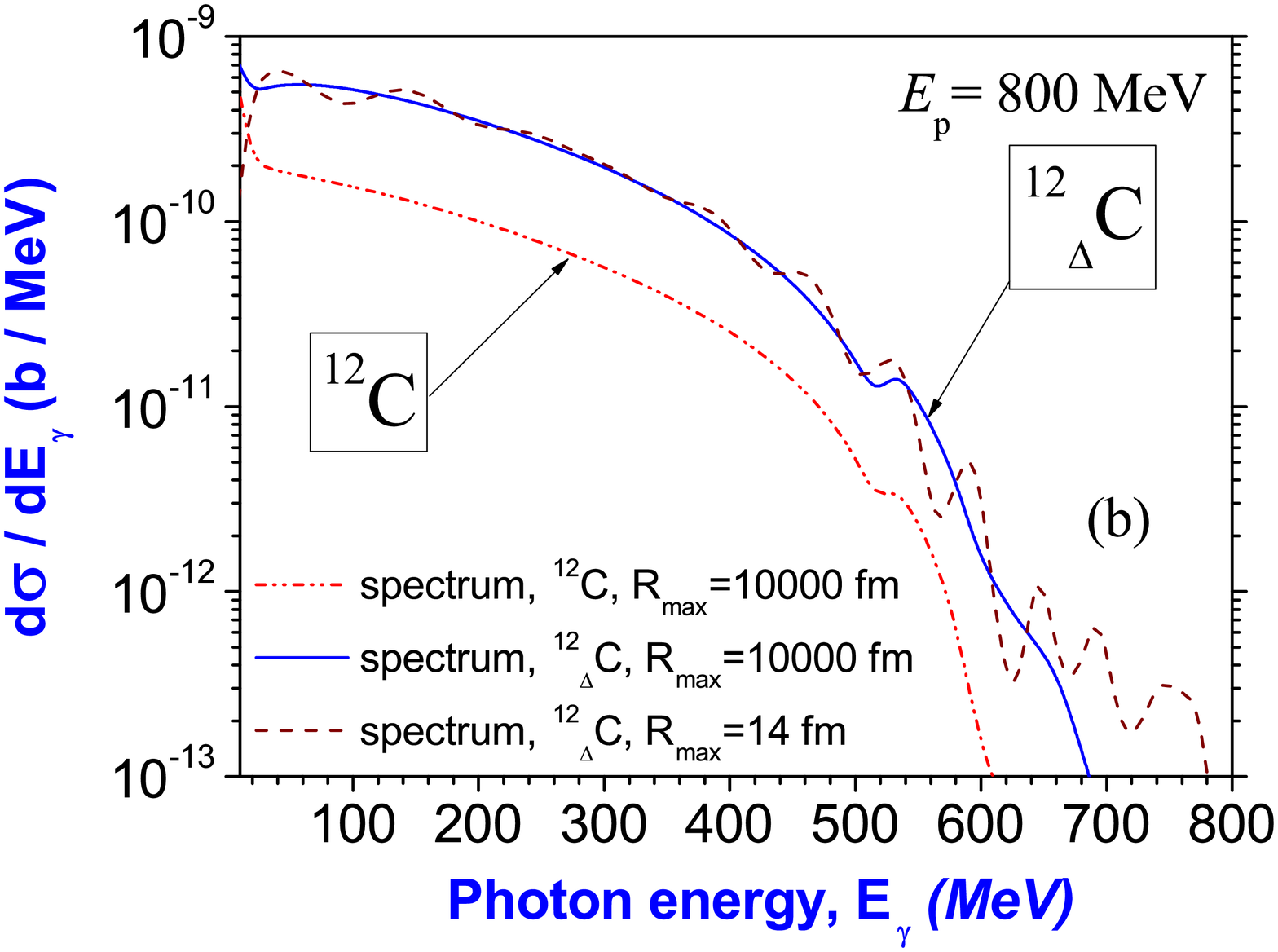}}
% \hspace{-1mm}\includegraphics[width=90mm]{Figures/fig.6b_isobara_p-12C-isobara_r2=14fm_800MeV_spectra.eps}}
\vspace{-4mm}
\caption{\small (Color online)
Panel (a):
The calculated bremsstrahlung contributions in $\alpha$ decay of the \isotope[226]{Ra} nucleus.
Panel (b):
The calculated bremsstrahlung spectra in the scattering of protons on
the \isotope[12]{C} and \isotope[12][\Delta]{C} nuclei at energy of proton beam of $E_{\rm p}=800$~MeV
at different space region of integration.
% Here,
%
One can see that the spectrum for at $R_{\rm max}=14$~fm at high energy region (red dashed line) is larger than
the spectra at full space region (brown solid line).
\label{fig.6}}
\end{figure}

Analyzed maximal enhancement of bremsstrahlung emission presented above is difficult to realize in experiments as condition $Z > N$ is not satisfied for stable nuclei.
Moreover, $\Delta$-resonance is shortly living state of baryon (mean lifetime is about $5.58 \cdot 10^{-24}$~sec).
However, one can remind possibility of emission of bremsstrahlung photons during tunneling in $\alpha$ decay of heavy nuclei analyzed in
Refs.~\cite{Maydanyuk.2006.EPJA,Maydanyuk.2008.EPJA,Maydanyuk.2008.MPLA,Maydanyuk.2009.NPA}.
As it was shown in Ref.~\cite{Maydanyuk.2008.MPLA} (see Fig.~4 in that paper for \isotope[214]{Po} and \isotope[226]{Ra}), exclusion of possibility of emission of photons from the external space region outside potential barrier of $\alpha$ decay does not decrease the full bremsstrahlung spectrum, but gives opposite effect. In particular, there is destructive interference between bremsstrahlung emission from tunneling region and bremsstrahlung emission from the external region outside barrier.
In Fig.~\ref{fig.6}~(a) we reproduce such an effect for $\alpha$ decay of the \isotope[226]{Ra} nucleus.
In this figure one can see that emission from the tunneling region (see green dashed line in that figure) is larger than the full bremsstrahlung emission (see blue solid line in that figure) at higher energies of photon. But the full bremsstrahlung emission is in good agreement with experimental data that confirms existence of such an effect of destructive interference.
This picture is general for bremsstrahlung in the $\alpha$ decay of different nuclei.

This property can be used in our current research of study of $\Delta$-resonance in nuclei.
$\Delta$-resonance is shortly lived baryon, so % in approximation
it is formed in nucleus-target only during propagation of the scattered proton (from beam) through the space region of this nucleus.
So, we will take into account only this space region of proton-nucleus system in calculation of matrix elements of emission and we will estimate the bremsstrahlung spectra.
% estimate bremsstrahlung photons emitted only during propagation of proton (of beam) through the space region of the nucleus-target in scattering,
% when $\Delta$-resonance is shortly lived in nucleus.
%
Energy of proton in beam is much larger than barrier of the proton nucleus potential.
However, in calculation of the bremsstrahlung matrix elements we use two wave functions of proton-nucleus scattering --- these are wave functions in states before and after emission of photon.
In particular, after emission of photon, energy of relative motion of proton concerning to nucleus is reduced on the energy of the photon emitted.
So, at enough high energies of the emitted photons wave function in the final state reaches under-barrier energies with tunneling through the barrier.
So, in high energy region of photons we have also phenomenon of destructive interference between contributions of emission from the tunneling region and from the external region outside barrier.
In particular, on the basis of this logic one can suppose that the spectrum for the shorty lived $\Delta$-resonance in nucleus-target should be larger in the high energy photon region than spectrum of the full emission in the previous figures.
Results of such calculations of the spectrum for \isotope[12][\Delta]{C} in comparison with previous results are shown in Fig.~\ref{fig.6}~(b).
In next Fig.~\ref{fig.7} we present similar calculations of the spectra for the \isotope[40][\Delta]{Ca} and \isotope[208][\Delta]{Pb} nuclei.
\begin{figure}[htbp]
\centerline{\includegraphics[width=90mm]{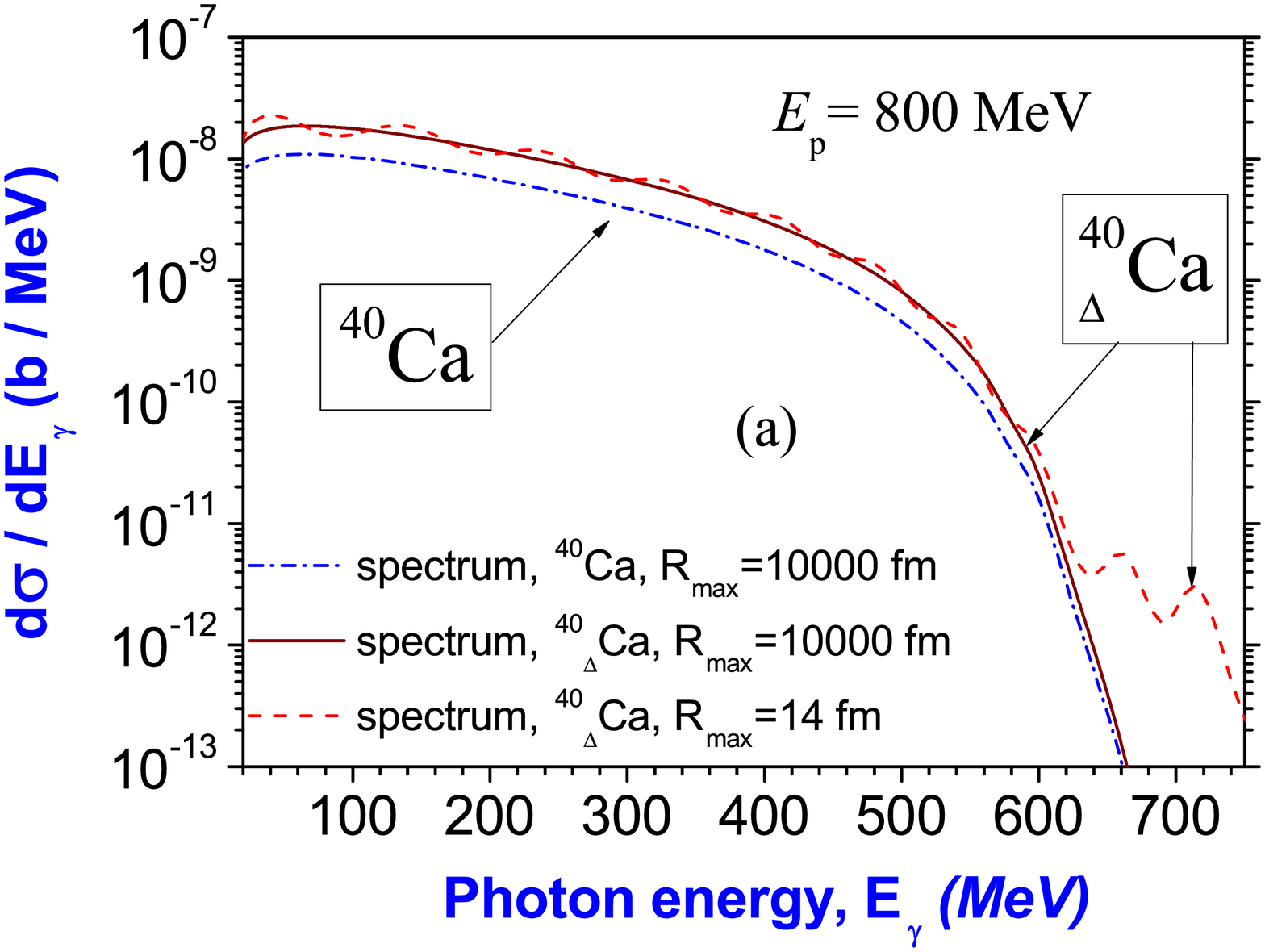}
\hspace{-1mm}\includegraphics[width=90mm]{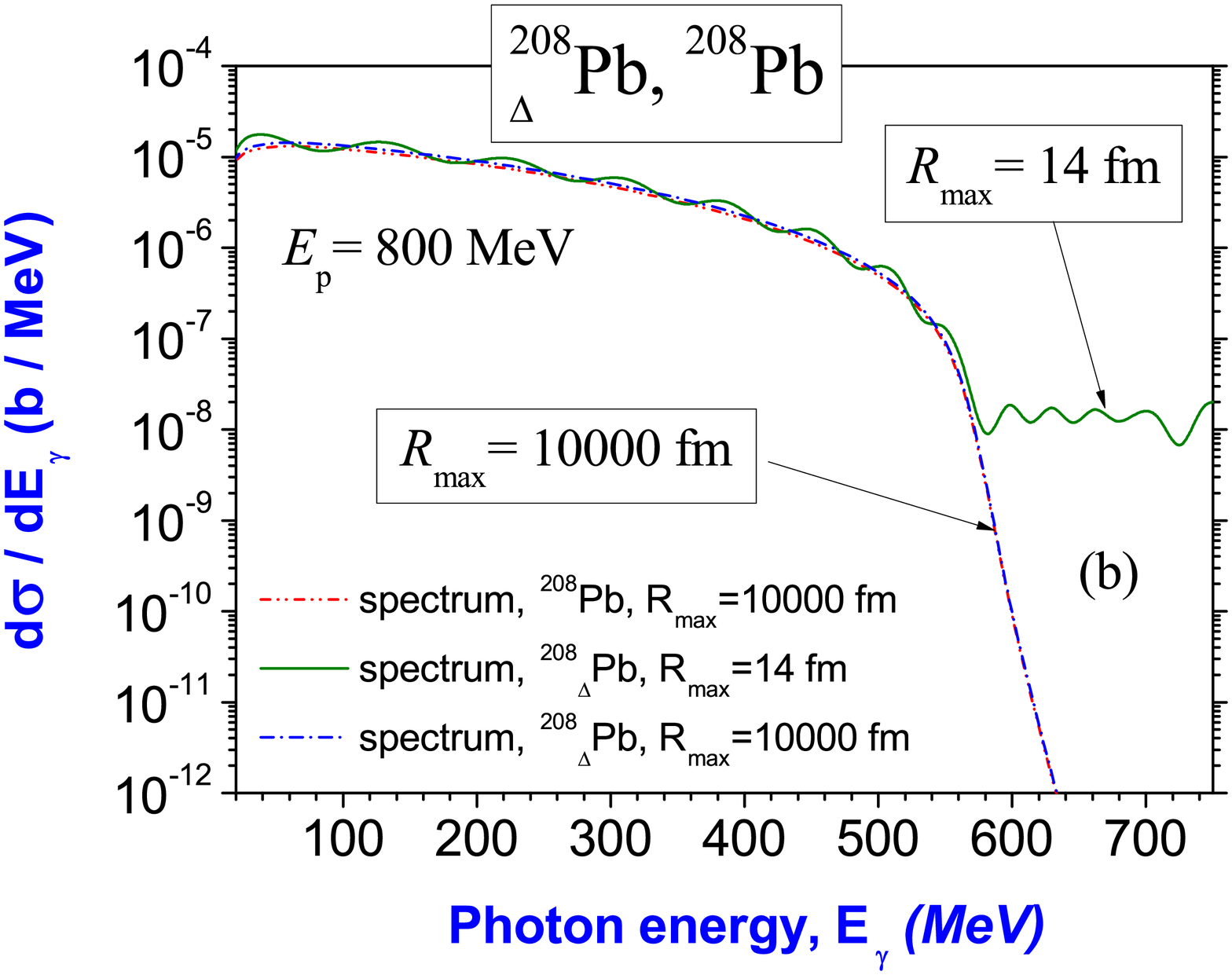}}
\vspace{-4mm}
\caption{\small (Color online)
The calculated bremsstrahlung spectra in the scattering of protons on
the \isotope[40]{Ca} and \isotope[40][\Delta]{Ca} nuclei (a) and
the \isotope[208]{Pb} and \isotope[208][\Delta]{Pb} nuclei (b)
at energy of proton beam of $E_{\rm p}=800$~MeV
at different space regions of integration.
% Here,
%
One can see that the spectrum for at $R_{\rm max}=14$~fm at high energy region % (red dashed line)
is larger than
the spectra at full space region. % (brown solid line).
\label{fig.7}}
\end{figure}
Calculations in these figures confirm our logic above.

% Note that inclusion of further decay of $\Delta$-resonance in nucleus
% can destroy continuation of wave function of relative motion between proton and nucleus-target.
% So, inclusion of emission of photons from the external region outside nucleus-target in the proton-nucleus scattering,
% than transition destroy continuation of full wave function of relative motion, and

% Now if to compare the spectra for normal nuclei with the spectra for nuclei with the shortly lived $\Delta$-resonance,
% then one can see essential difference between these spectra at high energy region of photons.
% This property can be used for proposal for future experiments with measurements of photons,
% as tools to distinguish process of formation of $\Delta$-resonance in the nucleus-target.

%-----------------------------------------------------------------------------------------------------------------------

%-----------------------------------------------------------------------------------------------------------------------
\subsection{Correction of the spectra on the basis of difference in masses for nucleon and $\Delta$-resonance
\label{sec.analysis.4}}

The $\Delta$-resonance is nearly~300 MeV heavier than the nucleon.
This mass difference is of the same order of magnitude as the photon and incoming-proton energies under study.
One can analyze if to neglect this difference in masses is a good approximation.

\begin{figure}[htbp]
\centerline{\includegraphics[width=90mm]{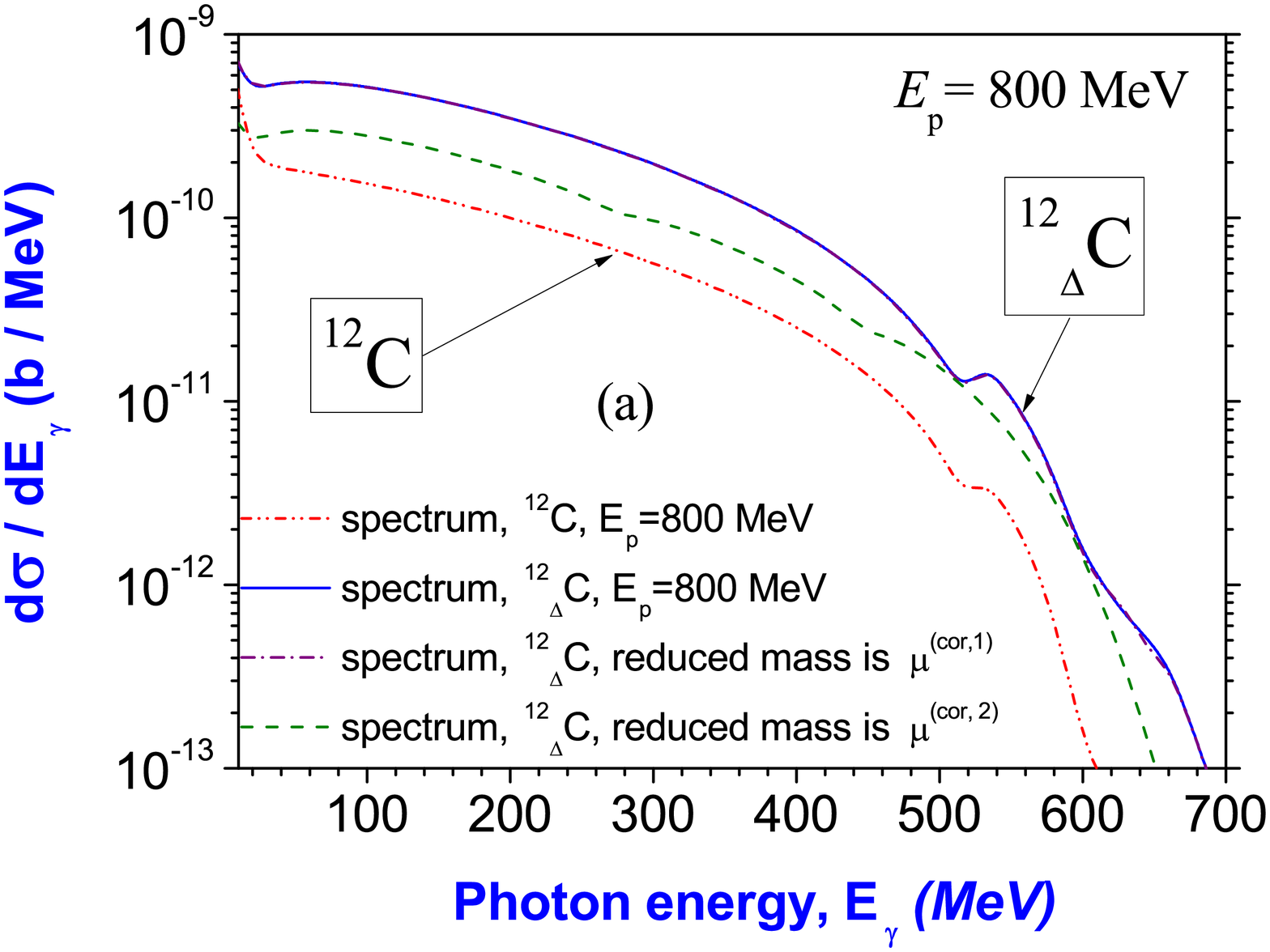}
\hspace{-1mm}\includegraphics[width=90mm]{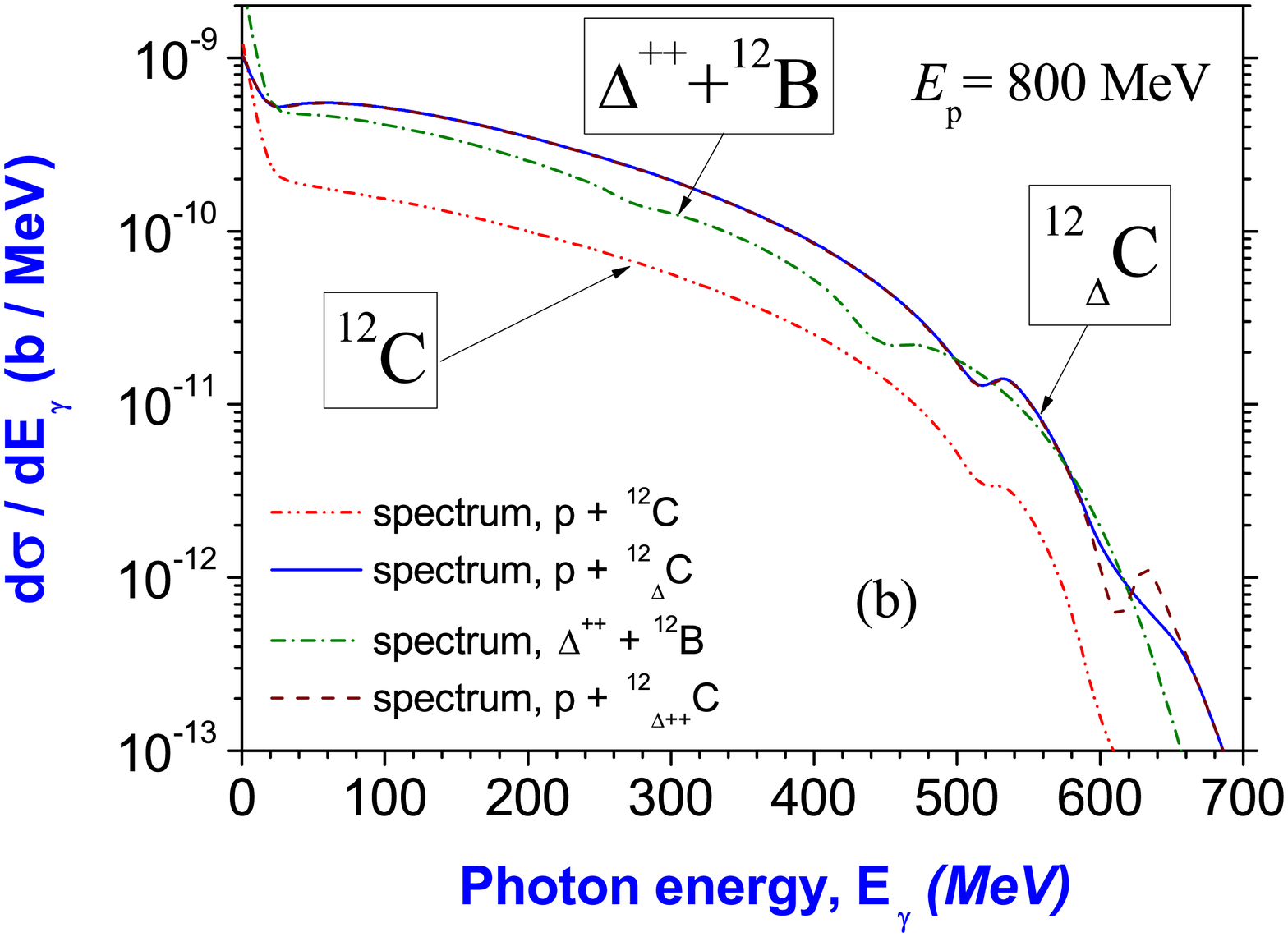}}
\vspace{-4mm}
\caption{\small (Color online)
The calculated full bremsstrahlung spectra % (with coherent and incoherent terms)
in the scattering of protons off the \isotope[12]{C}, \isotope[12][\Delta^{+}]{C}, \isotope[12][\Delta^{++}]{C} nuclei (a)
and in the scattering $\Delta^{++} + \isotope[12]{B}$ (b)
at energy of beam of $E=800$~MeV.
[Panel (a)]: Corrections of the spectra on the basis of taking into account difference between masses of $\Delta$-resonance and nucleon in calculations.
[Panel (b)]: The new spectra at transition $pp \to \Delta^{++} n$ in the nuclear system (see text for explanation).
\label{fig.8}}
\end{figure}

Reduced mass of proton-nucleus system is a main parameter, which is changed after inclusion of such a correction (difference between masses for proton and $\Delta$-resonance).
% difference between masses for proton and delta-resonance plays role in calculations of reduced mass of proton-nucleus system.
From previous study we conclude that relative motion of proton (in beam) and nucleus is the most important process forming the largest emission of photons.
Here, reduced mass is important parameter (it is used in calculations of wave functions of relative motion for states before and after emission of photons, also in formulas for operator of emission of photons).
One can consider two following cases.
\begin{itemize}
\item
One of protons of nucleus is changed to $\Delta$ resonance. We obtain new reduced mass as $\mu^{\rm (cor,\, 1)} = m_{p}\, m_{A}^{\rm (cor)} / (m_{\rm p} + m_{A}^{\rm (cor)})$,
where $m_{A}^{\rm (cor)} = m_{\Delta} + (Z-1)m_{p} + (A-Z)\, m_{\rm n}$.

\item
Proton from beam is changed to $\Delta$ resonance.
We obtain new reduced mass as $\mu^{\rm (cor,\, 2)} = m_{\Delta} m_{A} / (m_{\Delta} + m_{A})$.
% where $m_{A} = m_{\Delta} + (Z-1)m_{p} + (A-Z)\, m_{\rm n}$.

\end{itemize}
Here, $m_{\rm p}$, $m_{\rm n}$, $m_{\Delta}$ are masses of proton, neutron and $\Delta$-resonance,
$m_{A}$ is mass of nucleus,
$Z$ and $A$ are numbers of protons and nucleons in nucleus.
Calculations of the bremsstrahlung spectra on the basis of these two corrected reduced masses are presented in the new Fig.~\ref{fig.8}~(a) for \isotope[12][\Delta]{Ca}
[for comparison, two old spectra from Fig.~\ref{fig.4}~(a) are used also].
The spectrum for the first case almost coincides with old spectrum for \isotope[12][\Delta]{Ca} [see the purple dash-dotted line in Fig.~\ref{fig.8}~(a)] (difference between two spectra is in 2-nd or 3-d digits).
But, the spectrum for the second case is visibly different from the previous spectrum for \isotope[12][\Delta]{Ca} [see the green dashed line in Fig.~\ref{fig.8}~(a)].
Difference between old spectra [in Fig.~\ref{fig.4}~(a)] is larger.

Another transition $pN \to \Delta + N$ is also possible in this problem, when the final nucleon can be different from the initial one (e.g. $pp \to \Delta^{++} n$).
To analyze this transition, one can consider two following processes.
\begin{itemize}
\item
Two protons of nucleus are changed to $\Delta^{++}$ resonance and neutron in this nucleus.

% We obtain new reduced mass as $\mu^{\rm (cor,\, 1)} = m_{p}\, m_{A}^{\rm (cor)} / (m_{\rm p} + m_{A}^{\rm (cor)})$,
% where $m_{A}^{\rm (cor)} = m_{\Delta} + (Z-1)m_{p} + (A-Z)\, m_{\rm n}$.

\item
Proton from beam is changed to $\Delta^{++}$ resonance, and one neutron of nucleus is changed to proton.
% (nucleus is changed to \isotope[12][\Delta^{++}]{B}).

% We obtain new reduced mass as $\mu^{\rm (cor,\, 2)} = m_{\Delta} m_{A} / (m_{\Delta} + m_{A})$.
% % where $m_{A} = m_{\Delta} + (Z-1)m_{p} + (A-Z)\, m_{\rm n}$.
\end{itemize}
%
% Here, $m_{\rm p}$, $m_{\rm n}$, $m_{\Delta}$ are masses of proton, neutron and $\Delta$-resonance,
% $m_{A}$ is mass of nucleus,
% $Z$ and $A$ are numbers of protons and nucleons in nucleus.

One could suppose that the second case gives essential changes of the bremsstrahlung spectrum.
Interesting analysis can be obtained from such modifications of that reaction under study.
Electric charge of the scattered fragment is increased almost twice!
In result, effective electric charge $Z_{\rm eff}$ of the nuclear system is increased almost 4 times in dependence on the nucleus-target
(i.e., $Z_{\rm eff}^{\rm (new)} = 1.304065$ and $Z_{\rm eff}^{\rm (old)} = 0.3039928$ for $p + \isotope[12]{C}$).
This effective charge is included to the matrix element of coherent bremsstrahlung emission [see Eqs.~(24)--(28)]. So, this contribution is increased almost 4 times.
However, incoherent bremsstrahlung contribution is larger than coherent one. And this incoherent term does not include effective charge. So, summarized full spectrum is not deformed much.
This situation is shown by new green dash-dotted line in new Fig.~\ref{fig.8}~(b).
By simple words, one can suppose that magnetic field of nuclear system is more important that electric field in emission of photons.

The first process is calculated for $p + \isotope[12][\Delta^{++}]{C}$ at energy of beam $E_{\rm p} = 800$~MeV and shown in Fig.~\ref{fig.8}~(b) by new brown dashed line.
One can see that this spectrum is very close to blue solid line in this figure.

\begin{figure}[htbp]
\centerline{\includegraphics[width=90mm]{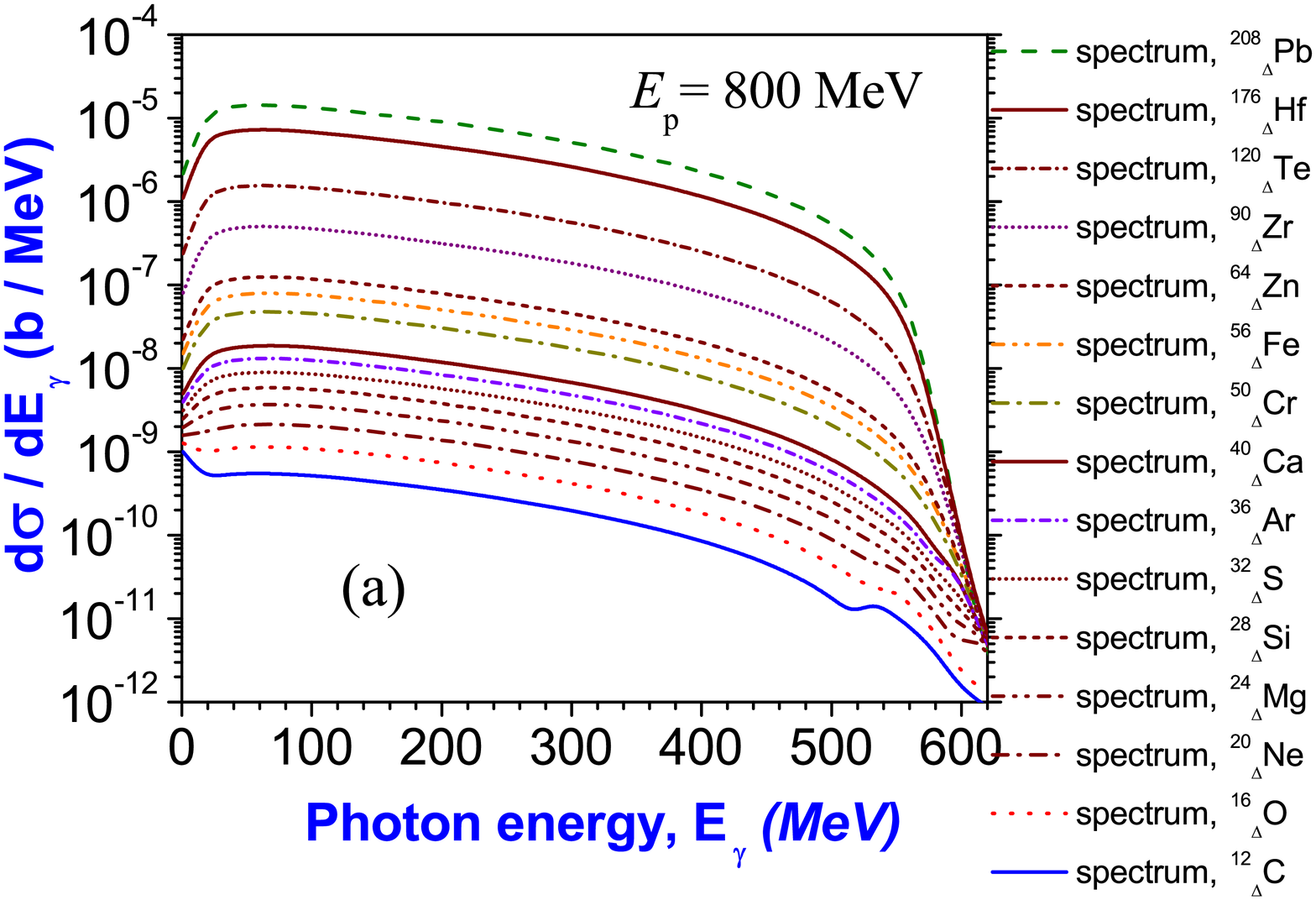}
\hspace{-1mm}\includegraphics[width=90mm]{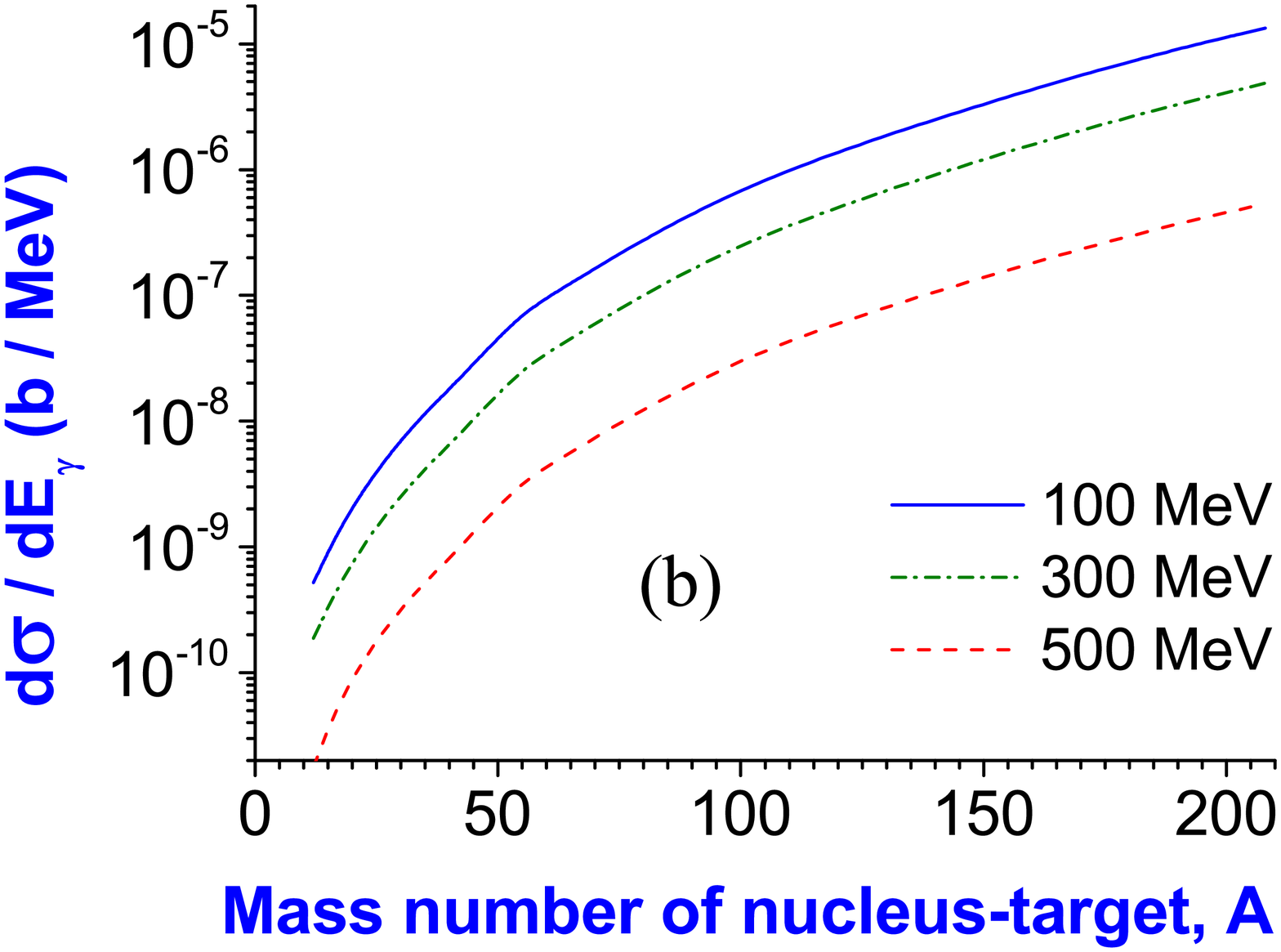}}
\vspace{-4mm}
\caption{\small (Color online)
The calculated full bremsstrahlung spectra % (with coherent and incoherent terms)
in dependence on mass of nucleus-target (with included $\Delta$-resonance)
in the scattering of protons % off the \isotope[12]{C}, \isotope[40]{Ca}, \isotope[197]{Au}, \isotope[208]{Pb} nuclei
at energy of proton beam of $E_{\rm p}=800$~MeV
\label{fig.9}}
\end{figure}

Also we add calculations of the bremsstrahlung spectra in dependence on mass of nucleus-target (where $\Delta$-resonance is included).
Such calculations are presented in Fig.~\ref{fig.9}.
One can see that the cross-section of bremsstrahlung emission is increased monotonously at increasing of mass of nucleus-target.
% (changes of the spectra for difference isotopes exist also, but I omit this analysis).
Note that wave function of relative motion (for states before emission of photons and after such emission) is calculated on the basis of proton-nucleus potential,
which is determined concerning to numbers of protons and neutrons in nucleus-target, according to Ref.~\cite{Becchetti.1969.PR}
(that potential is also extrapolated for region of light nuclei).
Dependence of the bremsstrahlung spectra on the numbers of protons and neutrons in the nucleus-target can be explained by the following reasons:
(1) Calculations of proton-nucleus wave functions on the basis of such a potential,
(2) Dependence of operator of emission of photon on the numbers of protons and neutrons in the nucleus-target.

% From such results one can conclude that the bremsstrahlung emission after formation of $\Delta$-resonance from one of nucleons of nucleus is changed not much for light and heavy nuclei, in general.
% *******************************************************************************************************************

% *******************************************************************************************************************
\section{Conclusions and perspective
\label{sec.conclusions}}

In this paper we investigate emission of the bremsstrahlung photons in the scattering of protons off nuclei at the $\Delta$-resonance energy region.
A special focus in this research is directed on question of how much the bremsstrahlung spectrum is changed after transition of one nucleon in nucleus to $\Delta$-resonance.
For this research, we improve our previous bremsstrahlung formalism
% (see Refs.~\cite{Maydanyuk_Zhang.2015.PRC,Maydanyuk.2012.PRC,Maydanyuk_Zhang_Zou.2016.PRC,Liu_Maydanyuk_Zhang_Liu.2019.PRC.hypernuclei,Maydanyuk_Zhang_Zou.2019.PRC.microscopy} and reference therein),
(see Ref.~\cite{Maydanyuk_Zhang_Zou.2019.PRC.microscopy}, reference therein),
including new aspects for $\Delta$-resonance in nucleus-target.
On such a basis we estimate the spectra of the bremsstrahlung photons and find the following.

\begin{itemize}
\item
For start of analysis, we calculate the bremsstrahlung spectra in the scattering of protons on the \isotope[12]{C}, \isotope[40]{Ca}, \isotope[208]{Pb} nuclei at energy of proton beam $E_{\rm p}$ of 800~MeV (see Fig.~\ref{fig.3}).
In these calculations we include coherent and incoherent bremsstrahlung contributions,
we test this formalism and calculations for the \isotope[197]{Au} nucleus at $E_{\rm p}=190$~MeV on the experimental data~\cite{Goethem.2002.PRL}
(see Fig.~\ref{fig.1}~(a)).
We estimate that incoherent emission is essentially more intensive than coherent one
(see Fig.~\ref{fig.2}, ratio between such contributions is about $10^{6}$--$10^{7}$ for $p + \isotope[197]{Au}$ at $E_{\rm p}=190$~MeV,
this result is in agreement with results in Refs.~\cite{Maydanyuk_Zhang.2015.PRC,Maydanyuk_Zhang_Zou.2016.PRC,Liu_Maydanyuk_Zhang_Liu.2019.PRC.hypernuclei}).
This incoherent contribution is directly dependent on magnetic moments of nucleons of nucleus-target
[see Eqs.~(\ref{eq.resultingformulas.1}), (\ref{eq.resultingformulas.2}), (\ref{eq.resultingformulas.5})].
This confirms our supposition that inclusion of incoherent processes in study of $\Delta$-resonance in proton-nucleus scattering is important (as can change picture of the studied process).
This aspect has never been studied before, % (for example, see Ref.~\cite{Gil_Oset.1998.PLB.v416}),
so it is one of aims of this paper.

\item
We show that increasing of energy of protons beam in nuclear scattering increases intensity of bremsstrahlung emission [see Fig.~\ref{fig.1}~(b) for comparison,
test with experimental data].
Bremsstrahlung emission is larger for heavier nuclei, difference between the spectra for light and heavy nuclei is essential
(see Fig.~\ref{fig.3} for \isotope[12]{C}, \isotope[40]{Ca}, \isotope[197]{Au}, \isotope[208]{Pb},
difference is about $10^{5}$ times between the spectra for \isotope[208]{Pb} and \isotope[12]{C} at $E_{\rm p} = 800$~MeV).

\item
We analyze coherent and incoherent contributions, electric and magnetic contributions in the full bremsstrahlung for different nuclei and energies of proton beam [see Eqs.~(\ref{eq.resultingformulas.1})--(\ref{eq.resultingformulas.2})].
In the coherent bremsstrahlung, the magnetic emission % based on $M_{p}^{(M,\, {\rm dip})}$
is almost the same as electric emission % based on $M_{p}^{(E,\, {\rm dip})}$
[$\sigma_{\rm mag}^{\rm (coh)} / \sigma_{\rm el}^{\rm (coh)} = 3.3213$ for \isotope[197]{Au} at $E_{\rm p} = 190$~MeV for 10--180~MeV of photons;
see Eq.~(\ref{eq.analysis.1.1})].
In the incoherent bremsstrahlung, role of background emission based on $M_{k}$ is a little larger than magnetic contribution based on $M_{\Delta M}$
($\sigma_{\rm background}^{\rm (incoh)} / \sigma_{\rm mag}^{\rm (incoh)} = 4.04$ for \isotope[197]{Au} at $E_{\rm p} = 190$~MeV for 10--180~MeV of photons).
Ratio between incoherent emission and coherent emission is increased at increasing of energy of photon emitted
(see Fig.~\ref{fig.2}, for \isotope[197]{Au} at $E_{\rm p}=190$~MeV).

\item
For inclusion of
transition from proton of nucleus-target to $\Delta^{+}$-resonance ($p\,N \to \Delta^{+} N$)
% formation of $\Delta$-resonance in nucleus-target
in the proton-nucleus scattering to the model,
we use scheme of the coherent processes described in Ref.~\cite{Gil_Oset.1998.PLB.v416}.
%
% We estimate change in the bremsstrahlung emission after such an inclusion to the model.
% We calculate the spectra at the modified model, where transition of one nucleon on $\Delta$-resonance takes place in the nucleus-target.
%
We find that emission of bremsstrahlung photons in that reaction with one $\Delta$-resonance in nucleus-target is more intensive than for normal nucleus (see Fig.~\ref{fig.4} for \isotope[12]{C}, \isotope[12][\Delta]{C} and \isotope[40]{Ca}, \isotope[40][\Delta]{Ca} at $E_{\rm p}=800$~MeV).
This result is indication of reinforcement of bremsstrahlung in proton-nucleus scattering.
We find that difference between
the spectra for normal nuclei and nuclei with included $\Delta$-resonance is larger for more light nuclei,
but the spectra are larger for heavier nuclei
(see Fig.~\ref{fig.4} for \isotope[12]{C}, \isotope[40]{Ca} in comparison with \isotope[12][\Delta]{C}, \isotope[40][\Delta]{Ca}).

\item
In order to find nuclei with maximal reinforcement of bremsstrahlung due to transition $p\,N \to \Delta^{+} N$, we performed analysis and obtained condition~(\ref{eq.resultingformulas.2}) determining ratio between protons and neutrons for the normal nucleus.
On the example for \isotope[10][\Delta^{+}]{C}, we show
that difference between the spectra for \isotope[10]{C} and \isotope[10][\Delta^{+}]{C} is larger essentially
than difference between the spectra for \isotope[12]{C} and \isotope[12][\Delta^{+}]{C} (see Fig.~\ref{fig.5}).
However, stable nuclei (like \isotope[12]{C}, \isotope[40]{Ca}, \isotope[208]{Pb}) do not satisfy to that condition~(\ref{eq.resultingformulas.2}).
As parameter $\bar{\mu}_{\rm pn}$ has the highest influence on reinforcement of bremsstrahlung of such a type, it confirms importance of incoherent processes in this research
(i.e., there is no enhancement of bremsstrahlung due to transition $p\,N \to \Delta^{+} N$, if incoherent processes are not included to the model).

\item
We take into account that
$\Delta$-resonance is shortly lived baryon, and
it is formed in nucleus-target only during propagation of the scattered proton (from beam) through the space region of this nucleus.
On such a basis, we calculate new bremsstrahlung spectra and find
% we will take into account only this space region of proton-nucleus system in calculation of matrix elements of emission and we will estimate the bremsstrahlung spectra.
that the spectrum for the shorty lived $\Delta$-resonance in the nucleus-target should be larger in the high energy photon region than the spectrum without creation of $\Delta$-resonance in nucleus-target
(see Figs.~\ref{fig.6}~(b), \ref{fig.7} for \isotope[12][\Delta]{C}, \isotope[40][\Delta]{Ca}, \isotope[208][\Delta]{Pb}).
This effect has the same origin as phenomenon of destructive interference between emission from tunneling region and external region investigated for bremsstrahlung in the $\alpha$ decay
(see Fig.~\ref{fig.6}~(a) for \isotope[226]{Ra}, also Ref.~\cite{Maydanyuk.2008.MPLA}).

\end{itemize}
%-----------------------------------------------------------------------------------------------------------------------

%-----------------------------------------------------------------------------------------------------------------------
\noindent
If to compare the spectra for normal nuclei with the spectra for nuclei with the shortly lived $\Delta$-resonance,
then one can see essential difference between these spectra at high energy region of photons.
This property can be used for proposal for future experiments with measurements of photons,
as tools to distinguish process of formation of $\Delta$-resonance in the nucleus-target.
%-----------------------------------------------------------------------------------------------------------------------

%-----------------------------------------------------------------------------------------------------------------------
% (note that calculations in Ref.~\cite{Gil_Oset.1998.PLB.v416} were predictions, not tested on the bremsstrahlung experimental data).
% Processes with emission of photons in scattering of protons on the \isotope[12]{C}, \isotope[40]{Ca}, \isotope[208]{Pb} nuclei
% in the $\Delta$-resonance energy region were analyzed~\cite{Gil_Oset.1998.PLB.v416}
% This incoherent bremsstrahlung is highly dependent on magnetic moments of nucleons of the scattered nuclear fragments,
% but those estimations were obtained for magnetic moments of nucleon for vacuum.
%-----------------------------------------------------------------------------------------------------------------------

%-----------------------------------------------------------------------------------------------------------------------
\section*{Acknowledgements
\label{sec.acknowledgements}}

Author is highly appreciated to
Profs. Pengming Zhang and Liping Zou for useful discussions.
% Prof.~A.~G.~Magner for useful discussions concerning to magnetic moments of nucleons in nuclear matter an in vacuum.
%
% Prof.~M.~Ya.~Amusia for his interesting insight on general aspects of bremsstrahlung topic,
% Prof.~V.~S.~Vasilevsky for interesting discussions concerning to peculiarities of Dirac equation,
% Profs.~V.~A.~Plujko, S.~N.~Fedotkin, A.~G.~Magner, F.~A.~Ivanyuk, A.~P.~Ilyin, A.~Ya.~Dzyublik for fruitful and useful discussions.
% S.~P.~M. thanks the Institute of Modern Physics of Chinese Academy of Sciences for warm hospitality and support.
% This work was supported by the National Natural Science Foundation of China (Grant Nos. 11975320 and 11805242).

% 11427904 and 11535016).
% the Major State Basic Research Development Program in China (No. 2015CB856903),
% the National Natural Science Foundation of China (Grant Nos. 11575254, 11447105 and 11175215),
% the Chinese Academy of Sciences fellowships for researchers from developing countries (No. 2014FFJA0003).
% *******************************************************************************************************************

% *******************************************************************************************************************
% \newpage
\appendix
\section{Matrix element of emission
\label{sec.app.short.2}}

% We define matrix element of emission of the bremsstrahlung photons, using the wave functions $\Psi_{i}$ and $\Psi_{f}$ of
% the full nuclear system in states before emission of photons ($i$-state) and after such emission ($f$-state), as
%
% \begin{equation}
%   F = \langle \Psi_{f} |\, \hat{H}_{\gamma} |\, \Psi_{i} \rangle.
% \label{eq.app.short.2.7.1}
% \end{equation}
%
% Нам нужно проинтегрировать такой матричный элемент по всем независимым переменным.
% Эти переменные --- это пространственные координаты $\vb{R}$, $\vb{r}$, $\rhobf_{Am}$.
% Мы должны учесть пространственное представление для всех используемых импульсов $\vu{P}$, $\vu{p}$, $\vb{\tilde{p}}_{A m}$ (как
% In this matrix element we should integrate over all independent variables.
% These variables are space variables $\vb{R}$, $\vb{r}$, $\rhobf_{Am}$.
% We should take into account space representation of all used moments $\vu{P}$, $\vu{p}$, $\vb{\tilde{p}}_{A m}$ (as
% $\vu{P} = -i\hbar\, \vb{d/dR}$,
% $\vu{p} = -i\hbar\, \vb{d/dr}$,
% $\vb{\tilde{p}}_{A m} = -i\hbar\, \vb{d/d} \rhobf_{Am}$).

In this Appendix we calculate the matrix element of emission.
Substituting formulas (\ref{eq.2.5.2})--(\ref{eq.2.5.7}) for operator of emission to (\ref{eq.13.1.1}), we obtain:
\begin{equation}
  \langle \Psi_{f} |\, \hat{H}_{\gamma} |\, \Psi_{i} \rangle \;\; = \;\;
  \sqrt{\displaystyle\frac{2\pi\, c^{2}}{\hbar w_{\rm ph}}}\,
  \Bigl\{ M_{P} + M_{p} + M_{k} + M_{\Delta E} + M_{\Delta M} \Bigr\},
\label{eq.app.short.2.7.2}
\end{equation}
where
\begin{equation}
\begin{array}{lcl}
\vspace{1mm}
  M_{P} & = &
  \sqrt{\displaystyle\frac{\hbar w_{\rm ph}} {2\pi c^{2}}}\;
  \mel{\Psi_{f}\,} {\hat{H}_{P}} {\,\Psi_{i}}\; = \\
  % \biggl\langle \Psi_{f}\,} \biggl|\, \hat{H}_{P} \biggr|\, \Psi_{i} \biggr\rangle\; = \\

  & = &
  -\, \displaystyle\frac{1}{m_{A} + m_{\rm p}}\,
  \displaystyle\sum\limits_{\alpha=1,2}
  \biggl\langle \Psi_{f}\, \biggl|\,
    2\, \mu_{N}\, m_{\rm p}\;
    e^{-i\, \vb{k_{\rm ph}} \vb{R}}
    \biggl\{
      z_{\rm p}\, e^{-i\, c_{A}\, \vb{k_{\rm ph}} \vb{r}} +
        % \displaystyle\sum_{i=1}^{4} z_{i}\, e^{-i\, \vb{k_{\rm ph}} \rhobf_{\alpha i}} +
      e^{i\, c_{\rm p}\, \vb{k_{\rm ph}} \vb{r} }
        \displaystyle\sum_{j=1}^{A} z_{j}\, e^{-i\, \vb{k_{\rm ph}} \rhobf_{Aj}}
    \biggr\}\, \vb{e}^{(\alpha)} \cdot \vu{P} + \\
  & + &
    i\:\mu_{N}\,
    e^{-i\, \vb{k_{\rm ph}} \vb{R}}\,
    \biggl\{
      e^{-i\, c_{A}\, \vb{k_{\rm ph}} \vb{r}}\, \mu_{\rm p}\, m_{\rm p}\, \sigmabf  +
      % e^{-i\, c_{A}\, \vb{k_{\rm ph}} \vb{r}}\, \displaystyle\sum_{i=1}^{4} \mu_{i}\, m_{\alpha i}\, e^{-i\, \vb{k_{\rm ph}} \rhobf_{\alpha i}}\, \sigmabf  +
      e^{i\, c_{\rm p}\, \vb{k_{\rm ph}} \vb{r}}\,
        \displaystyle\sum_{j=1}^{A} \mu_{j}\, m_{Aj}\, e^{-i\, \vb{k_{\rm ph}} \rhobf_{Aj}}\, \sigmabf
  \biggr\}\, \cdot \bigl[ \vu{P} \cp \vb{e}^{(\alpha)} \bigr]\,
  \biggr|\, \Psi_{i} \biggr\rangle,
\end{array}
\label{eq.app.short.2.7.3.a}
\end{equation}
\begin{equation}
\begin{array}{lcl}
\vspace{1mm}
  M_{p} & = &
  \sqrt{\displaystyle\frac{\hbar w_{\rm ph}} {2\pi c^{2}}}\;
  \mel{\Psi_{f}\,} {\hat{H}_{p}} {\,\Psi_{i}}\; = \\

  & = &
  - \displaystyle\sum\limits_{\alpha=1,2}
    \biggl\langle
      \Psi_{f}\,
    \biggl|\,
  2\, \mu_{N}\,  m_{\rm p}\,
  e^{-i\, \vb{k_{\rm ph}} \vb{R}}
  \biggl\{
    e^{-i\, c_{A} \vb{k_{\rm ph}} \vb{r}}\, \displaystyle\frac{z_{\rm p}}{m_{\rm p}} -
    e^{i\, c_{\rm p} \vb{k_{\rm ph}} \vb{r}}\,  \displaystyle\frac{1}{m_{A}}\, \displaystyle\sum_{j=1}^{A} z_{j}\, e^{-i\, \vb{k_{\rm ph}} \rhobf_{Aj}}
  \biggr\}\; \vb{e}^{(\alpha)} \cdot \vu{p}\; + \\

  & + &
  i\,\mu_{N}\,
  e^{-i\, \vb{k_{\rm ph}} \vb{R}}
  \biggl\{
    e^{-i\, c_{A} \vb{k_{\rm ph}} \vb{r}} \displaystyle\frac{1}{m_{\rm p}}\,
      \mu_{\rm p}\, m_{\rm p}\, \sigmabf
    % e^{-i\, c_{A} \vb{k_{\rm ph}} \vb{r}} \displaystyle\frac{1}{m_{\alpha}}\,
    % \displaystyle\sum_{i=1}^{4} \mu_{i}\, m_{\alpha i}\; e^{-i\, \vb{k_{\rm ph}} \rhobf_{\alpha i}}\, \sigmabf
    -
    e^{i\, c_{\rm p} \vb{k_{\rm ph}} \vb{r}} \displaystyle\frac{1}{m_{A}}
      \displaystyle\sum_{j=1}^{A} \mu_{j}\, m_{Aj}\; e^{-i\, \vb{k_{\rm ph}} \rhobf_{Aj}}\, \sigmabf \biggr\}
      \cdot \bigl[ \vu{p} \times \vb{e}^{(\alpha)} \bigr]
  \biggr|\, \Psi_{i}\, \biggr\rangle ,
\end{array}
\label{eq.app.short.2.7.3.b}
\end{equation}
\begin{equation}
\begin{array}{lcl}
\vspace{1mm}
  M_{k} & = &
  \sqrt{\displaystyle\frac{\hbar w_{\rm ph}} {2\pi c^{2}}}\;
  \mel{\Psi_{f}\,} {\hat{H}_{k}} {\,\Psi_{i}}\; =  \\
& = &
  i\, \hbar\, \mu_{N}\,
  \displaystyle\sum\limits_{\alpha=1,2}
  \biggl\langle \Psi_{f}\, \biggl|\,
    e^{-i\, \vb{k_{\rm ph}} \vb{R}}
  \biggl\{
    e^{-i\, c_{A}\, \vb{k_{\rm ph}} \vb{r}}\,
    % \displaystyle\sum_{i=1}^{4}
      \mu_{\rm p}\, \sigmabf +
    e^{i\, c_{\rm p}\, \vb{k_{\rm ph}} \vb{r}}\,
    \displaystyle\sum_{j=1}^{A}
      \mu_{j}\, e^{-i\, \vb{k_{\rm ph}} \rhobf_{Aj}}\, \sigmabf
  \biggr\}
  \cdot \bigl[ \vb{k_{\rm ph}} \cp \vb{e}^{(\alpha)} \bigr]
  \biggr|\, \Psi_{i} \biggr\rangle,
\end{array}
\label{eq.app.short.2.7.3.c}
\end{equation}
\begin{equation}
\begin{array}{lcl}
\vspace{0.1mm}
  M_{\Delta E} & = &
  \sqrt{\displaystyle\frac{\hbar w_{\rm ph}} {2\pi c^{2}}}\: \mel{\Psi_{f}\,} {\hat{H}_{\gamma E}} {\,\Psi_{i}} =

  -\, \displaystyle\sum\limits_{\alpha=1,2} \vb{e}^{(\alpha)}
    \biggl\langle \Psi_{f}\, \biggl|\, 2\, \mu_{N}\, e^{-i\, \vb{k_{\rm ph}}\vb{R}}\; \times \\

  & \times &
  \biggl\{
    % e^{-i\, c_{A}\, \vb{k_{\rm ph}} \vb{r}}\,
    % \displaystyle\sum_{i=1}^{3} \displaystyle\frac{z_{i} m_{\rm p}}{m_{\alpha i}}\, e^{-i\, \vb{k_{\rm ph}} \rhobf_{\alpha i}}\, \vb{\tilde{p}}_{\alpha i} +
    e^{i\, c_{\rm p}\, \vb{k_{\rm ph}} \vb{r}}\,
      \displaystyle\sum_{j=1}^{A-1} \displaystyle\frac{z_{j} m_{\rm p}}{m_{Aj}}\, e^{-i\, \vb{k_{\rm ph}} \rhobf_{Aj}}\, \vb{\tilde{p}}_{Aj} -

%     \displaystyle\frac{m_{\rm p}}{m_{\alpha}}\, e^{-i\, c_{A}\, \vb{k_{\rm ph}} \vb{r}}\,
%     \displaystyle\sum_{i=1}^{4} z_{i}\, e^{-i\, \vb{k_{\rm ph}} \rhobf_{\alpha i}}\, \Bigl( \displaystyle\sum_{k=1}^{n-1} \vb{\tilde{p}}_{\alpha k} \Bigr) +

    \displaystyle\frac{m_{\rm p}}{m_{A}}\,
    e^{i\, c_{\rm p}\, \vb{k_{\rm ph}} \vb{r}}\, \displaystyle\sum_{j=1}^{A} z_{j}\, e^{-i\, \vb{k_{\rm ph}} \rhobf_{Aj}}
    \Bigl( \displaystyle\sum_{k=1}^{A-1} \vb{\tilde{p}}_{Ak} \Bigr)\,
  \biggr\}
  \biggr|\, \Psi_{i} \biggr\rangle,
\end{array}
\label{eq.app.short.2.7.3.d}
\end{equation}
\begin{equation}
\begin{array}{lll}
\vspace{0.1mm}
  & M_{\Delta M} =
  \sqrt{\displaystyle\frac{\hbar w_{\rm ph}} {2\pi c^{2}}}\: \mel{\Psi_{f}\,} {\hat{H}_{\gamma M}} {\,\Psi_{i}}\; = % \\
  -\, i\, \mu_{N}\, \displaystyle\sum\limits_{\alpha=1,2}
    \biggl\langle \Psi_{f}\, \biggl|\, e^{-i\, \vb{k_{\rm ph}} \vb{R}}\; \times \\

% \vspace{0.3mm}
  \times &
  \biggl\{
    % e^{-i\, \vb{k_{\rm ph}} c_{A}\, \vb{r}}\,
    % \displaystyle\sum_{i=1}^{n-1} \mu_{i}\, e^{-i\, \vb{k_{\rm ph}} \rhobf_{\alpha i}}\, \sigmabf \cdot \bigl[ \vb{\tilde{p}}_{\alpha i} \times \vb{e}^{(\alpha)} \bigr] +
    e^{i\, \vb{k_{\rm ph}} c_{\rm p}\, \vb{r}}\,
      \displaystyle\sum_{j=1}^{A-1} \mu_{j}\, e^{-i\, \vb{k_{\rm ph}} \rhobf_{Aj}}\, \sigmabf \cdot \bigl[ \vb{\tilde{p}}_{Aj} \times \vb{e}^{(\alpha)} \bigr] - % \\

  % e^{-i\, \vb{k_{\rm ph}} c_{A}\, \vb{r}}\,
  % \displaystyle\sum_{i=1}^{n}
  %   \mu_{i}\, \displaystyle\frac{m_{\alpha i}}{m_{\alpha}}\,
  % e^{-i\, \vb{k_{\rm ph}} \rhobf_{\alpha i}}\, \displaystyle\sum_{k=1}^{n-1} \sigmabf \cdot \bigl[ \vb{\tilde{p}}_{\alpha k} \times \vb{e}^{(\alpha)} \bigr] +

    e^{i\, \vb{k_{\rm ph}} c_{\rm p}\, \vb{r}}\,
    \displaystyle\sum_{j=1}^{A}
    \mu_{j}\, \displaystyle\frac{m_{Aj}}{m_{A}}\,
    e^{-i\, \vb{k_{\rm ph}} \rhobf_{Aj}}\, \displaystyle\sum_{k=1}^{A-1} \sigmabf \cdot \bigl[ \vb{\tilde{p}}_{Ak} \times \vb{e}^{(\alpha)} \bigr]
  \biggr\}
  \biggr|\, \Psi_{i} \biggr\rangle.
\end{array}
\label{eq.app.short.2.7.3.e}
\end{equation}
% -----------------------------------------------------------------------------------------------------------------------

% -----------------------------------------------------------------------------------------------------------------------
% \newpage
\subsection{Integration over space variable $\vb{R}$
\label{sec.app.short.2.8}}

In Sec.~\ref{sec.13} we defined the wave function of the full nuclear system as
\[
  \Psi =
  \Phi (\vb{R}) \cdot
  \Phi_{\rm p - nucl} (\vb{r}) \cdot
  \psi_{\rm nucl} (\beta_{A}).
  % \psi_{\alpha} (\beta_{\alpha}),
\]
Here, $\Phi (\vb{R})$ is wave function describing evolution of center of mass of the full nuclear system.
We rewrite this function as
\begin{equation}
\begin{array}{lcl}
  \Psi = \Phi (\vb{R}) \cdot F (\vb{r}, \beta_{A}, \beta_{\rm p}), &
  F (\vb{r}, \beta_{A}, \beta_{\rm p}) =
  \Phi_{\rm p - nucl} (\vb{r}) \cdot
  \psi_{\rm nucl} (\beta_{A}),
  % \psi_{\rm p} (\beta_{\rm p}),
\end{array}
\label{eq.app.short.2.8.1}
\end{equation}
We shall assume approximated form for the function $\Phi_{\bar{s}}$ before and after emission of photon as
\begin{equation}
  \Phi_{\bar{s}} (\vb{R}) =  e^{-i\,\vb{K}_{\bar{s}}\cdot\vb{R}},
\label{eq.app.short.2.8.2}
\end{equation}
%
% где $\bar{s} = i$ или $f$ (индексы $i$ и $f$ обозначают начальное состояние, т.е. состояние до излучения фотона,
% и конечное состояние, т.е. состояние после излучения фотона),
% $\vb{K}_{s}$ --- импульс полной системы~\cite{Kopitin.1997.YF}.
% В наших предыдущих работах, мы изучали распады ядер, где мы выбирали
%
where $\bar{s} = i$ or $f$ (indexes $i$ and $f$ denote the initial state, i.e. the state before emission of photon,
and the final state, i.e. the state after emission of photon),
$\vb{K}_{s}$ is momentum of the total system. %~\cite{Kopitin.1997.YF}.
% In our previous works, we studied decays of nuclei where we assumed
%
% \begin{equation}
%   \vb{K}_{i} = 0
% \label{eq.app.short.2.8.3}
% \end{equation}
%
% as we considered the $\alpha$-decaying nuclear system before emission of photons as not moving in the laboratory frame.
% However, for $\alpha$-nucleus scattering $\vb{K}_{i} \ne 0$ is more reasonable
% (it seems for $\alpha$ decay we should also use $\vb{K}_{i} \ne 0$, that should be taken into account in further study).
%
% (as we considered a system of daughter nucleus and emitted fragment before emission of photons as not moved in the laboratory frame).
% However, for scattering $\vb{K}_{i} \ne 0$ is more reasonable.
%
% \vspace{1.5mm}
% \noindent
% \textcolor[rgb]{0.00,0.00,1.00}{\textbf{It seems for $\alpha$ decay we should also use $\vb{K}_{i} \ne 0$, that should be taken into account in further study.}}
% \vspace{2mm}
%
% Сперва мы вычислим матричный элемент $M_{p}$ [начиная с формулы~(\ref{eq.app.short.2.7.3.b})]:
Let us calculate the matrix elements [starting from Eq.~(\ref{eq.app.short.2.7.3.b})]:
\begin{equation}
\begin{array}{lcl}
\vspace{-0.1mm}
  M_{P} & = & \hbar\, (2\pi)^{3}\, \delta (\vb{K}_{f} - \vb{K}_{i} - \vb{k}_{\rm ph})\; \times \\
\vspace{-0.1mm}
  & \times &
  \displaystyle\frac{1}{m_{A} + m_{\rm p}}\,
  \displaystyle\sum\limits_{\alpha=1,2}
  \biggl\langle F_{f}\, \biggl|\,
    2\, \mu_{N}\, m_{\rm p}\;
    \biggl\{
      % e^{-i\, c_{A}\, \vb{k_{\rm ph}} \vb{r}} \displaystyle\sum_{i=1}^{4} z_{i}\, e^{-i\, \vb{k_{\rm ph}} \rhobf_{\alpha i}} +
      z_{\rm p}\, e^{-i\, c_{A}\, \vb{k_{\rm ph}} \vb{r}} +
      e^{i\, c_{\rm p}\, \vb{k_{\rm ph}} \vb{r} } \displaystyle\sum_{j=1}^{A} z_{j}\, e^{-i\, \vb{k_{\rm ph}} \rhobf_{Aj}}
    \biggr\}\, \vb{e}^{(\alpha)} \cdot \vb{K}_{i}\; + \\

  & + &
    i\,\mu_{N}\,
    \biggl\{
      % e^{-i\, c_{A}\, \vb{k_{\rm ph}} \vb{r}}\, \displaystyle\sum_{i=1}^{4} \mu_{i}\, m_{\alpha i}\, e^{-i\, \vb{k_{\rm ph}} \rhobf_{\alpha i}}\, \sigmabf  +
      e^{-i\, c_{A}\, \vb{k_{\rm ph}} \vb{r}}\, \mu_{\rm p}\, m_{\rm p}\, \sigmabf  +
      e^{i\, c_{\rm p}\, \vb{k_{\rm ph}} \vb{r}}\, \displaystyle\sum_{j=1}^{A} \mu_{j}\, m_{Aj}\, e^{-i\, \vb{k_{\rm ph}} \rhobf_{Aj}}\, \sigmabf
  \biggr\}\, \cdot \bigl[ \vb{K}_{i} \cp \vb{e}^{(\alpha)} \bigr]\,
  \biggr|\, F_{i} \biggr\rangle.
\end{array}
\label{eq.app.short.2.8.12}
\end{equation}
\begin{equation}
\begin{array}{lll}
\vspace{-0.1mm}
  & & M_{p} = -\, (2\pi)^{3} \delta (\vb{K}_{f} - \vb{K}_{i} - \vb{k}_{\rm ph})\; % \times \\
% \vspace{-0.1mm}
%   & \times &
  \displaystyle\sum\limits_{\alpha=1,2}
  \biggl\langle F_{f}\, \biggl|\,
  2\, \mu_{N}\,  m_{\rm p}\,
  \Bigl\{
    % e^{-i\, c_{A} \vb{k_{\rm ph}} \vb{r}}\, \displaystyle\frac{1}{m_{\alpha}}\, \displaystyle\sum_{i=1}^{4} z_{i}\, e^{-i\, \vb{k_{\rm ph}} \rhobf_{\alpha i}} -
    e^{-i\, c_{A} \vb{k_{\rm ph}} \vb{r}}\, \displaystyle\frac{z_{\rm p}}{m_{\rm p}}  -
    e^{i\, c_{\rm p} \vb{k_{\rm ph}} \vb{r}}\,  \displaystyle\frac{1}{m_{A}}\, \displaystyle\sum_{j=1}^{A} z_{j}\, e^{-i\, \vb{k_{\rm ph}} \rhobf_{Aj}}
  \Bigr\}\; \vb{e}^{(\alpha)} \cdot \vu{p}\; + \\
  & + &
  i\,\mu_{N}\,
  \Bigl\{
    % e^{-i\, c_{A} \vb{k_{\rm ph}} \vb{r}} \displaystyle\frac{1}{m_{\alpha}}\,
    % \displaystyle\sum_{i=1}^{4}
    %   \mu_{i}\, m_{\alpha i}\;
    %   e^{-i\, \vb{k_{\rm ph}} \rhobf_{\alpha i}}\, \sigmabf -

    e^{-i\, c_{A} \vb{k_{\rm ph}} \vb{r}}\, \mu_{\rm p}\, \sigmabf -

    e^{i\, c_{\rm p} \vb{k_{\rm ph}} \vb{r}} \displaystyle\frac{1}{m_{A}}
    \displaystyle\sum_{j=1}^{A}
      \mu_{j}\, m_{Aj}\;
      e^{-i\, \vb{k_{\rm ph}} \rhobf_{Aj}}\, \sigmabf \Bigr\}
    \cdot \bigl[ \vu{p} \times \vb{e}^{(\alpha)} \bigr]\,
  \biggr|\, F_{i}\, \biggr\rangle.
\end{array}
\label{eq.app.short.2.8.7}
\end{equation}
\begin{equation}
\begin{array}{lcl}
\vspace{-0.1mm}
  M_{k} & = & (2\pi)^{3} \delta (\vb{K}_{f} - \vb{K}_{i} - \vb{k}_{\rm ph})\, \mu_{N}\; \times \\
\vspace{-0.1mm}
  & \times &
  i\, \hbar\,
  \displaystyle\sum\limits_{\alpha=1,2}
  \biggl\langle F_{f}\, \biggl|\,
    % e^{-i\, \vb{k_{\rm ph}} \vb{R}}
  \biggl\{
    % e^{-i\, c_{A}\, \vb{k_{\rm ph}} \vb{r}}\, \displaystyle\sum_{i=1}^{4} \mu_{i}\, e^{-i\, \vb{k_{\rm ph}} \rhobf_{\alpha i}}\, \sigmabf +

    e^{-i\, c_{A}\, \vb{k_{\rm ph}} \vb{r}}\, \mu_{\rm p}\, \sigmabf +

    e^{i\, c_{\rm p}\, \vb{k_{\rm ph}} \vb{r}}\,
    \displaystyle\sum_{j=1}^{A}
      \mu_{j}\, e^{-i\, \vb{k_{\rm ph}} \rhobf_{Aj}}\, \sigmabf
  \biggr\}
  \cdot \bigl[ \vb{k_{\rm ph}} \cp \vb{e}^{(\alpha)} \bigr]
  \biggr|\, F_{i} \biggr\rangle,
\end{array}
\label{eq.app.short.2.8.8}
\end{equation}
\begin{equation}
\begin{array}{lcl}
\vspace{-0.1mm}
  M_{\Delta E} & = & -\, (2\pi)^{3} \delta (\vb{K}_{f} - \vb{K}_{i} - \vb{k}_{\rm ph})\; \times \\
\vspace{-0.1mm}
  & \times &
  2\, \mu_{N}
  \displaystyle\sum\limits_{\alpha=1,2} \vb{e}^{(\alpha)}
  \biggl\langle F_{f}\, \biggl|\,
    e^{i\, c_{\rm p}\, \vb{k_{\rm ph}} \vb{r}}\,
    \displaystyle\sum_{j=1}^{A-1} \displaystyle\frac{z_{j} m_{\rm p}}{m_{Aj}}\, e^{-i\, \vb{k_{\rm ph}} \rhobf_{Aj}}\, \vb{\tilde{p}}_{Aj} -

    \displaystyle\frac{m_{\rm p}}{m_{A}}\,
    e^{i\, c_{\rm p}\, \vb{k_{\rm ph}} \vb{r}}\,
    \displaystyle\sum_{j=1}^{A} z_{j}\, e^{-i\, \vb{k_{\rm ph}} \rhobf_{Aj}} \Bigl( \displaystyle\sum_{k=1}^{A-1} \vb{\tilde{p}}_{Ak} \Bigr)\,
  \biggr|\, F_{i} \biggr\rangle,
\end{array}
\label{eq.app.short.2.8.9}
\end{equation}
\begin{equation}
\begin{array}{lll}
\vspace{0.2mm}
  & M_{\Delta M} = -\, i\, (2\pi)^{3}\, \delta (\vb{K}_{f} - \vb{K}_{i} - \vb{k}_{\rm ph})\,\mu_{N}\; \times \\

  \times &
    \displaystyle\sum\limits_{\alpha=1,2} \biggl\langle F_{f}\, \biggl|\;
    e^{i\, \vb{k_{\rm ph}} c_{\rm p}\, \vb{r}}\,
    \displaystyle\sum_{j=1}^{A-1}
      \mu_{j}\, e^{-i\, \vb{k_{\rm ph}} \rhobf_{Aj}}\, \sigmabf \cdot \bigl[ \vb{\tilde{p}}_{Aj} \times \vb{e}^{(\alpha)} \bigr] -

    e^{i\, \vb{k_{\rm ph}} c_{\rm p}\, \vb{r}}\,
    \displaystyle\sum_{j=1}^{A}
      \mu_{j}\, \displaystyle\frac{m_{Aj}}{m_{A}}\,
      e^{-i\, \vb{k_{\rm ph}} \rhobf_{Aj}}\,
      \displaystyle\sum_{k=1}^{A-1} \sigmabf \cdot \bigl[ \vb{\tilde{p}}_{Ak} \times \vb{e}^{(\alpha)} \bigr]
  \biggr|\, F_{i} \biggr\rangle,
\end{array}
\label{eq.app.short.2.8.10}
\end{equation}
In these formulas we have integration over space variables $\vb{r}$, $\rhobf_{Aj}$ ($j = 1 \ldots A-1$),
and for proton-nucleus scattering for $M_{P}$ we have $\vb{K}_{i} \ne 0$ and obtain property:
\begin{equation}
\begin{array}{lcl}
  \vu{P} \Psi_{i} = \vu{P} \Phi_{i} (\vb{R}) F_{i} = -\,\hbar\, \vb{K}_{i}\, \Phi_{\bar{s}} (\vb{R}) F_{i}, &
  \vb{K}_{i} \ne 0.
\end{array}
\label{eq.app.short.2.8.11}
\end{equation}
% -----------------------------------------------------------------------------------------------------------------------

% -----------------------------------------------------------------------------------------------------------------------
\subsection{Electric and magnetic form factors
\label{sec.app.short.2.9}}

\subsubsection{Calculations of matrix element $M_{p}$
\label{sec.app.short.2.9.a}}

Substituting explicit formulation (\ref{eq.app.short.2.8.1}) for wave function $F (\vb{r}, \beta_{A}, \beta_{\rm p})$ to the obtained matrix element (\ref{eq.app.short.2.8.7}),
we calculate it
\begin{equation}
\begin{array}{lll}
\vspace{-0.1mm}
  M_{p} & = &
  i \hbar\, (2\pi)^{3} \delta (\vb{K}_{f} - \vb{K}_{i} - \vb{k}_{\rm ph}) \cdot
  \displaystyle\sum\limits_{\alpha=1,2}
  \displaystyle\int\limits_{}^{}
    \Phi_{\rm p - nucl, f}^{*} (\vb{r})\;
    e^{i\, \vb{k}_{\rm ph} \vb{r}}\; \times \\
\vspace{0.5mm}
  & \times &
  \biggl\{
  2\, \mu_{N}\,  m_{\rm p}\,
  \Bigl[
    e^{-i\, c_{A} \vb{k_{\rm ph}} \vb{r}}\, \displaystyle\frac{F_{{\rm p},\, {\rm el}}}{m_{\rm p}} -
    e^{i\, c_{\rm p} \vb{k_{\rm ph}} \vb{r}}\,  \displaystyle\frac{F_{A,\, {\rm el}}}{m_{A}}
  \Bigr]\,
  e^{- i\, \vb{k}_{\rm ph} \vb{r}} \cdot
  \vb{e}^{(\alpha)}\, \vb{\displaystyle\frac{d}{dr}}\; + \\
% \vspace{-0.1mm}
  & + &
  i\,\mu_{N}\,
  \Bigl[
    e^{-i\, c_{A} \vb{k_{\rm ph}} \vb{r}}\, \vb{F}_{{\rm p},\, {\rm mag}} -
    e^{i\, c_{\rm p} \vb{k_{\rm ph}} \vb{r}}\, \vb{F}_{A,\, {\rm mag}}
  \Bigr]\,
  e^{- i\, \vb{k}_{\rm ph} \vb{r}} \cdot
  \Bigl[ \vb{\displaystyle\frac{d}{dr}} \times \vb{e}^{(\alpha)} \Bigr]\,
  \biggr\} \cdot
  \Phi_{\rm p - nucl, i} (\vb{r})\; \vb{dr}.
\end{array}
\label{eq.app.short.2.9.10}
\end{equation}
Here, we define \emph{electric and magnetic form factors} of the proton of scattering and nucleus as
\begin{equation}
\begin{array}{lll}
\vspace{1mm}
  F_{{\rm p},\, {\rm el}} (\vb{k}_{\rm ph}) & = &
    % \displaystyle\sum\limits_{n=1}^{4}
    %   \Bigl\langle \psi_{\alpha, f} (\beta_{\alpha})\, \Bigl|\, z_{n}\, e^{-i \vb{k}_{\rm ph} \rhobf_{\alpha n} } \Bigr|\,  \psi_{\alpha, i} (\beta_{\alpha}) \Bigr\rangle , \\
    \Bigl\langle \psi_{{\rm p}, f} (\beta_{\rm p})\, \Bigl|\, z_{\rm p}\, \Bigr|\,  \psi_{{\rm p}, i} (\beta_{\rm p}) \Bigr\rangle \equiv z_{\rm p}, \\

\vspace{1mm}
  F_{A,\, {\rm el}} (\vb{k}_{\rm ph}) & = &
    \displaystyle\sum\limits_{m=1}^{A}
    \Bigl\langle \psi_{\rm nucl, f} (\beta_{A}) \Bigl|\,
      z_{m}\, e^{-i \vb{k}_{\rm ph} \rhobf_{A m} }
    \Bigr|\, \psi_{\rm nucl, i} (\beta_{A}) \Bigr\rangle , \\

\vspace{1mm}
  \vb{F}_{{\rm p},\, {\rm mag}} (\vb{k}_{\rm ph}) & = &
    \Bigl\langle \psi_{\alpha, f} (\beta_{\alpha})\, \Bigl|\,
      \mu_{\rm p}\, \sigmabf
    \Bigr| \psi_{\alpha, i} (\beta_{\alpha}) \Bigr\rangle, \\

  \vb{F}_{A,\, {\rm mag}} (\vb{k}_{\rm ph}) & = &
    \displaystyle\frac{1}{m_{A}}
    \displaystyle\sum_{j=1}^{A}
    \Bigl\langle \psi_{\rm nucl, f} (\beta_{A})\, \Bigl|\,
        \mu_{j}\, m_{Aj}\; e^{-i\, \vb{k_{\rm ph}} \rhobf_{Aj}}\, \sigmabf
    \Bigr| \psi_{\rm nucl, i} (\beta_{A}) \Bigr\rangle\,
\end{array}
\label{eq.app.short.2.9.7}
\end{equation}
and we take into account normalization condition for wave functions as
\begin{equation}
\begin{array}{lll}
  \Bigl\langle \psi_{\rm nucl, f} (\beta_{A}) \Bigr|\, \psi_{\rm nucl, i} (\beta_{A}) \Bigr\rangle = 1, &
  \Bigl\langle \psi_{{\rm p}, f} (\beta_{\rm p})\, \Bigl|\, \psi_{{\rm p}, i} (\beta_{\rm p}) \Bigr\rangle = 1.
\end{array}
\label{eq.app.short.2.9.5}
\end{equation}
% -----------------------------------------------------------------------------------------------------------------------

% -----------------------------------------------------------------------------------------------------------------------
\subsubsection{Calculations for the matrix elements $M_{\Delta E}$ and $M_{\Delta M}$
\label{sec.2.9.b}}

For $M_{\Delta E}$ and $M_{\Delta M}$ we obtained solutions (\ref{eq.app.short.2.8.9}) and (\ref{eq.app.short.2.8.10}).
Using logic in the previous subsection, we transform these solutions as
\begin{equation}
\begin{array}{lll}
\vspace{-0.1mm}
  M_{\Delta E} & = &
  -\, (2\pi)^{3} \delta (\vb{K}_{f} - \vb{K}_{i} - \vb{k}_{\rm ph}) \cdot
  2\, \mu_{N}
  \displaystyle\sum\limits_{\alpha=1,2} \vb{e}^{(\alpha)}
  \displaystyle\int\limits_{}^{}
    \Phi_{\rm p - nucl, f}^{*} (\vb{r})\; \times \\
  & \times &
  \biggl\{
    e^{i\, c_{\rm p}\, \vb{k_{\rm ph}} \vb{r}}\, \vb{D}_{A 1,\, {\rm el}} -
    \displaystyle\frac{m_{\rm p}}{m_{A}}\, e^{i\, c_{\rm p}\, \vb{k_{\rm ph}} \vb{r}}\, \vb{D}_{A 2,\, {\rm el}}
  \biggr\} \cdot
  \Phi_{\rm p - nucl, i} (\vb{r})\; \vb{dr},
\end{array}
\label{eq.app.short.2.9.b.1}
\end{equation}
\begin{equation}
\begin{array}{lll}
\vspace{-0.1mm}
  M_{\Delta M} & = &
  -\, i\, (2\pi)^{3} \delta (\vb{K}_{f} - \vb{K}_{i} - \vb{k}_{\rm ph}) \cdot \mu_{N}\,
  \displaystyle\sum\limits_{\alpha=1,2}
  \displaystyle\int\limits_{}^{}
    \Phi_{\rm p - nucl, f}^{*} (\vb{r})\; \times \\

  & \times &
  \biggl\{
    e^{i\, c_{\rm p}\, \vb{k_{\rm ph}} \vb{r}}\; D_{A 1,\, {\rm mag}} (\vb{e}^{(\alpha)}) -
    e^{i\, c_{\rm p}\, \vb{k_{\rm ph}} \vb{r}}\; D_{A 2,\, {\rm mag}} (\vb{e}^{(\alpha)})
  \biggr\} \cdot
  \Phi_{\rm p - nucl, i} (\vb{r})\; \vb{dr},
\end{array}
\label{eq.app.short.2.9.b.2}
\end{equation}
where
\begin{equation}
\begin{array}{lll}
\vspace{1mm}
  \vb{D}_{A 1,\, {\rm el}} = &
    \displaystyle\sum\limits_{i=1}^{A-1}
      \displaystyle\frac{z_{j} m_{\rm p}}{m_{Aj}}\,
      \Bigl\langle \psi_{A, f} (\beta_{A})\, \Bigl|\,
        e^{-i \vb{k}_{\rm ph} \rhobf_{Aj}} \vb{\tilde{p}}_{Aj}
      \Bigr|\,  \psi_{Aj} (\beta_{A}) \Bigr\rangle , \\

  \vb{D}_{A 2,\, {\rm el}} = &
    \displaystyle\sum\limits_{i=1}^{A}
      z_{j}\,
      \Bigl\langle \psi_{A, f} (\beta_{A})\, \Bigl|\,
        e^{-i \vb{k}_{\rm ph} \rhobf_{Aj}}
        \Bigl( \displaystyle\sum_{k=1}^{A-1} \vb{\tilde{p}}_{Ak} \Bigr)
      \Bigr|\,  \psi_{Aj} (\beta_{A}) \Bigr\rangle,
\end{array}
\label{eq.app.short.2.9.b.3}
\end{equation}
\begin{equation}
\begin{array}{lll}
\vspace{1mm}
  D_{A 1,\, {\rm mag}} (\vb{e}^{(\alpha)}) = &
    \displaystyle\sum\limits_{j=1}^{A-1}
    \mu_{j}\,
    \Bigl\langle \psi_{A, f} (\beta_{A})\, \Bigl|\,
      e^{-i\, \vb{k_{\rm ph}} \rhobf_{Aj}}\; \sigmabf \cdot
      \bigl[ \vb{\tilde{p}}_{Aj} \times \vb{e}^{(\alpha)} \bigr]
    \Bigr|\,  \psi_{Aj} (\beta_{A}) \Bigr\rangle, \\

  D_{A 2,\, {\rm mag}} (\vb{e}^{(\alpha)}) = &
    \displaystyle\sum\limits_{j=1}^{A}
      \mu_{j}\,
      \displaystyle\frac{m_{Aj}}{m_{A}}\,
      \Bigl\langle \psi_{A, f} (\beta_{A})\, \Bigl|\,
      e^{-i\, \vb{k_{\rm ph}} \rhobf_{Aj}}\; \sigmabf \cdot
      \Bigl[ \Bigl( \displaystyle\sum_{k=1}^{A-1} \vb{\tilde{p}}_{Ak} \Bigr) \times \vb{e}^{(\alpha)} \Bigr]
      \Bigr|\,  \psi_{A, i} (\beta_{A}) \Bigr\rangle.
\end{array}
\label{eq.app.short.2.9.b.4}
\end{equation}
% -----------------------------------------------------------------------------------------------------------------------

% -----------------------------------------------------------------------------------------------------------------------
\subsubsection{Calculations for the matrix elements $M_{k}$
\label{sec.2.9.c}}

For $M_{k}$ we obtained solution (\ref{eq.app.short.2.8.8}).
Using logic in the previous subsection, we transform these solutions as
\begin{equation}
\begin{array}{lcl}
\vspace{-0.1mm}
  M_{k} & = &
  i\, \hbar\, (2\pi)^{3} \delta (\vb{K}_{f} - \vb{K}_{i} - \vb{k}_{\rm ph}) \cdot \mu_{N}\,
  \displaystyle\sum\limits_{\alpha=1,2}
    \bigl[ \vb{k_{\rm ph}} \cp \vb{e}^{(\alpha)} \bigr]
  \displaystyle\int\limits_{}^{}
    \Phi_{{\rm p - nucl,} f}^{*} (\vb{r})\; \times \\
\vspace{-0.1mm}
  & \times &
  \biggl\{
    e^{-i\, c_{A}\, \vb{k_{\rm ph}} \vb{r}}\, \vb{D}_{{\rm p},\, {\rm k}} +
    e^{i\, c_{\rm p}\, \vb{k_{\rm ph}} \vb{r}}\, \vb{D}_{A,\, {\rm k}}
  \biggr\} \cdot
  \Phi_{{\rm p - nucl,} i} (\vb{r})\; \vb{dr},
\end{array}
\label{eq.app.short.2.9.c.1}
\end{equation}
where
\begin{equation}
\begin{array}{lll}
\vspace{1mm}
  \vb{D}_{{\rm p},\, {\rm k}} = &
  \mu_{\rm p}\,
    \Bigl\langle \psi_{{\rm p}, f} (\beta_{\rm p})\, \Bigl|\,
      \sigmabf
    \Bigr|\,  \psi_{{\rm p}, i} (\beta_{\rm p}) \Bigr\rangle , \\

  \vb{D}_{A,\, {\rm k}} = &
  \displaystyle\sum\limits_{j=1}^{A}
    \mu_{j}\,
    \Bigl\langle \psi_{A, f} (\beta_{A})\, \Bigl|\,
      e^{-i\, \vb{k_{\rm ph}} \rhobf_{Aj}}\, \sigmabf
    \Bigr|\,  \psi_{A, j} (\beta_{A}) \Bigr\rangle.
\end{array}
\label{eq.app.short.2.9.c.2}
\end{equation}
% -----------------------------------------------------------------------------------------------------------------------

% -----------------------------------------------------------------------------------------------------------------------
\subsubsection{Calculations for the matrix elements $M_{P}$
\label{sec.2.9.d}}

For $M_{P}$ we obtained solution (\ref{eq.app.short.2.8.12}).
Using a logic in the previous subsection, we transform these solutions as
\begin{equation}
\begin{array}{lcl}
\vspace{-0.1mm}
  M_{P} & = &
  \hbar\, (2\pi)^{3} \delta (\vb{K}_{f} - \vb{K}_{i} - \vb{k}_{\rm ph}) \cdot
  \displaystyle\frac{1}{m_{A} + m_{\rm p}}\,
  \displaystyle\sum\limits_{\alpha=1,2}
  \displaystyle\int\limits_{}^{}
    \Phi_{{\rm p - nucl,} f}^{*} (\vb{r})\; \times \\

\vspace{-0.1mm}
  & \times &
  \biggl\{
    2\, \mu_{N}\, m_{\rm p}\;
    \Bigl[
      e^{-i\, c_{A}\, \vb{k_{\rm ph}} \vb{r}} F_{{\rm p},\, {\rm el}} +
      e^{i\, c_{\rm p}\, \vb{k_{\rm ph}} \vb{r}} F_{A,\, {\rm el}}
    \Bigr]\, \vb{e}^{(\alpha)} \cdot \vb{K}_{i} + \\

  & + &
  i\,\mu_{N}\,
  \Bigl[
    m_{\rm p}\, e^{-i\, c_{A}\, \vb{k_{\rm ph}} \vb{r}}\, \vb{F}_{{\rm p},\, {\rm mag}} +
    m_{A}\, e^{i\, c_{\rm p}\, \vb{k_{\rm ph}} \vb{r}}\, \vb{F}_{A,\, {\rm mag}}
  \Bigr] \cdot
    \bigl[ \vb{K}_{i} \cp \vb{e}^{(\alpha)} \bigr]
  \biggr\} \cdot
  \Phi_{{\rm p - nucl,} i} (\vb{r})\; \vb{dr}.
\end{array}
\label{eq.app.short.2.9.d.3}
\end{equation}
where
\begin{equation}
\begin{array}{lll}
\vspace{1mm}
  D_{{\rm p}, P\, {\rm el}} = &
  % \displaystyle\sum\limits_{i=1}^{4} z_{i}\, \Bigl\langle \psi_{\alpha, f} (\beta_{\alpha})\, \Bigl|\,
  %     e^{-i\, \vb{k_{\rm ph}} \rhobf_{\alpha i}} \Bigr|\,  \psi_{\alpha, i} (\beta_{\alpha}) \Bigr\rangle = F_{\alpha,\, {\rm el}}, \\
    \Bigl\langle \psi_{{\rm p}, f} (\beta_{\rm p})\, \Bigl|\, z_{\rm p}\, \Bigr|\,  \psi_{{\rm p}, i} (\beta_{\rm p}) \Bigr\rangle \equiv z_{\rm p} =
  F_{{\rm p},\, {\rm el}}, \\

\vspace{1mm}
  D_{A,P\, {\rm el}} = &
    \displaystyle\sum\limits_{m=1}^{A}
    \Bigl\langle \psi_{\rm nucl, f} (\beta_{A}) \Bigl|\,
      z_{m}\, e^{-i \vb{k}_{\rm ph} \rhobf_{A m} }
    \Bigr|\, \psi_{\rm nucl, i} (\beta_{A}) \Bigr\rangle =
  F_{A,\, {\rm el}}, \\

\vspace{1mm}
  \vb{D}_{{\rm p}, P\, {\rm mag}} = &
  % \displaystyle\sum\limits_{i=1}^{4}
  %   \mu_{i}\, m_{\alpha i}\,
  %   \Bigl\langle \psi_{\alpha, f} (\beta_{\alpha})\, \Bigl|\,
  %     e^{-i\, \vb{k_{\rm ph}} \rhobf_{\alpha i}}\, \sigmabf
  %   \Bigr|\,  \psi_{\alpha, i} (\beta_{\alpha}) \Bigr\rangle = \vb{F}_{\alpha,\, {\rm mag}}, \\
    \Bigl\langle \psi_{\alpha, f} (\beta_{\alpha})\, \Bigl|\,
      \mu_{\rm p}\, \sigmabf
    \Bigr| \psi_{\alpha, i} (\beta_{\alpha}) \Bigr\rangle =
  \vb{F}_{{\rm p},\, {\rm mag}}, \\

  \vb{D}_{A,P\, {\rm mag}} = &
    \displaystyle\frac{1}{m_{A}}
    \displaystyle\sum_{j=1}^{A}
    \Bigl\langle \psi_{\rm nucl, f} (\beta_{A})\, \Bigl|\,
        \mu_{j}\, m_{Aj}\; e^{-i\, \vb{k_{\rm ph}} \rhobf_{Aj}}\, \sigmabf
    \Bigr| \psi_{\rm nucl, i} (\beta_{A}) \Bigr\rangle =
  \vb{F}_{A,\, {\rm mag}}.
\end{array}
\label{eq.app.short.2.9.d.2}
\end{equation}
% *******************************************************************************************************************

% *******************************************************************************************************************
\subsection{Effective electric charge and effective magnetic moment of the full system
\label{sec.app.short.2.10}}

Let us consider the first two terms inside the first brackets in the matrix element $M_{p}$ in form~(\ref{eq.app.short.2.9.10}).
In the first approximation, electrical form factors tend to electric charges of proton and the nucleus.
We write
\begin{equation}
\begin{array}{lll}
  e^{-i\, c_{A} \vb{k_{\rm ph}} \vb{r}}\, \displaystyle\frac{1}{m_{\rm p}}\, F_{{\rm p},\, {\rm el}} -
  e^{i\, c_{\rm p} \vb{k_{\rm ph}} \vb{r}}\,  \displaystyle\frac{1}{m_{A}}\, F_{A,\, {\rm el}} =

  \displaystyle\frac{1}{\mu} \cdot
  \Bigl[
    e^{-i\, c_{A} \vb{k_{\rm ph}} \vb{r}}\, \displaystyle\frac{m_{A}}{m_{\rm p} + m_{A}}\, F_{{\rm p},\, {\rm el}} -
    e^{i\, c_{\rm p} \vb{k_{\rm ph}} \vb{r}}\,  \displaystyle\frac{m_{\rm p}}{m_{\rm p} + m_{A}}\, F_{A,\, {\rm el}}
  \Bigr],
\end{array}
\label{eq.app.short.2.10.1}
\end{equation}
where
\begin{equation}
  \mu = \displaystyle\frac{m_{\rm p}\, m_{A}}{m_{\rm p} + m_{A}}
\label{eq.app.short.2.10.2}
\end{equation}
%
% --- это \textcolor[rgb]{1.00,0.00,0.00}{\textbf{\emph{приведенная масса системы из протона рассеяния и ядра-мишени}}}.
% Итак, мы ввели новое определение \textcolor[rgb]{1.00,0.00,0.00}{\textbf{\emph{эффективного электрического заряда}}} полной системы как
is reduced mass of system of the scattered proton and the nucleus.
We introduce definitions of effective electric charge and effective magnetic moment of the full system as
\begin{equation}
\begin{array}{lll}
  Z_{\rm eff} (\vb{k}_{\rm ph}, \vb{r}) =
  e^{i\, \vb{k_{\rm ph}} \vb{r}}\,
  \Bigl[
    e^{-i\, c_{A} \vb{k_{\rm ph}} \vb{r}}\, \displaystyle\frac{m_{A}}{m_{\rm p} + m_{A}}\, F_{{\rm p},\, {\rm el}} -
    e^{i\, c_{\rm p} \vb{k_{\rm ph}} \vb{r}}\,  \displaystyle\frac{m_{\rm p}}{m_{\rm p} + m_{A}}\, F_{A,\, {\rm el}}
  \Bigr],
\end{array}
\label{eq.app.short.2.10.3}
\end{equation}
\begin{equation}
\begin{array}{lll}
  \textbf{M}_{\rm eff} (\vb{k}_{\rm ph}, \vb{r}) =
  e^{i\, \vb{k_{\rm ph}} \vb{r}}\, \displaystyle\frac{1}{m_{\rm p}}
  \Bigl[
    e^{-i\, c_{A} \vb{k_{\rm ph}} \vb{r}}\, \displaystyle\frac{m_{A}}{m_{\rm p} + m_{A}}\, m_{\rm p}\, \vb{F}_{{\rm p},\, {\rm mag}} -
    e^{i\, c_{\rm p} \vb{k_{\rm ph}} \vb{r}}\,  \displaystyle\frac{m_{\rm p}}{m_{\rm p} + m_{A}}\, m_{A}\, \vb{F}_{A,\, {\rm mag}}
  \Bigr].
\end{array}
\label{eq.app.short.2.10.4}
\end{equation}
We rewrite expression (\ref{eq.app.short.2.9.10}) for $M_{p}$ via effective electric charge and magnetic moment in a compact form as
\begin{equation}
\begin{array}{lll}
\vspace{-0.2mm}
  M_{p} & = &
  i \hbar\, (2\pi)^{3} \delta (\vb{K}_{f} - \vb{K}_{i} - \vb{k}_{\rm ph}) \cdot
  \displaystyle\sum\limits_{\alpha=1,2}
  \displaystyle\int\limits_{}^{}
    \Phi_{\rm p - nucl, f}^{*} (\vb{r})\;
    e^{-i\, \vb{k}_{\rm ph} \vb{r}}\; \times \\
\vspace{0.3mm}
  & \times &
  \biggl\{
  2\, \mu_{N}\,  m_{\rm p} \cdot
  \displaystyle\frac{1}{\mu}\, Z_{\rm eff} (\vb{k}_{\rm ph}, \vb{r}) \cdot
  \vb{e}^{(\alpha)}\, \vb{\displaystyle\frac{d}{dr}} +

  i\, \mu_{N} \cdot
  \displaystyle\frac{m_{\rm p}}{\mu}\, \vb{M}_{\rm eff} (\vb{k}_{\rm ph}, \vb{r}) \cdot
  \Bigl[ \vb{\displaystyle\frac{d}{dr}} \times \vb{e}^{(\alpha)} \Bigr]
  \biggr\} \cdot
  \Phi_{\rm p - nucl, i} (\vb{r})\; \vb{dr}\; = \\

\vspace{-0.2mm}
  & = &
  i \hbar\, (2\pi)^{3}\: \displaystyle\frac{m_{\rm p}}{\mu}\, \mu_{N}\: \delta (\vb{K}_{f} - \vb{K}_{i} - \vb{k}_{\rm ph})
  \displaystyle\sum\limits_{\alpha=1,2}
  \displaystyle\int\limits_{}^{}
    \Phi_{\rm p - nucl, f}^{*} (\vb{r})\;
    e^{-i\, \vb{k}_{\rm ph} \vb{r}}\; \times \\
  & \times &
  \biggl\{
  2\, Z_{\rm eff} (\vb{k}_{\rm ph}, \vb{r}) \cdot
  \vb{e}^{(\alpha)}\, \vb{\displaystyle\frac{d}{dr}} +

  i\,
  \vb{M}_{\rm eff} (\vb{k}_{\rm ph}, \vb{r}) \cdot
  \Bigl[ \vb{\displaystyle\frac{d}{dr}} \times \vb{e}^{(\alpha)} \Bigr]
  \biggr\} \cdot
  \Phi_{\rm p - nucl, i} (\vb{r})\; \vb{dr}.
\end{array}
\label{eq.app.short.2.10.6}
\end{equation}
We have obtained the final formula for the matrix element, where we have our new introduced effective electric charge and magnetic moment of the full nuclear system (of the proton of scattering and nucleus-target).
% *******************************************************************************************************************

% *******************************************************************************************************************
% \newpage
\subsection{Integration over momentum $K_{f}$
\label{sec.app.short.2.12}}

We will determine cross-section of emission of photons, which does not depend on vector $\vb{K}_{f}$.
%  (импульса полной ядерной системы после излучения фотона в лабораторной системе отсчета).
All degrees of freedom, related with $\vb{K}_{f}$, should be removed and we integrate all matrix elements over vector $\mathbf{K}_{f}$.
Taking into account property:
%
% Мы будем определять сечения излучения фотонов, не зависящие от вектора $\vb{K}_{f}$ (импульса полной ядерной системы после излучения фотона в лабораторной системе отсчета).
% Значит все степени свободы, связанные с $\mathbf{K}_{f}$, надо исключить и поэтому мы проинтегрируем все матричные элементы по всевозможным значениям вектора $\vb{K}_{f}$.
% Учитывая свойство:
%
\begin{equation}
\begin{array}{lll}
  \displaystyle\int \delta (\vb{K}_{i} - \vb{K}_{f} - \vb{k})\; \vb{dK}_{f} = 1, &
  \vb{K}_{i} = \vb{K}_{f} + \vb{k},
\end{array}
\label{eq.app.short.2.12.1}
\end{equation}
we integrate each matrix element as
\begin{equation}
  M_{s} \Rightarrow \displaystyle\int M_{s} (\vb{K}_{f})\; \vb{dK}_{f},
\label{eq.app.short.2.12.2}
\end{equation}
%
% где $s$ --- индекс, указывающий на матричные элементы $M_{p}$, $M_{P}$, $M_{k}$, $M_{\Delta E}$ и $M_{\Delta M}$.
% В частности, из (\ref{eq.app.short.2.10.6}), (\ref{eq.app.short.2.9.b.3}), (\ref{eq.app.short.2.9.c.1}), (\ref{eq.app.short.2.9.d.1}), (\ref{eq.app.short.2.9.d.2}) получим:
where $s$ is index indicating the matrix elements $M_{p}$, $M_{P}$, $M_{k}$, $M_{\Delta E}$ and $M_{\Delta M}$.
In particular, from (\ref{eq.app.short.2.10.6}), (\ref{eq.app.short.2.9.b.3}), (\ref{eq.app.short.2.9.c.1}), (\ref{eq.app.short.2.9.d.1}), (\ref{eq.app.short.2.9.d.2}) we obtain:
\begin{equation}
\begin{array}{lll}
\vspace{-0.2mm}
  M_{p} & = &
  i \hbar\, (2\pi)^{3}\: \displaystyle\frac{m_{\rm p}}{\mu}\, \mu_{N}
  % \delta (\vb{K}_{f} - \vb{K}_{i} - \vb{k}_{\rm ph})
  \displaystyle\sum\limits_{\alpha=1,2}
  \displaystyle\int\limits_{}^{}
    \Phi_{\rm p - nucl, f}^{*} (\vb{r})\;
    e^{-i\, \vb{k}_{\rm ph} \vb{r}}\; \times \\
  & \times &
  \biggl\{
  2\, Z_{\rm eff} (\vb{k}_{\rm ph}, \vb{r}) \cdot
  \vb{e}^{(\alpha)}\, \vb{\displaystyle\frac{d}{dr}} +

  i\,
  \vb{M}_{\rm eff} (\vb{k}_{\rm ph}, \vb{r}) \cdot
  \Bigl[ \vb{\displaystyle\frac{d}{dr}} \times \vb{e}^{(\alpha)} \Bigr]
  \biggr\} \cdot
  \Phi_{\rm p - nucl, i} (\vb{r})\; \vb{dr}.
\end{array}
\label{eq.app.short.2.12.3}
% \label{eq.app.short.2.10.6}
\end{equation}
\begin{equation}
\begin{array}{lcl}
\vspace{-0.1mm}
  M_{P} & = &
  % \delta (\vb{K}_{f} - \vb{K}_{i} - \vb{k}_{\rm ph}) \cdot
  \displaystyle\frac{\hbar\, (2\pi)^{3}}{m_{A} + m_{\rm p}}\,
  \displaystyle\sum\limits_{\alpha=1,2}
  \displaystyle\int\limits_{}^{}
    \Phi_{\rm p - nucl, f}^{*} (\vb{r})\; % \times \\

% \vspace{-0.1mm}
%   & \times &
  \biggl\{
    2\, \mu_{N}\, m_{\rm p}\;
    \Bigl[
      e^{-i\, c_{A}\, \vb{k_{\rm ph}} \vb{r}} F_{{\rm p}, P\, {\rm el}} +
      e^{i\, c_{\rm p}\, \vb{k_{\rm ph}} \vb{r} } F_{A,P\, {\rm el}}
    \Bigr]\, \vb{e}^{(\alpha)} \cdot \vb{K}_{i} + \\

  & + &
  i\,\mu_{N}\,
  \Bigl[
    m_{\rm p}\, e^{-i\, c_{A}\, \vb{k_{\rm ph}} \vb{r}}\, \vb{F}_{{\rm p},P\, {\rm mag}} +
      % \displaystyle\sum_{i=1}^{4} \mu_{i}\, m_{\alpha i}\, e^{-i\, \vb{k_{\rm ph}} \rhobf_{\alpha i}}\, \sigmabf  +
    m_{A}\, e^{i\, c_{\rm p}\, \vb{k_{\rm ph}} \vb{r}}\, \vb{F}_{A,P\, {\rm mag}}
      % \displaystyle\sum_{j=1}^{A} \mu_{j}\, m_{Aj}\, e^{-i\, \vb{k_{\rm ph}} \rhobf_{Aj}}\, \sigmabf
  \Bigr]\, \cdot
    \bigl[ \vb{K}_{i} \cp \vb{e}^{(\alpha)} \bigr]
  \biggr\} \cdot
  \Phi_{\rm p - nucl, i} (\vb{r})\; \vb{dr},
\end{array}
\label{eq.app.short.2.12.4}
% \label{eq.app.short.2.9.d.1}
\end{equation}
\begin{equation}
\begin{array}{lcl}
% \vspace{-0.1mm}
  M_{k} & = &
  i\, \hbar\, (2\pi)^{3}
  % \delta (\vb{K}_{f} - \vb{K}_{i} - \vb{k}_{\rm ph})
  \cdot \mu_{N}\,
  \displaystyle\sum\limits_{\alpha=1,2}
    \bigl[ \vb{k_{\rm ph}} \cp \vb{e}^{(\alpha)} \bigr]
  \displaystyle\int\limits_{}^{}
    \Phi_{{\rm p - nucl,} f}^{*} (\vb{r})\; % \times \\
% \vspace{-0.1mm}
%   & \times &
  \biggl\{
    e^{-i\, c_{A}\, \vb{k_{\rm ph}} \vb{r}}\, \vb{D}_{{\rm p},\, {\rm k}} +
    e^{i\, c_{\rm p}\, \vb{k_{\rm ph}} \vb{r}}\, \vb{D}_{A,\, {\rm k}}
  \biggr\} \cdot
  \Phi_{{\rm p - nucl,} i} (\vb{r})\; \vb{dr},
\end{array}
\label{eq.app.short.2.12.5}
% \label{eq.app.short.2.9.c.1}
\end{equation}
\begin{equation}
\begin{array}{lll}
% \vspace{-0.1mm}
  M_{\Delta E} & = &
  -\, (2\pi)^{3}
  % \delta (\vb{K}_{f} - \vb{K}_{i} - \vb{k}_{\rm ph}) \cdot
  2\, \mu_{N}
  \displaystyle\sum\limits_{\alpha=1,2} \vb{e}^{(\alpha)}
  \displaystyle\int\limits_{}^{}
    \Phi_{\rm p - nucl, f}^{*} (\vb{r})\; % \times \\
%   & \times &
  \biggl\{
    e^{i\, c_{\rm p}\, \vb{k_{\rm ph}} \vb{r}}\, \vb{D}_{A 1,\, {\rm el}} -
    \displaystyle\frac{m_{\rm p}}{m_{A}}\, e^{i\, c_{\rm p}\, \vb{k_{\rm ph}} \vb{r}}\, \vb{D}_{A 2,\, {\rm el}}
  \biggr\} \cdot
  \Phi_{\rm p - nucl, i} (\vb{r})\; \vb{dr},
\end{array}
\label{eq.app.short.2.12.6}
% \label{eq.app.short.2.9.b.1}
\end{equation}
\begin{equation}
\begin{array}{lll}
% \vspace{-0.1mm}
  M_{\Delta M} & = &
  -\, i\, (2\pi)^{3}
  % \delta (\vb{K}_{f} - \vb{K}_{i} - \vb{k}_{\rm ph})
  \cdot \mu_{N}\,
  \displaystyle\sum\limits_{\alpha=1,2}
  \displaystyle\int\limits_{}^{}
    \Phi_{\rm p - nucl, f}^{*} (\vb{r})\; % \times \\

%   & \times &
  \biggl\{
    e^{i\, c_{\rm p}\, \vb{k_{\rm ph}} \vb{r}}\; D_{A 1,\, {\rm mag}} (\vb{e}^{(\alpha)}) -
    e^{i\, c_{\rm p}\, \vb{k_{\rm ph}} \vb{r}}\; D_{A 2,\, {\rm mag}} (\vb{e}^{(\alpha)})
  \biggr\} \cdot
  \Phi_{\rm p - nucl, i} (\vb{r})\; \vb{dr}.
\end{array}
\label{eq.app.short.2.12.7}
% \label{eq.app.short.2.9.b.2}
\end{equation}

% *******************************************************************************************************************

% *******************************************************************************************************************
% \newpage
\section{Calculations of matrix elements % in dipole approximation
\label{sec.app.3}}

\subsection{Matrix elements of coherent emission on the basis of $M_{p}$
\label{sec.2.13}}

% В этом разделе мы рассмотрим разные приближения, упрощающие вычисления на компьютере, но снижающие точность определения спектра.
% В (\ref{eq.2.12.3}) на стр.~\pageref{eq.2.12.3} мы получили выражение для матричного элемента, который можно считать когерентным:

% In (\ref{eq.2.12.3}) we obtained matrix element, which can be considered as coherent.
Rewrite Eq.~(\ref{eq.app.2.12.3}) as
\begin{equation}
  M_{p} = M_{p}^{(E)} + M_{p}^{(M)},
\label{eq.2.13.1.2}
\end{equation}
where
\begin{equation}
\begin{array}{lll}
\vspace{-0.2mm}
  M_{p}^{(E)} & = &
  i \hbar\, (2\pi)^{3}\, \displaystyle\frac{2\, \mu_{N}\,  m_{\rm p}}{\mu}\;
  \displaystyle\sum\limits_{\alpha=1,2}
    \vb{e}^{(\alpha)}
    \biggl\langle\: \Phi_{\rm p - nucl, f} (\vb{r})\; \biggl|\,
    Z_{\rm eff} (\vb{k}_{\rm ph}, \vb{r}) \,
    e^{-i\, \vb{k}_{\rm ph} \vb{r}}\;
    \vb{\displaystyle\frac{d}{dr}}
    \biggr|\: \Phi_{\rm p - nucl, i} (\vb{r})\: \biggr\rangle, \\

  M_{p}^{(M)} & = &
  -\, \hbar\, (2\pi)^{3}\, \displaystyle\frac{\mu_{N}\,  m_{\rm p}}{\mu}\;
  \displaystyle\sum\limits_{\alpha=1,2}
  \biggl\langle\: \Phi_{\rm p - nucl, f} (\vb{r})\; \biggl|\,
  \vb{M}_{\rm eff} (\vb{k}_{\rm ph}, \vb{r}) \cdot
  e^{-i\, \vb{k}_{\rm ph} \vb{r}} \cdot
  \Bigl[ \vb{\displaystyle\frac{d}{dr}} \times \vb{e}^{(\alpha)} \Bigr]
  \biggr|\: \Phi_{\rm p - nucl, i} (\vb{r})\: \biggr\rangle.
\end{array}
\label{eq.2.13.1.3}
\end{equation}
%-----------------------------------------------------------------------------------------------------------------------

%-----------------------------------------------------------------------------------------------------------------------
% \subsubsection{Dipole approximation of effective electric charge
% Дипольное приближение для эффективного электрического заряда
% \label{sec.2.13.2}}

The effective electric charge (\ref{eq.app.2.10.3})
% Эффективный заряд системы (\ref{eq.2.10.3}) на стр.~\pageref{eq.2.10.3}
%
\[
\begin{array}{lll}
  Z_{\rm eff} (\vb{k}_{\rm ph}, \vb{r}) =
  e^{i\, \vb{k_{\rm ph}} \vb{r}}\,
  \Bigl[
    e^{-i\, c_{A} \vb{k_{\rm ph}} \vb{r}}\, \displaystyle\frac{m_{A}}{m_{\rm p} + m_{A}}\, F_{{\rm p},\, {\rm el}} -
    e^{i\, c_{\rm p} \vb{k_{\rm ph}} \vb{r}}\,  \displaystyle\frac{m_{\rm p}}{m_{\rm p} + m_{A}}\, F_{A,\, {\rm el}}
  \Bigr]
\end{array}
% \label{eq.2.10.3}
\]
in the first approximation $\exp(i\vb{k_{\rm ph}} \vb{r}) \to 1$ (we call it as dipole for the effective electric charge) is
% In particular, in the first approximation (called as \definition{dipole}) we obtain:
% в первом приближении $\exp(i\vb{k_{\rm ph}} \vb{r}) \to 1$ (т.~е. при $\vb{k_{\rm ph}} \vb{r} \to 0$, которое назовем \definition{монопольным} в применении к эффективному заряду) приобретет вид:
%
\begin{equation}
  Z_{\rm eff}^{\rm (dip)} (\vb{k}_{\rm ph}) = \displaystyle\frac{m_{A}\, z_{\rm p} - m_{\rm p}\, Z_{\rm A}(\vb{k}_{\rm ph})}{m_{\rm p} + m_{A}}.
\label{eq.2.13.2.1}
\end{equation}
In this approximation the effective charge is independent on relative distance between proton and nucleus.
% Видно, что в таком приближении эффективный заряд становится независимым от относительного расстояния между протоном и центром масс ядра (и его можно вынести вне матричного элемента).
%
We neglect relative displacements of nucleons of nucleus inside its space region, and
form factor of nucleus is just summation of electric charges of nucleons of nucleus:
\begin{equation}
  Z_{\rm A} (\vb{k}_{\rm ph}) \to
  \Bigl\langle \psi_{\rm nucl, f} (\rhobf_{A1} \ldots \rhobf_{AA-1})\: \Bigl|\;
    \displaystyle\sum\limits_{j=1}^{A}
      z_{Aj}\:
  \Bigr|\: \psi_{\rm nucl, i} (\rhobf_{A1} \ldots \rhobf_{AA-1})\, \Bigr\rangle =
  \displaystyle\sum\limits_{j=1}^{A}\, z_{Aj} = z_{\rm A}
\label{eq.2.13.2.2}
\end{equation}
as functions $\psi_{\rm nucl, s}$ are normalized. We write
\begin{equation}
  Z_{\rm eff}^{\rm (dip, 0)} = \displaystyle\frac{m_{A}\, z_{\rm p} - m_{\rm p}\, z_{\rm A}}{m_{\rm p} + m_{A}},
\label{eq.2.13.2.3}
\end{equation}
\begin{equation}
\begin{array}{llll}
% \vspace{-0.2mm}
  M_{p}^{(E,\, {\rm dip},0)} =
  i \hbar\, (2\pi)^{3} \displaystyle\frac{2\, \mu_{N}\,  m_{\rm p}}{\mu}\;
  Z_{\rm eff}^{\rm (dip, 0)}\;
  \displaystyle\sum\limits_{\alpha=1,2}
    \vb{e}^{(\alpha)} \cdot \vb{I}_{1}, &
    % \biggl\langle\: \Phi_{\rm p - nucl, f} (\vb{r})\; \biggl|\, e^{-i\, \vb{k}_{\rm ph} \vb{r}}\; \vb{\displaystyle\frac{d}{dr}} \biggr|\: \Phi_{\rm p - nucl, i} (\vb{r})\: \biggr\rangle.

  \vb{I}_{1} =
    \biggl\langle\: \Phi_{\rm p - nucl, f} (\vb{r})\; \biggl|\,
    e^{-i\, \vb{k}_{\rm ph} \vb{r}}\;
    \vb{\displaystyle\frac{d}{dr}}
    \biggr|\: \Phi_{\rm p - nucl, i} (\vb{r})\: \biggr\rangle,
\end{array}
\label{eq.2.13.2.4}
\end{equation}
%
% and also
% и также
%
% \begin{equation}
%   p_{p}^{(E,\, {\rm dip},0)} =
%   -\, \displaystyle\frac{m_{\rm p}}{e} \cdot M_{p}^{\rm (dip,\, 0)} =
%   -\, \displaystyle\frac{m_{\rm p}}{e} \cdot
%   i \hbar\, (2\pi)^{3} \displaystyle\frac{2\, \mu_{N}\,  m_{\rm p}}{\mu}\;
%   Z_{\rm eff}^{\rm (dip, 0)}\;
%   \displaystyle\sum\limits_{\alpha=1,2}
%     \vb{e}^{(\alpha)}
%     \biggl\langle\: \Phi_{\rm p - nucl, f} (\vb{r})\; \biggl|\,
%     e^{-i\, \vb{k}_{\rm ph} \vb{r}}\;
%     \vb{\displaystyle\frac{d}{dr}}
%     \biggr|\: \Phi_{\rm p - nucl, i} (\vb{r})\: \biggr\rangle.
% \label{eq.2.13.2.5}
% \end{equation}
%
where upper index \emph{``dip''} denotes inclusion of approximations above.
% Здесь дополнительным верхним индексом \emph{``dip''} мы обозначили включение выбранных дипольных приближений.
% Учитывая явный вид ядерного магнетона $\mu_{N} = e\hbar / (2m_{\rm p}c)$, перепишем последнюю формулу так:
%
% \begin{equation}
% \begin{array}{lll}
% \vspace{0.2mm}
%   M_{p}^{(E,\, {\rm dip},0)} =
%   i \hbar^{2}\, (2\pi)^{3}
%   \displaystyle\frac{e}{\mu c}\;
%   Z_{\rm eff}^{\rm (dip, 0)}\;
%   \displaystyle\sum\limits_{\alpha=1,2} \vb{e}^{(\alpha)} \cdot \vb{I}_{1}, &
%
%   \vb{I}_{1} =
%     \biggl\langle\: \Phi_{\rm p - nucl, f} (\vb{r})\; \biggl|\,    e^{-i\, \vb{k}_{\rm ph} \vb{r}}\;
%     \vb{\displaystyle\frac{d}{dr}}  \biggr|\: \Phi_{\rm p - nucl, i} (\vb{r})\: \biggr\rangle.
% \end{array}
% \label{eq.2.13.2.6}
% \end{equation}
%-----------------------------------------------------------------------------------------------------------------------

%-----------------------------------------------------------------------------------------------------------------------
% \subsubsection{Dipole approximation of effective magnetic moment
% Дипольное приближение для эффективного магнитного момента
% \label{sec.2.13.5}}

The effective magnetic moment of system (\ref{eq.app.2.10.4})
% Эффективный магнитный момент системы (\ref{eq.2.10.4}) на стр.~\pageref{eq.2.10.4}
%
%
% \[
% \begin{array}{lll}
%   \textbf{M}_{\rm eff} (\vb{k}_{\rm ph}, \vb{r}) =
%   e^{i\, \vb{k_{\rm ph}} \vb{r}}\, \displaystyle\frac{1}{m_{\rm p}}
%   \Bigl[
%     e^{-i\, c_{A} \vb{k_{\rm ph}} \vb{r}}\, \displaystyle\frac{m_{A}}{m_{\rm p} + m_{A}}\, m_{\rm p}\, \vb{F}_{{\rm p},\, {\rm mag}} -
%     e^{i\, c_{\rm p} \vb{k_{\rm ph}} \vb{r}}\,  \displaystyle\frac{m_{\rm p}}{m_{\rm p} + m_{A}}\, m_{A}\, \vb{F}_{A,\, {\rm mag}}
%   \Bigr].
% \end{array}
% % \label{eq.2.10.4}
% \]
%
in the first approximation $\exp(i\vb{k_{\rm ph}} \vb{r}) \to 1$ (we call it as dipole in application to the effective magnetic moment) obtains a form
% в первом приближении $\exp(i\vb{k_{\rm ph}} \vb{r}) \to 1$ (т.~е. при $\vb{k_{\rm ph}} \vb{r} \to 0$, которое назовем \definition{дипольным} в применении к магнитному моменту) приобретет вид:
%
\begin{equation}
\begin{array}{lll}
  \vb{M}_{\rm eff} (\vb{k}_{\rm ph}, \vb{r}) =
  \displaystyle\frac{1}{m_{\rm p}}
  \displaystyle\frac{m_{\rm p}\, m_{A}}{m_{\rm p} + m_{A}}\,
  \Bigl[ \vb{F}_{{\rm p},\, {\rm mag}} - \vb{F}_{A,\, {\rm mag}} \Bigr] =

  \displaystyle\frac{\mu}{m_{\rm p}}\, \Bigl[ \vb{F}_{{\rm p},\, {\rm mag}} (\vb{k}_{\rm ph}) - \vb{F}_{A,\, {\rm mag}} (\vb{k}_{\rm ph}) \Bigr].
\end{array}
\label{eq.2.13.5.1}
\end{equation}
In such an approximation, the effective magnetic moment is not dependent on relative distance between proton and nucleus.
Using Eq.~(\ref{eq.app.2.9.7}) for form factor of nucleus $\vb{F}_{A,\, {\rm mag}} (\vb{k}_{\rm ph})$, we write
%
% Мы вновь получили в таком приближении независимость эффективного момента от относительного расстояния между протоном и ядром.
% Используя явный вид (\ref{eq.2.9.7}) для форм-фактора ядра $\vb{F}_{A,\, {\rm mag}} (\vb{k}_{\rm ph})$ [см. стр.~\pageref{eq.2.9.7}], считая его независимым от расстояний нуклонов к центру ядра, запишем:
%
\begin{equation}
\begin{array}{lll}
\vspace{1mm}
  \vb{F}_{{\rm p},\, {\rm mag}} & = &
    \Bigl\langle \psi_{{\rm p}, f} (\beta_{\rm p})\, \Bigl|\,
      \mu_{\rm p}\, \sigmabf
    \Bigr| \psi_{{\rm p}, i} (\beta_{\rm p}) \Bigr\rangle =
    \vb{M}_{\rm p}, \\

\vspace{0.2mm}
  \vb{F}_{A,\, {\rm mag}}^{\rm (dip)} & = &
    \displaystyle\frac{1}{m_{A}}
    \displaystyle\sum_{j=1}^{A}
    \Bigl\langle \psi_{\rm nucl, f} (\beta_{A})\, \Bigl|\,
        \mu_{j}\, m_{Aj}\; e^{-i\, \vb{k_{\rm ph}} \rhobf_{Aj}}\, \sigmabf
    \Bigr| \psi_{\rm nucl, i} (\beta_{A}) \Bigr\rangle \to \\
  & \to &
    \displaystyle\frac{1}{m_{A}}
    \displaystyle\sum_{j=1}^{A}
      \Bigl\langle \psi_{\rm nucl, f} (\beta_{A})\, \Bigl|\, \mu_{j}\, m_{Aj}\, \sigmabf \Bigr| \psi_{\rm nucl, i} (\beta_{A}) \Bigr\rangle =
    \vb{M}_{A}.
\end{array}
\label{eq.2.13.5.2}
\end{equation}
Here we introduced magnetic moment of nucleus $\vb{M}_{A}$ (without characteristics of the emitted photon). Write
% Здесь мы ввели магнитный момент ядра $\vb{M}_{A}$ (без включения характеристик излученных фотонов) и записали модификацию магнитного момента протона $\vb{M}_{\rm p}$. Запишем:
%
\begin{equation}
  \vb{M}_{\rm eff}^{\rm (dip,0)} = \displaystyle\frac{\mu}{m_{\rm p}}\, \Bigl[\vb{M}_{\rm p} - \vb{M}_{A} \Bigr],
\label{eq.2.13.5.3}
\end{equation}
\begin{equation}
\begin{array}{lll}
  M_{p}^{(M,\, {\rm dip},0)} & = &
  -\, \hbar\, (2\pi)^{3}\, \displaystyle\frac{\mu_{N}\, m_{\rm p}}{\mu}\;
  \vb{M}_{\rm eff}^{\rm (dip, 0)}
  \displaystyle\sum\limits_{\alpha=1,2} \Bigl[ \vb{I}_{1} \times \vb{e}^{(\alpha)} \Bigr].
\end{array}
\label{eq.2.13.5.4}
\end{equation}
%
% This matrix element is calculated in Appendix~\ref{sec.app.matr_el.coh_mag}.
% According to formulas (\ref{eq.app.matr_el.coh_mag.7})--(\ref{eq.app.matr_el.coh_mag.8}) in p.~\pageref{eq.app.matr_el.coh_mag.8}, we obtain:
%
% \textcolor[rgb]{1.00,0.00,0.00}{\textbf{Вычисление этого матричного элемента приведено в Приложении~\ref{sec.app.matr_el.coh_mag} (см. стр.~\pageref{sec.app.matr_el.coh_mag}).
% Согласно формулам (\ref{eq.app.matr_el.coh_mag.8})--(\ref{eq.app.matr_el.coh_mag.9}) на стр.~\pageref{eq.app.matr_el.coh_mag.8}, получаем:
% }}
%
Following to logic of calculation of the matrix element $\vb{D}_{A1,\, {\rm el}} (\vb{k}_{\rm ph})$ is given in Appendic~C in Ref.~\cite{Maydanyuk_Zhang.2015.PRC},
we calculate
\begin{equation}
\begin{array}{lll}
  M_{p}^{(M,\, {\rm dip})} =
  \hbar\, (2\pi)^{3}\, \displaystyle\frac{\mu_{N}}{\mu} \cdot \alpha_{M} \cdot
  (\vb{e}_{\rm x} + \vb{e}_{\rm z})\,
  \displaystyle\sum\limits_{\alpha=1,2} \Bigl[ \vb{I}_{1} \times \vb{e}^{(\alpha)} \Bigr], &

  \alpha_{M} =
  \Bigl[
    Z_{\rm A} (\vb{k}_{\rm ph})\: m_{p}\, \bar{\mu}_{\rm pn} -
    z_{\rm p}\, m_{A}\, \mu_{\rm p}
  \Bigr] \cdot
  \displaystyle\frac{m_{p}}{m_{p} + m_{A}}
\end{array}
\label{eq.2.13.5.5}
% \label{eq.app.matr_el.coh_mag.8}
% \label{eq.app.matr_el.coh_mag.9}
\end{equation}
and
\begin{equation}
\begin{array}{lll}
\vspace{0.7mm}
  \vb{M}_{\rm eff} (\vb{k}_{\rm ph}, \vb{r}) & = &
  e^{i\, \vb{k_{\rm ph}} \vb{r}}\,
  \Bigl[
    e^{-i\, c_{A} \vb{k_{\rm ph}} \vb{r}} \cdot m_{A}\, \mu_{\rm p} \cdot z_{\rm p} -
    e^{i\, c_{p} \vb{k_{\rm ph}} \vb{r}} \cdot m_{p}\, \bar{\mu}_{\rm pn} \cdot Z_{\rm A} (\vb{k}_{\rm ph})
  \Bigr] \cdot
  \displaystyle\frac{m_{p}}{m_{p} + m_{A}}\; (\vb{e}_{\rm x} + \vb{e}_{\rm z}), \\

  \vb{M}_{\rm eff}^{\rm (dip)} (\vb{k}_{\rm ph}) & = &
  \Bigl[
    z_{\rm p}\, m_{A}\, \mu_{\rm p} -
    Z_{\rm A} (\vb{k}_{\rm ph})\: m_{p}\, \bar{\mu}_{\rm pn}
  \Bigr] \cdot
  \displaystyle\frac{m_{p}}{m_{p} + m_{A}}\; (\vb{e}_{\rm x} + \vb{e}_{\rm z}).
\end{array}
\label{eq.app.matr_el.coh_mag.7}
\end{equation}
% *******************************************************************************************************************

% *******************************************************************************************************************
\subsection{Matrix element of incoherent emission of magnetic type basing on $M_{\Delta M}$
\label{sec.2.15}}

We calculate the matrix element $M_{\Delta M}$ in Eq.~(\ref{eq.app.2.12.7}).
As functions $D_{A 1,\, {\rm mag}}$, $D_{A 2,\, {\rm mag}}$ do not depend on variable $\vb{r}$, we rewrite formula in Eq.~(\ref{eq.app.2.12.7}) as multiplication of two independent integrals as
\begin{equation}
\begin{array}{lll}
\vspace{1.0mm}
  M_{\Delta M} & = &
  -\, i\, (2\pi)^{3}\: \mu_{N}
  \displaystyle\sum\limits_{\alpha=1,2}\;
  \Bigl\{
  \Bigl\langle
    \Phi_{\rm p - nucl, f} (\vb{r})\; \Bigl|\, e^{i\, c_{\rm p}\, \vb{k_{\rm ph}} \vb{r}}\, \Bigr|\, \Phi_{\rm p - nucl, i} (\vb{r})\:
  \Bigr\rangle\,
  \Bigl( D_{A 1,\, {\rm mag}} (\vb{e}^{(\alpha)}) - D_{A 2,\, {\rm mag}} (\vb{e}^{(\alpha)}) \Bigr)
  \Bigr\}.
\end{array}
\label{eq.2.15.1}
\end{equation}
% -----------------------------------------------------------------------------------------------------------------------
%
% -----------------------------------------------------------------------------------------------------------------------
% После суммирования по спиновым состояниям, для четного числа спиновых состояний мы получим:
After summation over spin states, for even number of spin states we obtain:
\begin{equation}
\begin{array}{lll}
  D_{A 2,\, {\rm mag}} (\vb{e}^{(\alpha)}) = 0
\end{array}
\label{eq.2.15.2}
\end{equation}
%
% и матричный элемент (\ref{eq.2.15.1}) упрощается как
and the matrix element (\ref{eq.2.15.1}) is simplified as
\begin{equation}
\begin{array}{lll}
\vspace{1.0mm}
  M_{\Delta M} & = &
  -\, i\, (2\pi)^{3}\: \mu_{N}
  \displaystyle\sum\limits_{\alpha=1,2}\;
  \Bigl\{
  \Bigl\langle
    \Phi_{\rm p - nucl, f} (\vb{r})\; \Bigl|\, e^{i\, c_{\rm p}\, \vb{k_{\rm ph}} \vb{r}}\, \Bigr|\, \Phi_{\rm p - nucl, i} (\vb{r})\:
  \Bigr\rangle \cdot
  D_{A 1,\, {\rm mag}} (\vb{e}^{(\alpha)})
  \Bigr\}.
\end{array}
\label{eq.2.15.3}
\end{equation}
% -----------------------------------------------------------------------------------------------------------------------
%
% -----------------------------------------------------------------------------------------------------------------------
We calculate summation in form factors for even-even nuclei
\begin{equation}
\begin{array}{lll}
  \displaystyle\sum\limits_{\alpha=1,2} D_{A 1,\, {\rm mag}} (\vb{e}^{(\alpha)}) & = &
  -\, \displaystyle\frac{\hbar\, (A-1)}{2\,A}\; \bar{\mu}_{\rm pn}\, k_{\rm ph} \cdot Z_{\rm A} (\vb{k}_{\rm ph}),
\end{array}
\label{eq.2.15.4}
\end{equation}
%
% где $\bar{\mu}_{\rm pn} = \mu_{\rm n} + \kappa\,\mu_{\rm n}$,
% $\kappa = (A-N)/N$,
% $A$ и $N$ --- числа нуклонов и нейтронов в ядре.
% Также имеем решение:
where $\bar{\mu}_{\rm pn} = \mu_{\rm n} + \kappa\,\mu_{\rm n}$,
$\kappa = (A-N)/N$,
$A$ and $N$ are numbers of nucleons and neutrons in nucleus.
Note solution:
\begin{equation}
\begin{array}{lll}
  \vb{D}_{A 1,\, {\rm el}} = &
    \displaystyle\sum\limits_{j=1}^{A-1}
      \displaystyle\frac{z_{j} m_{\rm p}}{m_{Aj}}\,
      \Bigl\langle \psi_{A, f} (\beta_{A})\, \Bigl|\,
        e^{-i \vb{k}_{\rm ph} \rhobf_{Aj}} \vb{\tilde{p}}_{Aj}
      \Bigr|\,  \psi_{A, j} (\beta_{A}) \Bigr\rangle =
  \displaystyle\frac{\hbar}{2}\; \displaystyle\frac{A-1}{A}\; \vb{k}_{\rm ph}\; Z_{\rm A} (\vb{k}_{\rm ph}).
\end{array}
\label{eq.2.15.5}
\end{equation}
%-----------------------------------------------------------------------------------------------------------------------
%
%-----------------------------------------------------------------------------------------------------------------------
Using such a formulation, we find final solution for the matrix element (\ref{eq.2.15.3})
[assuming $\bar{s}_{k} = s_{k}$]:
\begin{equation}
\begin{array}{lll}
  M_{\Delta M} & = &
  i\, \hbar\, (2\pi)^{3}\, \mu_{N}\, f_{1} \cdot |\vb{k}_{\rm ph}| \cdot Z_{\rm A} (\vb{k}_{\rm ph}) \cdot I_{2},
  % \Bigl\langle \Phi_{\rm p - nucl, f} (\vb{r})\; \Bigl|\, e^{i\, c_{\rm p}\, \vb{k_{\rm ph}} \vb{r}}\, \Bigr|\, \Phi_{\rm p - nucl, i} (\vb{r})\: \Bigr\rangle,
\end{array}
\label{eq.2.15.6}
\end{equation}
where
\begin{equation}
\begin{array}{lll}
  I_{2} = \Bigl\langle \Phi_{\rm p - nucl, f} (\vb{r})\; \Bigl|\, e^{i\, c_{\rm p}\, \vb{k_{\rm ph}} \vb{r}}\, \Bigr|\, \Phi_{\rm p - nucl, i} (\vb{r})\: \Bigr\rangle, &
  f_{1} = \displaystyle\frac{A-1}{2A}\: \bar{\mu}_{\rm pn}.
\end{array}
\label{eq.2.15.7}
\end{equation}
In similar way, we obtain
(taking orthogonality of vectors $\vb{e}^{(\alpha)}$ and $\vb{k}_{\rm ph}$ into account):
\begin{equation}
\begin{array}{llllllll}
  \vb{e}^{(\alpha)} \cdot \vb{D}_{A 1,\, {\rm el}} = 0, &
  \vb{e}^{(\alpha)} \cdot \vb{D}_{A 2,\, {\rm el}} = 0, &
  M_{\Delta E} = 0.
\end{array}
\label{eq.2.14.7}
\end{equation}

% *******************************************************************************************************************

% *******************************************************************************************************************
\subsection{Matrix element of incoherent emission of magnetic type on the basis of $M_{k}$
\label{sec.2.16}}

We calculate the matrix element $M_{k}$ in Eq.~(\ref{eq.app.2.12.5}).
As functions $\vb{D}_{{\rm p},\, {\rm k}}$, $\vb{D}_{A,\, {\rm k}}$ do not depend on variable $\vb{r}$, we rewrite Eq.~(\ref{eq.app.2.12.5}) as multiplication of two independent integrals as
\begin{equation}
\begin{array}{lcl}
  M_{k} & = &
  i\, \hbar\, (2\pi)^{3}\, \mu_{N}\,
  \displaystyle\sum\limits_{\alpha=1,2}
    \bigl[ \vb{k_{\rm ph}} \cp \vb{e}^{(\alpha)} \bigr]\; \times \\

  & \times &
  \Bigl\{
    \vb{D}_{{\rm p},\, {\rm k}}\,
    \Bigl\langle
      \Phi_{\rm p - nucl, f} (\vb{r})\; \Bigl|\,
      e^{- i\, c_{A}\, \vb{k_{\rm ph}} \vb{r}}\; \Bigl|\,
      \Phi_{\rm p - nucl, i} (\vb{r})\;
    \Bigr\rangle +

    \vb{D}_{A,\, {\rm k}}\,
    \Bigl\langle
      \Phi_{\rm p - nucl, f} (\vb{r})\; \Bigl|\,
      e^{i\, c_{\rm p}\, \vb{k_{\rm ph}} \vb{r}}\; \Bigl|\,
      \Phi_{\rm p - nucl, i} (\vb{r})\;
    \Bigr\rangle
  \Bigr\}.
\end{array}
\label{eq.2.16.1}
\end{equation}
% -----------------------------------------------------------------------------------------------------------------------
%
% -----------------------------------------------------------------------------------------------------------------------
Following formalism in Ref.~\cite{Maydanyuk_Zhang.2015.PRC}, we obtain:
% This form factor is calculated in Appendix~\ref{sec.app.matr_el.k.1} [see Eq.~(\ref{eq.app.matr_el.k.1.4.5}) in p.~\pageref{eq.app.matr_el.k.1.4.5}]:
% Расчет этих форм-факторов приведен в Приложении~\ref{sec.app.matr_el.k.1}. Согдасно (\ref{eq.app.matr_el.k.1.4.5}) и (\ref{eq.app.matr_el.k.1.4.6}) [см. стр.~\pageref{eq.app.matr_el.k.1.4.5}], имеем:
%
\begin{equation}
\begin{array}{lll}
  \displaystyle\sum\limits_{\alpha=1,2} \bigl[ \vb{k_{\rm ph}} \cp \vb{e}^{(\alpha)} \bigr] \cdot \vb{D}_{A,\, {\rm k}} =
  - k_{\rm ph}\; \bar{\mu}_{\rm pn}\; Z_{\rm A} (\vb{k}_{\rm ph}), &

  \displaystyle\sum\limits_{\alpha=1,2} \bigl[ \vb{k_{\rm ph}} \cp \vb{e}^{(\alpha)} \bigr] \cdot \vb{D}_{p,\, {\rm k}} =
  - k_{\rm ph}\; \mu_{\rm p}\; z_{\rm p}.
\end{array}
\label{eq.2.16.3}
% \label{eq.app.matr_el.k.1.4.5}
\end{equation}
Taking these properties into account, from Eq.~(\ref{eq.2.16.1}) we obtain
for incoherent magnetic emission
\begin{equation}
\begin{array}{lcl}
  M_{k} & = &
  -\, i\, \hbar\, (2\pi)^{3}\, \mu_{N}\: k_{\rm ph}\:
    \mu_{\rm p}\: z_{\rm p} \cdot I_{3} -
    % \Bigl\langle \Phi_{\rm p - nucl, f} (\vb{r})\; \Bigl|\, e^{- i\, c_{A}\, \vb{k_{\rm ph}} \vb{r}}\; \Bigl|\, \Phi_{\rm p - nucl, i} (\vb{r})\; \Bigr\rangle -
  \displaystyle\frac{\bar{\mu}_{\rm pn}}{f_{1}} \cdot M_{\Delta M},
\end{array}
\label{eq.2.16.5}
\end{equation}
where
\begin{equation}
\begin{array}{lll}
  I_{3} =
  \Bigl\langle
      \Phi_{\rm p - nucl, f} (\vb{r})\; \Bigl|\,
      e^{- i\, c_{A}\, \vb{k_{\rm ph}} \vb{r}}\; \Bigl|\,
      \Phi_{\rm p - nucl, i} (\vb{r})\;
  \Bigr\rangle, &

  \displaystyle\frac{\bar{\mu}_{\rm pn}}{f_{1}} =
  % \displaystyle\frac{\bar{\mu}_{\rm pn}}{\displaystyle\frac{A-1}{2A}\: \bar{\mu}_{\rm pn}} =
  \displaystyle\frac{2\, A}{A-1}.
\end{array}
\label{eq.2.16.6}
\end{equation}
% *******************************************************************************************************************

% *******************************************************************************************************************
%-----------------------------------------------------------------------------------------------------------------------

\end{document}